\magnification=1200
\parskip=10pt plus 5pt
\parindent=14pt
\baselineskip=18pt
\input mssymb
\def\crosspart{{\not\mathrel{}\mkern-0.1mu}}
\pageno=0
\footline={\ifnum \pageno <1 \else \hss \folio \hss \fi}
\line{\hfil{DAMTP-R/94/49}}
\line{\hfil{(Updated Version)}}
\line{\hfil{January, 1996~~~~~}}
\vskip .3in
\centerline{\bf CONNECTIONS AND GENERALIZED GAUGED TRANSFORMATIONS}
\vskip .3in
\centerline{Simon Davis}
\vskip .3in
\centerline{Department of Applied Mathematics and Theoretical Physics}
\vskip 1pt
\centerline{University of Cambridge}
\vskip 1pt
\centerline{Silver Street, Cambridge CB3 9EW}
\vskip .3in
{\bf Abstract.}  With the standard fibre being a coset manifold, the
transformation of a connection form in a fibre bundle under the action of the
isometry group includes a dependence on the fibre coordinate.  Elimination of
the fibre coordinate from the transformation rule implies that the standard
fibre is a Lie group and that the bundle is a principal bundle.  The dependence
on the fibre coordinate is considered in the examples of the SO(4) action on
an $S^3$ bundle and the SO(8) action on an $S^7$ bundle.  The nonlinear SU(4)
action on an $S^7$ bundle is applied to the dimensional reduction of
11-dimensional supergravity and ten-dimensional superstring theory to four
dimensions.  A principle, consistent with higher-dimensional superstring
theory,
is suggested to explain the types of gauge interactions that arise in the
standard model based on the geometry of the internal symmetry spaces.  It is
shown why a Lie group structure is required for vector bosons in pure gauge
theories and that the application of division algebras to force unification
must
begin with the fermions comprising the elementary particle multiplets of the
standard model. Gauge transformations in quantum principal bundles, using
generalizations of left and right multiplication and connection forms, are
shown to satisfy conditions similar to those in classical gauge theories.
\vfill
\eject

\noindent {\bf 1. Introduction}

It has recently been shown that superstring theory possesses improved
finiteness properties at each order [1][2][3] and at large orders in the series
expansion for scattering amplitudes [4], suggesting that a consistent quantum
theory containing general relativity in the low-energy limit has been obtained.
 Phenomenologically relevant models of elementary particle
interactions might also be included for a suitable choice of superstring
vacuum [5].  A principle for selecting a specific superstring vacuum would
be needed as a theoretical basis for any choice of ground state.  One
possibility, motivated by topology change arising in the path integral
for quantum gravity [6], would involve an integration over different
background geometries, dominated by a sum over solutions to the string
equations of
motion, with the geometries  weighted by a factor involving the
analogue of an action on the space of renormalizable 2-dimensional field
theories, similar to the
\hfil\break
Zamalodchikov c-function [7], or a string field theory
action based on the noncommutative star product [8].  Alternatively, one
might wish to consider a principle directly constraining geometries
associated with a Kaluza-Klein unification of general relativity with
the elementary interactions.

The attempt to unify the elementary particle interactions with gravity has
gradually led to the development of new theories, which may be regarded as
generalizations of standard gauge theories [8][9][10][11].  String field
theory,
with a BRST symmetry group, and rational conformal field theories with a
quantum group symmetry, are examples of theories involving generalizations of
the Yang-Mills gauge group.

A formulation of generalized gauged theories can be achieved by considering a
larger category of bundles than the principal bundles.
In a previous investigation [12], a physical requirement imposed on the
transformation rule of the connection form for a general bundle led to
constraints on the types of geometries corresponding to Yang-Mills theories.
Furthermore, the category of principal bundles is selected uniquely by these
conditions.

One consequence of this result is that bundles with an $S^7$ fibre do not
directly correspond to Yang-Mills theories.  This provides an explanation for
the lack of gauge invariance of an octonionic generalization of Yang-Mills
theory [13] and the absence of a gauge principle in anti-symmetric tensor
theories based on the octonion algebra [14].  This is particularly relevant
for an approach to force unification utilizing an identification of the
internal symmetries with the different division algebras [15][16] and
associating the number of types of elementary particle interactions with the
number of division algebras, related mathematically to the parallelizability of
the spheres [17][18][19].

As the bundles with $S^3$ and $S^7$ fibres are of interest
theoretically, a closer study of transformations of the connection form
for bundles with the structure groups given by the isometry groups of
these spheres is initiated in this paper.  Several results are obtained
with regard to the fibre coordinate dependence of the transformation rule
of the connection form for these bundles.  Special use is made of the
properties of the fibres in the analysis of the generalized gauge
 transformations.  Similar considerations are applied to the newly developed
quantum principal bundles [20][21].  Implications for theories with generalized
gauge invariance are discussed.

\noindent {\bf 2. Gauge Potentials and Connections in Fibre Bundles}

Yang-Mills theories can be interpreted  mathematically in terms of principal
fibre bundles $(P, \pi, M)$ with fibre F being diffeomorphic to the structure
group G.  A choice of connection determines the decomposition of the tangent
space $T_p$, at the point p in the bundle space P, into a vertical subspace
$V_p$, which is isomorphic to the Lie algebra of G, and a horizontal subspace
$H_p$, which is invariant under the right action of the group.
In the formulation of Dieudonne [22], a principal connection
C: $T(M)\times P\to T(P)$
satisfies
$$\eqalign{(i)&~~T(\pi)(C(h_x, p_x))~=~h_x~~~~~o_P(C(h_x, p_x))~=~p_x~~~~~
x\in M,~h_x\in T_x(M),~p_x\in P_x
\cr
(ii)&~~h_x\to C(h_x, p_x)~is~linear~in~h_x
\cr
(iii)&~~C(h_x, p_x\cdot a)~=~C(h_x, p_x)\cdot a
\cr}
\eqno(1)$$
It may also be regarded as a choice of horizontal subspace $H_p \subset T_p(P)$
such that
$$\eqalign{(i)&~~T_p~=~V_p\oplus H_p
\cr
(ii)&~~\forall~ a \in G,~p\in P~~(R_a)_\ast H_p~=~H_{pa}
\cr
(iii)&~~H_p~depends~differentiably~on~p
\cr}
\eqno(2)$$
There is a basis for the Lie algebra ${\cal G}~=~T_e(G)$ consisting of
left-invariant vector fields $\xi_i$.  The basis $\{\xi_i\}$ can be mapped onto
a basis $\{\xi_i^\ast\}$ of $V_p$ since $e^{t\xi_i}$ is a one-parameter
subgroup of G, which acts on P, the bundle space, by right multiplication
$$(\xi_i^\ast)_p\cdot f~=~{d\over {dt}}f(p_t)\bigr|_{t=0}
\eqno(3)$$
The connection 1-form is a surjective linear mapping of T(P) onto ${\cal G}$.
$$\omega_p:\zeta_p~\to~t_p^{-1}(\zeta_p~-~C(T(\pi)\cdot \zeta_p, p))~~~~~
t_p:\xi_i\to (\xi_i^\ast)_p
\eqno(4)$$
If $r_x \in P$ such that $\pi(r_x)~=~x$ and $s \in G$, then
$$\eqalign{C(T(\pi)\cdot (\zeta_{r_x}\cdot s), r_x\cdot s)
{}~&=~C(T(\pi)\cdot \zeta_{r_x\cdot s}, r_x\cdot s)
\cr
{}~&=~C(T(\pi)\cdot \zeta_{r_x},
r_x\cdot s)
{}~=~C(T(\pi)\cdot \zeta_{r_x}, r_x)\cdot s
\cr}
\eqno(5)$$
As $t_{r_x\cdot s}(u)~=~(t_{r_x}(ad(s)\cdot u))\cdot s$,
$$\omega(r_x\cdot s)(\zeta_{r_x}\cdot s)~=~t_{r_x\cdot s}^{-1}(\zeta_{r_x\cdot
s}~-~
C(T(\pi)\cdot \zeta_{r_x\cdot s}, r_x\cdot s))
{}~=~ad(s^{-1})\omega(r_x)\cdot \zeta_{r_x}
\eqno(6)$$
and $(R_a)^\ast \omega~=~ad(a^{-1})\omega$.

For a general bundle $(E, M, \pi, F)$ and a $C^\infty$ section s of E over
$x \in M$, the connection $C(h_x, s(x))$ satisfies the conditions (1), and
since E can be identified with the trivial bundle $M \times F$ locally, in a
neighbourhood of x,
$$C(h_x, s(x))=C(h_x, y)=((x,y),~(h_x, C_x(h_x,y)))~~~~~~~~~~~~~~y \in F
\eqno(7)$$
where $C_x: T_x(M) \times F \to T(M \times F)\vert_x$, which is a bilinear
mapping when F is a vector space.  The covariant derivative then can be defined
in terms of sections of E
$$\tau_{s(x)}^{-1}(\nabla_{h_x}\cdot s)~=~T_x(s)\cdot h_x~-~C(h_x,s(x))
\in T_{s(x)}(E)
\eqno(8)$$
where $\tau_{s(x)}: T_{s(x)}(E_x) \to E_x$ is a tangent space mapping.

Returning to the consideration of principle bundles, $\omega$ takes its values
in ${\cal G}$ so that
\hfil\break
 $\omega~=~\omega^i
\xi_i$ where $\omega^i$ are real-valued 1-forms.
 The existence of a global
section is equivalent to trivializability of the bundle, while the choice of a
global section is a choice of gauge.  The gauge potential
$A_\mu~=~A_\mu^i\xi_i$, defined on the base manifold M rather than the bundle
space E, are the components of the connection 1-form.  Let
$\{U_\alpha\}$ be an open covering of M such that $\pi^{-1}(U_\alpha)\simeq
U_\alpha\times G$ corresponding to the trivialization $\psi:\pi^{-1}(U_\alpha)
\to U_\alpha \times G$,
$\psi(p)~=~(\psi(p), \phi(p)),~\phi(pa)~=~\phi(p)\cdot a$.  Then there exist a
preferred set of local sections $\sigma_\alpha(x)~=~p\cdot\phi_\alpha^{-1}(p)
{}~=~\psi_\alpha^{-1}(x,e)$ defining the 1-form
$\omega_\alpha=(\sigma_\alpha)^\ast \omega$ and the gauge potential on M
$$\omega_\alpha(\xi_\mu)~=~\omega((\sigma_\alpha)_\ast\cdot \xi_\mu)~=~
\omega^i((\sigma_\alpha)_\ast\cdot \xi_\mu)\xi_i~=~A_{(\alpha)\mu}^i
\xi_i~~~~~\xi_\mu \in T_x(M)
\eqno(9)$$
A choice of section corresponds to a choice of coordinate system in E and a
gauge transformation implies a change of coordinates.  If $\sigma_\alpha^\prime
(x)~=~\sigma_\alpha(x)\cdot g(x)$, then $A_\mu^{(\sigma^\prime)}~=~
g^{-1}A_\mu^{(\sigma)}g+g^{-1} \partial_\mu g$.  The curvature form is
$\Omega~=~D\omega~=~d\omega+{1\over 2} [\omega, \omega]$.
Again,
\hfil\break
 $\Omega_\alpha~=~(\sigma_\alpha)^\ast \Omega~=~{1\over 2}F_{(\alpha)\mu\nu}
dx^\mu \wedge dx^\nu$ and $F_{\mu\nu}^{(\sigma^\prime)}~=~
g^{-1}F_{\mu\nu}^{(\sigma)} g$.   The Yang-Mills action can then be constructed
$$I~=~{1\over 4}~\int_M~\Omega\wedge \ast \Omega
\eqno(10)$$

Now suppose that $\sigma(x), \sigma^\prime(x)~=~\sigma(x)\cdot g(x) \in E_x$
and $\rho$ is a
representation of GL(F) in an associated vector bundle with total space E
and standard fibre F.  Then
$$\eqalign{\sigma_\ast^\prime \cdot \xi_\mu~&=~(R_{\rho(g)})_\ast
(\sigma_\ast\cdot \xi_\mu)~+~(\sigma(x))_\ast\cdot (L_{\rho(g)\ast}\cdot
\xi_\mu)
\in T_{\sigma^\prime(x)}(E)
\cr
\omega(\sigma_\ast^\prime\cdot \xi_\mu)~&=~\omega(R_{\rho(g)\ast}
\cdot(\sigma_\ast\cdot \xi_\mu))~+~\omega((\sigma(x))_\ast\cdot
(L_{(\rho(g)\ast}\cdot \xi_\mu))
\cr}
\eqno(11)$$
by the linearity of $\omega$, which follows from the linearity of $C(h_x, p_x)$
in the first argument.

Thus, when the fibre F is a vector space on which the structure group
G acts linearly, the connection in the principal bundle can be used to
define a connection in the associated vector bundle E(M, F, G, P).  Given an
exterior
covariant differential $\nabla_X$, the curvature may be obtained through the
commutator $[\nabla_X, \nabla_Y]$.

For an associated bundle with arbitrary fibre F, a connection can be introduced
so that the horizontal lift of any tangent vector $h_x\in T_x(M)$ can be
defined.  For each section $\sigma:M\to E$,
$\sigma_\ast h_x~-~C_x(X_x, \sigma(x)) \in V_{\sigma(x)}(E)$, where the
vertical
subspace $V_{\sigma(x)}(E)$ is the tangent space to the fibre through the point
 $\sigma(x)$ of the bundle E.   By analogy with principal bundles,  a
connection
1-form can be constructed, although it  would be necessary to compare vertical
tangents at different points on the fibre.  When the fibre may be identified
with a group manifold G, group multiplication carries tangent vectors from
one point of G to another in a unique way; all tangent vectors can then be
mapped to the tangent space at the identity element, where they can be
compared.
This allows one to deduce the transformation rule for the Yang-Mills field
$A_\mu$.  A similar method for transporting vectors on fibres which are not
vector spaces or group manifolds would be required for the transformation rule
of the connection form $\omega^{(\sigma)}$ under a change of section.

\noindent{\bf 3. A Modification of the Kaluza-Klein Ansatz}

A new perspective on the derivation of Yang-Mills theories is obtained using
Kaluza-Klein theory.  If the extra dimensions describe a coset manifold
G/H, insertion of the ansatz metric [23] of the total space
$$g_{{\hat {\mu}}{\hat {\nu}}}~=~\left(\matrix{g_{\mu\nu}(x)&
A_\mu^i(x)K_{i\alpha}(y)
                                          \cr
                                        A_\mu^i(x)K_{i\alpha}(y)&
\gamma_{\alpha\beta}(y)\cr}
                      \right)
\eqno(12)$$
where $K_{i\alpha}(y)$ represent the Killing vectors on G/H,
into the action
$$S~=~\int~d^{4+D}z~(det e)~e_{\hat a}^{\hat \mu}~e_{\hat b}^{\hat \nu}
                   ~R_{{\hat \mu}{\hat \nu}}^{[{\hat a}{\hat b}]}
                                                    ~~~~~~~~~~~z=(x,y)
\eqno(13)$$
and integration of the y-coordinates gives an action containing
four-dimensional gravity and Yang-Mills theory with gauge group G.
The general coordinate transformation of the off-diagonal components of the
metric $g_{\mu\alpha}$ implies the gauge transformation in four dimensions
$$A_\mu^i(x)~\to~A_\mu^i(x)~+~\partial_\mu \epsilon^i(x)~+~
{f^i}_{jk} \epsilon^j(x) A_\mu^k(x)
\eqno(14)$$
upon use of the Killing vector equation and the Lie derivative relations
{\it \$}$_{K_i}K_j~=~{f^i}_{jk}K_k$.

Suppose, however, that the off-diagonal components are replaced by $A_\mu^i(x)
E_{i\alpha}(y)$ where $\{E_i\}$ represent smooth non-vanishing, orthonormal
vector fields on G/H, a property which would require parallelizability of the
manifold and which holds for $S^7$.  Following the same procedure with the
vector fields $E_i$, one finds a transformation rule that contains a dependence
on y, because $[E_i, E_j]~=~{f^i}_{jk}(y)E_k$.

A possible resolution of this problem is to consider a fixed point y and
only infinitesimal transformations about this point.  This would imply that the
original action is only given by an integral over x-coordinates, rather than
over all of the dimensions.  Now, under a general coordinate transformation,
the higher-dimensional Lagrangian must vary into a total derivative
$\partial_{\hat \mu} W^{\hat\mu}~=~\partial_\mu W^\mu~+~\partial_\alpha
W^\alpha$.   The integral of this total derivative over all of the
dimensions of the total space would vanish, but it would be non-zero generally
if the integral was performed over the x-coordinates only (unless there exists
a point y on G/H such that $\partial_\alpha W^\alpha(y)~=~0$).  Thus, although
the gauge transformation rule would be consistent under these conditions,
the action would no longer be invariant.

\noindent{\bf 4. Gauge Transformation Constraints for an SO(4) Action on an
               $S^3$ Bundle}

Gauge transformations may be viewed as active transformations on the standard
fibre.  Let $\sigma(x)~=~y\in G$, $\sigma^\prime(x)~=~\sigma(x)\cdot g(x)~=~
y\cdot g\in G$.  If $\xi\in T_x(M)$,
$$\eqalign{\sigma_\ast^\prime \cdot \xi~&=~R_{g\ast}\sigma_\ast \cdot
\xi~+~L_{y\ast} g_\ast \cdot \xi
\cr
{}~&=~R_{g\ast}\sigma_\ast \cdot \xi~+~L_{(y\cdot g)\ast} L_{g^{-1}\ast}
(g_\ast
\cdot \xi)
\cr}
\eqno(15)$$
Given a connection form
$$\eqalign{\omega(\sigma_\ast^\prime\cdot \xi)~=~L_{(y\cdot g)\ast}^{-1} {\cal
V}
(\sigma_\ast^\prime \cdot \xi)~&=~L_{(y\cdot g)\ast}^{-1}
[{\cal V}(R_{g\ast} \sigma_\ast \cdot \xi)~+~{\cal V}(L_{(y\cdot g)\ast}
L_{g^{-1}\ast} g_\ast \cdot \xi)]
\cr
{}~&=~L_{(y\cdot g)\ast}^{-1} R_{g\ast} {\cal V}(\sigma_\ast\cdot \xi)
{}~+~L_{g^{-1}\ast} g_\ast \cdot \xi
\cr}
\eqno(16)$$
where ${\cal V}$ represents the projection onto the vertical subspace
$V_{(x,\sigma(x))}(E)~\sim~V_{\sigma(x)}(E)$.
Since $\omega(\sigma_\ast\cdot \xi)~=~L_{y\ast}^{-1} {\cal V}(\sigma_\ast \cdot
\xi)$
$$\eqalign{\omega(\sigma_\ast^\prime \cdot \xi)~&=~L_{(y\cdot g)\ast}^{-1}
R_{g\ast}
L_{y\ast} \omega(\sigma_\ast\cdot \xi)~+~L_{g^{-1}\ast} g_\ast \cdot \xi
\cr
{}~&=~ad(g^{-1}) \omega(\sigma_\ast\cdot\xi)~+~L_{g^{-1}\ast} g_\ast \cdot \xi
\cr}
\eqno(17)$$
Setting $\xi$ equal to $\partial_\mu$, one recovers the standard gauge
transformation law
$$A_\mu^\prime~=~ad(g^{-1})A_\mu~+~g^{-1}\partial_\mu g
\eqno(18)$$
The dependence on y in equation (16) has disappeared, which is necessary if the
theory is to be formulated in the four-dimensional space-time M.

The transformation group could be made larger than the standard fibre in the
principal bundle.  Consider, for example, a bundle which has standard fibre
$S^3$ and suppose that the group acting on this fibre is enlarged from SU(2)
to SO(4), with SO(4) acting on $S^3$ by right multiplication.
Two different homomorphisms $\iota_R,~ \iota_L: S^3 \to SO(4)$ are defined by
$y~=~(y_0, y_1, y_2, y_3)\in S^3$
$$\iota_R(y)~=~\left(\matrix{y_0& y_1&y_2&y_3
                             \cr
                             -y_1&y_0&-y_3& y_2
                              \cr
                              -y_2&y_3&y_0&-y_1
                              \cr
                              -y_3&-y_2&y_1&y_0
                               \cr}
                                        \right)
{}~~~~
\iota_L(y)~=~\left(\matrix{y_0&-y_1&-y_2&-y_3
                             \cr
                           y_1&y_0&-y_3&y_2
                              \cr
                            y_2&y_3&y_0&-y_1
                              \cr
                            y_3&-y_2&y_1&y_0
                               \cr}
                                       \right)
\eqno(19)$$
so that $y \cdot y^\prime~=~\iota_L(y)~y^\prime$, with $y^\prime$ as a
column vector, and $y^\prime \cdot y~=~ y^{\prime T}~\iota_R(y)$.  Recalling
further that $R_g y^\prime
{}~\equiv~y^\prime \cdot g~\leftrightarrow~y^{\prime T} \cdot R_g^T$, after
identifying the row vector $y^{\prime T}$ with the unit quaternion
$y^\prime \in S^3$,
$L_{(y\cdot g)\ast}^{-1} R_{g\ast} L_{y\ast}$ is the tangent mapping induced by
the SO(4) transformation $\iota_L(y \cdot g)^{-1} R_g \iota_L(y)$, or
equivalently,  $\iota_L(y)^T R_g^T [\iota_L(y\cdot g)^{-1}]^T$ acting
on four-component row vectors by right multiplication.  Since
SO(4) is locally isomorphic to $SU(2)\times SU(2)$,
$$R_g^T~=~\iota_R(g_0) \iota_L(g_0^\prime)^T~for~some~g_0,~g_0^\prime\in S^3
\eqno(20)$$
It then follows that
$$\eqalign{\iota_L(y)^T R_g^T [\iota_L(y\cdot g)^{-1}]^T~&=~
\iota_L(y)^T \iota_R(g_0) \iota_L(g_0^\prime)^T [\iota_L((g_0^\prime \cdot y)
\cdot
                                                            g_0)^{-1}]^T
\cr
{}~&=~\iota_R(g_0) \iota_L(y)^T \iota_L(g_0^\prime)^T
[\iota_L(g_0^\prime \cdot y)^{-1}]^T \iota_L(g_0^{-1})^T
\cr
{}~&=~\iota_R(g_0) \iota_L(g_0^{-1})^T
\cr}
\eqno(21)$$
which is independent of y.

The second term in equation (17), $L_{g^{-1}\ast} g_\ast \cdot \xi$ can be an
arbitrary element of $T_e(SO(4))$.
$$L_{g^{-1}\ast} g_\ast \cdot \xi~=~\left(\matrix{0&c_1&c_2&c_3&
                                               \cr
                                                -c_1&0&-c_6&c_5
                                                \cr
                                                -c_2&c_6&0&-c_4
                                                \cr
                                                 -c_3&-c_5&c_4&0
                                                 \cr}
                                                     \right)
\eqno(22)$$
Let $y\cdot g\equiv y^\prime=(y_0^\prime,y_1^\prime,y_2^\prime, y_3^\prime)$.
Then
$$\eqalign{\lambda_{y_\ast^\prime}\cdot L_{g^{-1}\ast} g_\ast \cdot \xi
{}~=~(-c_1y_1^\prime-c_2y_2^\prime-c_3y_3^\prime,~c_1y_0^\prime+c_6y_2^\prime-
c_5y_3^\prime,~&c_2y_0^\prime-c_6y_1^\prime+c_4y_3^\prime,
\cr
&~~~~~c_3y_0^\prime+c_5 y_1^\prime-c_4y_2^\prime)
\cr}
\eqno(23)$$
where $\lambda_{y^\prime\ast}$ is a map from tangent vectors in $T_e(SO(4))$
to tangent vectors in $T_{y^\prime}(S^3)$ induced by the mapping
$\lambda_{y^\prime}:~g~=~exp(c_I X_I)~\to~ y^\prime \cdot g~=~y^\prime \cdot
exp(c_I X_I)$ with $\{X_I\}$ being the generators of SO(4).
$\lambda_{y^\prime_\ast} \cdot L_{g^{-1}\ast}( g_\ast\cdot \xi) \cdot
\iota_L(y^{\prime  -1})^T$ is only independent of $y^\prime$ if
$c_1~=~c_4,~c_2~=~c_5,~c_3~=~c_6$.
So $L_{g^{-1}\ast}\cdot (g_\ast\cdot \xi)$ must be an element of an SU(2)
subalgebra,
and g(x) must be an element of an SU(2) subgroup to maintain y-independence of
the gauge transformation.

Let $\{X_i\},~i=1,~2,~3$ be the left-invariant basis of vector fields on $S^3$.
A section $\sigma(x)$ can also be regarded as a mapping from $g\in SO(4)$ to
$\sigma(x)\cdot g\in E$, so that it induces a tangent mapping from $T_e(SO(4))$
to $V_{\sigma(x)}(E)$, the vertical subspace of the tangent space of the
bundle.  The connection form is a rule for comparing vertical components of
 tangent vectors at different points on the fibre and is therefore an
isomorphism from $V_{\sigma(x)}(E)$ to $T_e(SO(4))\bigl|_{\{X_i\}}$.  While one
requires $\omega(\sigma_\ast\cdot \xi_\mu)\in T_e(SO(4))\bigl|_{\{X_i\}}$, for
an arbitrary section $\sigma^\prime(x)~=~\sigma(x)\cdot g(x)$,
$ad(g^{-1})[\omega(\sigma_\ast\cdot \xi_\mu)]~+~L_{g^{-1}\ast}(g_\ast\cdot
\xi_\mu)
\notin T_e(SO(4))\bigl|_{\{X_i\}}$.

To maintain $\omega(\sigma^\prime_\ast\cdot \xi_\mu)\in
T_e(SO(4))\bigl|_{\{X_i\}}$,
a compensating gauge transformation is needed.
Since
$$\eqalign{\omega[(\sigma(x)\cdot g(x)h^\prime(x))_\ast\cdot
\xi_\mu]~&=~ad(h^{\prime-1})
[ad(g^{-1})\omega(\sigma_\ast\cdot \xi_\mu)]
\cr
&+~ad(h^{\prime-1})
\omega[\sigma^\prime(x)_\ast
\cdot(L_{g^{-1}(x)\ast}(g_\ast\cdot \xi_\mu))]
+\omega[\sigma^\prime(x)_\ast\cdot(h^\prime_\ast
\cdot \xi_\mu)]
\cr
{}~&=~ad(h^{\prime -1})[\omega((\sigma(x)\cdot g)_\ast\cdot
\xi_\mu)]+\omega[\sigma^\prime(x)\cdot(h^\prime_\ast\cdot \xi_\mu)]
\cr}
\eqno(24)$$
one may define
$$\omega(\sigma^\prime_\ast\cdot\xi_\mu)~=~ad(h^{\prime -1}(\sigma, g))\{
ad(g^{-1})[\omega(\sigma_\ast\cdot \xi_\mu)]~+~L_{g^{-1}\ast}(g_\ast\cdot
\xi_\mu)\}
{}~+~L_{h^{\prime -1}(\sigma,g)\ast}(h^\prime(\sigma,g)_\ast\cdot \xi_\mu)
\eqno(25)$$
where $h^\prime(\sigma,g)$ is an element of the stability group of
$\sigma^\prime(x)$.

\noindent{\bf Theorem.}  There exists a unique $h^\prime(\sigma, g)$ such that
$\omega(\sigma^\prime_\ast\cdot \xi_\mu)\in T_e(SO(4))\bigl|_{\{X_i\}}$.
\hfil\break
{\bf Proof.}\hfil\break
${\underline {Existence}}$.  Let $g_0~=~exp(t_iX_i)$.
$$L_{g_0^{-1}\ast}({g_0}_\ast\cdot \xi_\mu)~=~exp(-t_iX_i)~\partial_\mu t_j~X_j
{}~exp(t_iX_i) \in T_e(SO(4))\bigl|_{\{X_i\}}
\eqno(26)$$
Now $\sigma(x)\cdot exp(t_iX_i)$ covers all of $S^3$.  Thus, if $\sigma^\prime
(x)~=~\sigma(x)\cdot g(x)$, in general, one can write $g(x)~=~exp(t_iX_i)\cdot
h^{\prime -1}$.  Therefore, there exists an $h^\prime$ such that
$ad((gh^\prime)^{-1})[\omega(\sigma_\ast\cdot\xi_\mu)]~+~L_{(gh^\prime)^{-1}}
((gh^\prime)_\ast\cdot\xi_\mu) \in T_e(SO(4))\bigl|_{\{X_i\}}$.  The result
then follows from
$$\eqalign{ad((gh^\prime)^{-1})[\omega(\sigma_\ast\cdot\xi_\mu)]
~+~L_{(gh^\prime)^{-1}\ast}
((gh^\prime)_\ast\cdot\xi_\mu)~&=~ad(h^{\prime
-1}(\sigma,g))\{ad(g^{-1})[\omega(\sigma_\ast\cdot\xi_\mu)]
\cr
&~~~~~~~~~~~~~~~~~~~~+~L_{g^{-1}\ast}(g_\ast\cdot
\xi_\mu)\}
\cr
&~~~~~~~~~~~~+~L_{h^{\prime
-1}(\sigma,g)\ast}(h^\prime(\sigma,g)_\ast\cdot\xi_\mu)
\cr}
\eqno(27)$$

\noindent${\underline {Uniqueness}}$.  Suppose $ad(h^{\prime\prime
-1})\{ad(g^{-1})
[\omega(\sigma_\ast\cdot \xi_\mu)]~+~L_{g^{-1}\ast}(g_\ast\cdot \xi_\mu)\}~+~
L_{h^{\prime\prime -1}\ast}(h^{\prime\prime}\cdot \xi_\mu)
\in T_e(SO(4))\bigl|_{\{X_i\}}$.  Let $h^{\prime\prime}~=~h^\prime h$.
Then this implies $L_{h^{-1}\ast}(h_\ast \cdot \xi_\mu)\in T_e(SO(4))
\bigl|_{\{X_i\}}$.  Take $h~=~exp(t_iX_i~+~{\tilde t}_i {\tilde X}_i)$.
$$\eqalign{L_{h^{-1}\ast}(h_\ast\cdot \xi_\mu)~&=~exp(-(t_i X_i~+~{\tilde t}_i
                                                     {\tilde X}_i))
                                    (\partial_\mu t_j X_j~+~\partial_\mu
                                         {\tilde t}_j {\tilde X}_j)
                                      exp(t_i X_i~+~{\tilde t}_i {\tilde X}_i)
\cr
{}~&=~(\partial_\mu t_j) exp(-(t_iX_i~+~{\tilde t}_i{\tilde X}_i)) X_j
    exp(t_i X_i~+~{\tilde t}_i {\tilde X}_i)
\cr
 ~&+~(\partial_\mu {\tilde t}_j) exp(-(t_i X_i~+~{\tilde t}_i {\tilde X}_i))
      {\tilde X}_j exp(t_i X_i~+~{\tilde t}_i {\tilde X}_i)
\cr}
\eqno(28)$$
As the generators of the two SU(2) subgroups commute, $[X_i, {\tilde
X}_j]~=~0$,
$$L_{h^{-1}\ast}(h_\ast\cdot \xi_\mu)~=~\partial_\mu t_j exp(-t_i X_i) X_j
exp(t_i X_i)~+~\partial_\mu {\tilde t}_j exp(-{\tilde t}_i {\tilde X}_i)
{\tilde X}_j exp({\tilde t}_i {\tilde X}_i)
\eqno(29)$$
and $L_{h^{-1}\ast}(h_\ast\cdot \xi_\mu) \in T_e(SO(4))\bigl|_{\{X_i\}}$
if and only if
$\partial_\mu {\tilde t}_j~=~0$.
We require that the relation
$$\eqalign{\omega(\sigma^\prime_\ast \cdot \xi_\mu)~=~ad(h^{\prime\prime -1})&
  \{ad(g^{-1})[\omega(\sigma_\ast\cdot\xi_\mu)]
\cr
{}~&+~L_{g^{-1}\ast}(g_\ast\cdot \xi_\mu)\}
{}~+~L_{h^{\prime\prime -1}\ast} (h^{\prime\prime}_\ast\cdot \xi_\mu)
\in T_e(SO(4)) \bigl|_{\{X_i\}}
\cr}
\eqno(30)$$
holds not just at a single point $x_0 \in M$ but for all x in some
neighbourhood
U of $x_0$.  Therefore $\partial_\mu {\tilde t}_j~=~0$ must hold in all of U,
implying that ${\tilde t}_j(x)~=~{\tilde t}_j(x_0)~=~constant$ in U
and $h(x)~=~exp(t_i(x) X_i~+~{\tilde t}_i(x_0){\tilde X}_i)$.
Both $h^\prime(x)$ and $h^{\prime\prime}(x)$  must stabilize $\sigma^\prime(x)
{}~=~\sigma(x)\cdot g(x)~\forall x\in U$ so that
$\sigma^\prime(x)\cdot h^{\prime\prime}(x)~=~\sigma^\prime(x) exp(t_i(x)X_i
{}~+~{\tilde t}_i(x_0) {\tilde X}_i)$.  Defining $\sigma^\prime(x)~=~
\sigma(x_0) exp(s_j(x) X_j),~s_j(x_0)~=~0$ where $\sigma(x_0)
exp(t_i(x_0)X_i~+~{\tilde t}_i(x_0){\tilde X}_i)~=~\sigma(x_0)$, it follows
that
$$\sigma^\prime(x)~exp(t_i(x)X_i~+~{\tilde t}_i(x_0) {\tilde X}_i)~=~
\sigma(x_0)~exp(-t_i(x_0) X_i)~exp( s_j(x) X_j)~exp(t_i(x) X_i)
\eqno(31)$$
Thus,
$$exp(-t_i(x_0)X_i)~exp(s_j(x)X_j)~exp(t_i(x)X_i)~=~ exp(s_j(x)X_j)
\eqno(32)$$
is required for the path $\sigma^\prime(x)~=~\sigma(x_0) exp(s_j(x) X_j)$ to be
stabilized by $h^{\prime\prime}(x)$.
Since
$$\eqalign{e^{s_j(x)X_j}e^{t_i(x)X_i}~&=~exp[(s_i(x)~+~t_i(x))X_i~+~{1\over 2}
                                                           s_i(x)
                                    t_j(x) c_{ijk}X_k~
\cr
&~~~~~~~~~+~{1\over {12}} s_i(x)
                                     s_j(x) t_k(x) c_{jkl}c_{ilm} X_m
\cr
&~~~~~~~~~+~
                                    {1\over {12}} s_i(x) t_j(x) t_l(x) c_{ijk}
                                     c_{klm} X_m~+~...~]
\cr
e^{t_j(x_0) X_j} e^{s_i(x)X_i}~&=~exp[(s_i(x)~+~t_i(x_0))X_i~+~{1\over 2}
                                       t_i(x_0)s_j(x)c_{ijk} X_k
\cr
&~~~~~~~~~+~
                                       {1\over {12}}t_i(x)t_j(x_0) s_k(x)
                                        c_{jkl}c_{ilm} X_m
\cr
&~~~~~~~~~+~{1\over {12}}t_i(x_0)s_j(x) s_l(x)
                                        c_{ijk}c_{klm} X_m~+~...~]
\cr}
\eqno(33)$$
Equating the powers gives
$$\eqalign{t_m(x)~+~{1\over 2}s_i(x)t_j(x)c_{ijm}~&+~{1\over
{12}}(s_i(x)t_j(x)t_l(x)
{}~+~t_i(x)s_j(x)s_l(x)) c_{ijk}c_{klm}~+~...
\cr
{}~&=~t_m(x_0)~-~{1\over 2}s_i(x)t_j(x_0) c_{ijm}~+~{1\over {12}}
   (s_i(x)t_j(x_0)t_l(x_0)
\cr
&~~~+~t_i(x_0)s_j(x)s_l(x))c_{ijk}c_{klm}~+~...
\cr}
\eqno(34)$$
Assuming that $s_i(x)$ is small in the neighbourhood U of $x_0$ and equating
terms with the same power of $s_i(x)$ gives
$$\eqalign{t_m(x)~&=~t_m(x_0)
\cr
{1\over 2}s_i(x)t_j(x) c_{ijm}~+~{1\over {12}}s_i(x)t_j(x)t_l(x) c_{ijk}
c_{klm}
{}~+~...~&=~-{1\over 2} s_i(x) t_j(x_0) c_{ijm}
\cr
&~~~~~~+~{1\over {12}}s_i(x) t_j(x_0)
           t_l(x_0) c_{ijk} c_{klm}
\cr
&~~~~~~+~...
\cr
&~\vdots
\cr}
\eqno(35)$$
Substituting the first equation into the second implies that
$s_i(x)t_j(x_0) c_{ijm}~+~...~=~0$ or equivalently that
$M_{ij}(x)t_j(x_0)~=~0$,
where $M_{ij}(x)$ is a matrix of rank 3, and consequently, $t_j(x_0)~=~0$.
Moreover, $exp({\tilde t}_i(x_0) {\tilde X}_i)$ does not stabilize
$\sigma^\prime(x)$ since it does not stabilize any point on $S^3$,
particularly $\sigma(x_0)$.  Thus, ${\tilde t}_i(x_0)~=~0$ and
$h(x)~=~Id$.  The last result also follows from the property of
$h(x)~=~exp(t_i(x_0)X_i~+~{\tilde t}_i(x_0) {\tilde X}_i)$
                   being a fixed element in SO(4), which cannot stabilize more
than one point on
the path $\sigma^\prime(x)~=~\sigma(x_0) exp(s_j(x)X_j)$, unless it is the
identity element, which is the only common element in each of the
stability subgroups of SO(4).  Therefore, $h^{\prime\prime}(x)~=~h^\prime(x)$
is unique.

Thus, one can make a comparison between the vector fields in the space
spanned by $\{X_i\}$ even under the action of an SO(4) transformation
for the $S^3$ bundle.  From equation (23), however, it can be seen that
independence of the term $L_{g^{-1}\ast} g_\ast \cdot \xi$ with respect
to the fibre coordinate requires that $g(x)$ must be an element of
an SU(2) subgroup.

\noindent{\bf 5. Gauge Transformation Constraints for an SO(8) Action on an
                  $S^7$ Bundle}

If the fibre is $S^7$, it admits a parallelism associated with the existence of
octonions as an 8-dimensional division algebra over the real numbers.
Consider the action of SO(8) by right multiplication on a trivial $S^7$ bundle.
Suppose $L_{g^{-1}\ast}g_\ast \cdot \xi_\mu$ is an arbitrary element of the
28-dimensional Lie algebra of SO(8).  Let $\iota_L$ an embedding of octonions
in SO(8), so that left multiplication by y is represented by right
 multiplication by the matrix $\iota_L(y)$.  Then, independence with respect to
the fibre coordinate of the inhomogeneous term in the gauge transformation
implies that $\lambda_{y\ast}\cdot L_{g^{-1}\ast}\cdot (g_\ast \cdot \xi)
\cdot (\iota_L(y^{-1})^T)_\ast$, or equivalently that the row vector
$y\cdot (d_{AB}J_{AB})\cdot (\iota_L(y^{-1})^T)$, where $J_{AB}$ are generators
of SO(8), is independent of y.
Since
$$\iota_L(y^{-1})^T~=~\left(\matrix{y_0&-y_1&-y_2&-y_3&-y_4&-y_5&-y_6&-y_7
                                      \cr
                                    y_1&y_0&-y_3&y_2&-y_5&y_4&-y_7&y_6
                                        \cr
                                     y_2&y_3&y_0&-y_1&-y_6&y_7&y_4&-y_5
                                         \cr
                                     y_3&-y_2&y_1&y_0&y_7&y_6&-y_5&-y_4
                                          \cr
                                     y_4&y_5&y_6&-y_7&y_0&-y_1&-y_2&y_3
                                           \cr
                                     y_5&-y_4&-y_7&-y_6&y_1&y_0&y_3&y_2
                                           \cr
                                      y_6&y_7&-y_4&y_5&y_2&-y_3&y_0&-y_1
                                          \cr
                                      y_7&-y_6&y_5&y_4&-y_3&-y_2&y_1&y_0
                                           \cr}
                                                 \right)
\eqno(36)$$
the vector $y\cdot(d_{AB}J_{AB})\cdot \iota_L(y^{-1})^T~=~(0~c_1~c_2~c_3~c_4
{}~c_5~c_6~c_7)$ where
$$\eqalignno{c_1~&=~d_{01}(y_0^2+y_1^2)+(d_{02}-d_{13})(y_1y_2+y_0y_3)+
(d_{03}+d_{12})
(y_1y_3-y_0y_2)
\cr
&~+~(d_{04}-d_{15})(y_1y_4+y_0y_5)
+(d_{05}+d_{14})(y_1y_5-y_0y_4)
+(d_{06}-d_{17})(y_1y_6+y_0y_7)
\cr
&~+~(d_{07}+d_{16})(y_1y_7-y_0y_6)
-d_{23}(y_2^2+y_3^2)+(-d_{24}-d_{35})(y_3y_4-y_2y_5)
\cr
&~+~(-d_{25}+d_{34})(y_3y_5
+y_2y_4)+(-d_{26}-d_{37})(y_3y_6-y_2y_7)
+(-d_{27}+d_{36})(y_3 y_7+y_2y_6)
\cr
&~-~d_{45}(y_4^2+y_5^2)+(-d_{46}-d_{57})(y_5y_6-y_4y_7)+(-d_{47}+d_{56})
(y_5y_7+y_4y_6)
\cr
&~-~d_{67}(y_6^2+y_7^2) &(37)
\cr
c_2~&=~(d_{01}+d_{23})(y_1y_2-y_0y_3)+d_{02}(y_0^2+y_2^2)+(d_{03}+d_{12})
(y_2y_3+y_0y_1)
\cr
&~+~(d_{04}-d_{26})(y_2y_4+y_0y_6)
+(d_{05}+d_{27})(y_2y_5-y_0y_7)+(d_{06}+d_{24})(y_2y_6-y_0y_4)
\cr
&~+~(d_{07}-d_{25})
(y_2y_7+y_0y_5)
+d_{13}(y_1^2+y_3^2)+(d_{14}-d_{36})(y_3y_4+y_1y_6)
\cr
&~+~(d_{15}+d_{37})
(y_3y_5-y_1y_7)
+(d_{16}+d_{34})(y_3 y_6-y_1y_4)
+(d_{17}-d_{35})(y_3y_7+y_1y_5)
\cr
&~+~(d_{45}-d_{67})(-y_4y_7-y_5y_6)-d_{46}
(y_4^2+y_6^2)+(d_{47}-d_{56})(-y_6y_7+y_4y_5)
\cr
&~+d_{57}(y_5^2+y_7^2) &(38)
\cr
c_3~&=~(d_{01}+d_{23})(y_1y_3+y_0y_2)+(d_{02}-d_{13})(y_2y_3-y_0y_1)+d_{03}
(y_0^2+y_3^2)
\cr
&~+~(d_{04}+d_{37})(y_3y_4-y_0y_7)
+(d_{05}+d_{36})(y_3y_5-y_0y_6)+(d_{06}-d_{35})(y_3y_6+y_0y_5)
\cr
&~+~(d_{07}-d_{34})
(y_3y_7+y_0y_4)
-d_{12}(y_1^2+y_2^2)+(d_{14}-d_{27})(-y_2y_4-y_1y_7)
\cr
&~+~(d_{15}-d_{26})
(-y_2y_5-y_1y_6)+(d_{16}+d_{25})(-y_2y_6+y_1y_5)
+(d_{17}+d_{24})(-y_2y_7+y_1y_4)
\cr
&~+~(d_{45}-d_{67})(y_5y_7-y_4y_6)+(d_{46}+d_{57})
(y_6y_7+y_4y_5)+d_{47}(y_4^2+y_7^2)
\cr
&~+d_{56}(y_5^2+y_6^2)
& (39)
\cr
c_4~&~=~(d_{01}+d_{45})(y_1y_4-y_0y_5)~+~(d_{02}+d_{46})(y_2y_4-y_0y_6)+(d_{03}
-d_{47})(y_3y_4+y_0y_7)
\cr
&~+~d_{04}(y_0^2+y_4^2)
+(d_{05}+d_{14})(y_4y_5+y_0y_1)+(d_{06}+d_{24})(y_4y_6+y_0y_2)
\cr
&~+~(d_{07}-d_{34})
(y_4y_7-y_0y_3)
+(d_{12}+d_{56})(y_2y_5-y_1y_6)
+(d_{13}-d_{57})(y_3y_5+y_1y_7)
\cr
&~+~d_{15}(y_1^2+y_5^2)+(d_{16}+d_{25})(y_5y_6+y_1
y_2)
+(d_{17}-d_{35})(d_5d_7-y_1y_3)
\cr
&~+~(d_{23}-d_{67})(y_3y_6+y_2y_7)+d_{26}(y_2^2+y_6^2)+(d_{27}-d_{36})(y_6y_7
-y_2y_3)
\cr
&~-~d_{37}(y_3^2+y_7^2)
&(40)
\cr
c_5~&~=(d_{01}+d_{45})(y_1y_5+y_0y_4)+(d_{02}-d_{57})(y_2y_5+y_0y_7)+(d_{03}-
d_{56})(y_3y_5+y_0y_6)
\cr
&~+~(d_{04}-d_{15})(y_4y_5-y_0y_1)
{}~+~d_{05}(y_0^2+y_5^2)+(d_{06}-d_{35})(y_5y_6-y_0y_3)
\cr
&~+~(d_{07}-d_{25})
(y_5y_7-y_0y_2)
+(d_{12}+d_{47})(-y_2y_4+y_1y_7)
+(d_{13}+d_{46})(-y_3y_4+y_1y_6)
\cr
&~-~d_{14}(y_1^2+y_4^2)+(d_{16}+d_{34})
(-y_4y_6-y_1y_3)
+(d_{17}+d_{24})(-y_4y_7-y_1y_2)
\cr
&~+~(d_{23}-d_{67})(-y_3y_7+y_2y_6)+(d_{26}+d_{37})(-y_6y_7-y_2y_3)
-d_{27}(y_2^2+y_7^2)
\cr
&~-~d_{36}(y_3^2+y_6^2)
&(41)
\cr
c_6~&=~(d_{01}+d_{67})(y_1y_6-y_0y_7)+(d_{02}+d_{46})(y_2y_6+y_0y_4)+(d_{03}
-d_{56})(y_3y_6-y_0y_5)
\cr
&~+~(d_{04}-d_{26})(y_4y_6-y_0y_2)
{}~+~(d_{05}+d_{36})(y_5y_6+y_0y_3)+d_{06}(y_0^2+y_6^2)
\cr
&~+~(d_{07}+d_{16})
(y_6y_7+y_0y_1)
+(d_{12}+d_{47})(y_2y_7+y_1y_4)
+(d_{13}-d_{57})(y_3y_7-y_1y_5)
\cr
&~+~(d_{14}-d_{27})(y_4y_7-y_1y_2)+
(d_{15}+d_{37})(y_5y_7+y_1y_3)+d_{17}(y_1^2+y_7^2)
\cr
&~+~(d_{23}-d_{45})(-y_3y_4-y_2y_5)-d_{24}(y_2^2+y_4^2)+(d_{25}-d_{34})
(-y_4y_5+y_2y_3)
\cr
&~+~d_{35}(y_3^2+y_5^2)
& (42)
\cr
c_7~&=~(d_{01}+d_{67})(y_1y_7+y_0y_6)+(d_{02}-d_{57})(y_2y_7-y_0y_5)+(d_{03}
-d_{47})(y_3y_7-y_0y_4)
\cr
&~+~(d_{04}+d_{37})(y_4y_7+y_0y_3)
{}~+~(d_{05}+d_{27})(y_5y_7+y_0y_2)+(d_{06}-d_{17})(y_6y_7-y_0y_1)
\cr
&~+~d_{07}
(y_0^2+y_7^2)
+(d_{12}+d_{56})(-y_2y_6-y_1y_5)
{}~+~(d_{13}+d_{46})(-y_3y_6-y_1y_4)
\cr
&~+~(d_{14}-d_{36})(-y_4y_6+y_1y_3)+(d_{15}
-d_{26})(-y_5y_6+y_1y_2)
-d_{16}(y_1^2+y_6^2)
\cr
&~+~(d_{23}-d_{45})(-y_2y_4+y_3y_5)+(d_{24}+d_{35})(y_4y_5+y_2y_3)+d_{25}
(y_2^2+y_5^2)
\cr
&~+~d_{34}(y_3^2+y_4^2)
& (43)
\cr}
$$
Altogether, independence of the $c_i$ with respect to y leads to 21 independent
constraints
$$\eqalign{d_{01}&~=~-d_{23}~=~-d_{45}~=~-d_{67}
\cr
d_{02}~&=~d_{13}~=~-d_{46}~=~d_{57}
\cr
d_{03}~&=~-d_{12}~=~d_{47}~=~d_{56}
\cr
d_{04}~&=~d_{15}~=~d_{26}~=~-d_{37}
\cr
d_{05}~&=~-d_{14}~=~-d_{27}~=~-d_{36}
\cr
d_{06}~&=~d_{17}~=~-d_{24}~=~d_{35}
\cr
d_{07}~&=~-d_{16}~=~d_{25}~=~d_{34}
\cr}
\eqno(44)$$
and $L_{g^{-1}\ast} g_\ast \cdot \xi$ is required to be in the
seven-dimensional
subspace spanned by the SO(8) generators $\{X_i\vert~i=1,...,7\}$.
This result can  be traced to the fact that although $\iota_R({\tilde g})$
and $\iota_L(y^{-1})^T$ do not commute as elements of SO(8),
$$y\cdot \iota_R({\tilde g})\cdot \iota_L(y^{-1})^T~=~y\cdot \iota_L(y^{-1})^T
\cdot\iota_R({\tilde g})
\eqno (45)$$
because any {\it two} elements of the octonions form an associative
algebra:
\hfil\break
 $y^{-1}(y\cdot {\tilde g})~=~(y^{-1}y)\cdot {\tilde g}$ for
$y,~{\tilde g} \in {\Bbb O}$.

Let $v_\mu~=~\sigma_\ast\cdot \xi_\mu \in T_{\sigma(x)}(E)$.  Then
$\omega(v_\mu)~=~L_{y\ast}^{-1}
\cdot {\cal  V} v_\mu$ and
$$\omega(R_{g\ast} v_\mu)~=~L_{(y\cdot g)\ast}^{-1} {\cal V}(R_{g\ast} v_\mu)
\eqno(46)$$
If $v_\mu$ lies in the vertical subspace of $T_{\sigma(x)}(E)$,
${\cal V}(R_{g\ast}v_\mu)~=~R_{g\ast} L_{y\ast} \omega(v_\mu)$.
Representing left multiplication by a unit octonion $L_y$ as right
multiplication by an SO(8) transformation, denoted by $R_{a_y}$, one
finds that
$$\omega(R_{g\ast} v_\mu)~=~R_{a_{(y\cdot g)}^{-1} \ast} R_{g\ast} R_{a_y \ast}
\omega(v_\mu)~=~R_{(a_y g a_{y\cdot g}^{-1})\ast} \omega (v_\mu)
\eqno(47)$$
Defining $a_y g a_{y\cdot g}^{-1}$ to be $h(y,g)\in H$, the stability subgroup
of SO(8) for the origin o in $S^7$, the connection form satisfies
$\omega(R_{g\ast} v_\mu)~=~R_{h(y,g)\ast}\omega(v_\mu)$.  Independence of the
homogeneous part of the gauge transformation with respect to the fibre
coordinate requires that h(y,g) does not depend on y.
Let $g~\leftrightarrow~R_g^T~=~(c_{ij}), i,~j~=~0,1,...,7$.
It can be shown that $h_{00}~=~1,~h_{0i}~=~0,~i~=~1,...,7$ and
$$\eqalign{h_{11}~&=~(y_0^2+y_1^2)(c_{00}c_{11}-c_{10}c_{01})+(y_0y_3+y_1y_2)
(c_{00}c_{21}-c_{10}c_{31}-c_{20}c_{01}+c_{30}c_{11})
\cr
&+(y_0y_2-y_1y_3)(-c_{00}c_{31}-c_{10}c_{21}+c_{20}c_{11}+c_{30}c_{01})
\cr
&+(y_0y_5+y_1y_4)(c_{00}c_{41}-c_{10}c_{51}-c_{40}c_{01}+c_{50}c_{11})
\cr
&+(y_0y_6-y_1y_7)(-c_{00}c_{71}-c_{10}c_{61}+c_{60}c_{11}+c_{70}c_{01})
+(y_2^2+y_3^2)(-c_{20}c_{31}+c_{30}c_{21})
\cr
&+(y_3y_4-y_2y_5)(-c_{20}c_{41}-c_{30}c_{51}+c_{40}c_{21}+c_{50}c_{31})
\cr
&+(y_2y_4+y_3y_5)(-c_{20}c_{51}+c_{30}c_{41}-c_{40}c_{31}+c_{50}c_{21})
\cr
&+(y_3y_6-y_2y_7)(-c_{20}c_{61}-c_{30}c_{71}+c_{60}c_{21}+c_{70}c_{31})
\cr
&+(y_3y_7+y_2y_6)(-c_{20}c_{71}+c_{30}c_{61}-c_{60}c_{31}+c_{70}c_{21})
\cr
&+(y_4^2+y_5^2)(-c_{40}c_{51}+c_{50}c_{41})+(y_5y_6-y_4y_7)(-c_{40}c_{61}
-c_{50}c_{71}+c_{60}c_{41}+c_{70}c_{51})
\cr
&+(y_5y_7+y_4y_6)(-c_{40}c_{71}+c_{50}c_{61}-c_{60}c_{51}+c_{70}c_{41})
+(y_6^2+y_7^2)(-c_{60}c_{71}+c_{70}c_{61})
\cr
&~+~...
\cr}
\eqno(48)$$
where the remaining terms are obtained up to a sign by interchanging
$(c_{i0},~c_{j1})$ with $(c_{i2},~c_{j3}),~(c_{i4},~c_{j5}),~(c_{i6},~c_{j7})$.
There are 60 independent constraints on the matrix elements $(c_{ij})$
following
from the independence of $h_{11}$ with respect to y. Further conditions may
be derived from the other $h_{ij}$.  Since there are 56 non-trivial elements
$h_{ij}$, at most 3360 constraints arise and many of these may be redundant.
Nevertheless, one would expect that there are no solutions to these
conditions, as there are only 36 defining relations for SO(8).  This
is confirmed by a calculation of $\iota_L(y)^T R_g^T [\iota_L(y\cdot
g)^{-1}]^T$
where
$$\eqalign{R_g^T~&=~e^{t_1 X_1}
\cr
{}~&=~\left(\matrix{cos~t_1&sin~t_1&0&0&0&0&0&0
                                    \cr
                                    -sin~t_1&cos~t_1&0&0&0&0&0&0
                                     \cr
                                     0&0&cos~t_1&-sin~t_1&0&0&0&0
                                      \cr
                                     0&0&sin~t_1&cos~t_1&0&0&0&0
                                       \cr
                                     0&0&0&0&cos~t_1&-sin~t_1&0&0
                                       \cr
                                      0&0&0&0&sin~t_1&cos~t_1&0&0&
                                       \cr
                                      0&0&0&0&0&0&cos~t_1&sin~t_1
                                       \cr
                                      0&0&0&0&0&0&-sin~t_1&cos~t_1
                                        \cr}
                                                \right)
\cr}
\eqno(49)$$
{}From equation (48), one finds that for this choice of $R_g^T$
$$\eqalign{h_{11}~&=~(y_0^2+y_1^2)(c_{00}c_{11}-c_{10}c_{01}
-c_{02}c_{03}-c_{04}c_{15}+c_{14}c_{05}-c_{06}c_{17}+c_{16}c_{07})
\cr
&+(y_2^2+y_3^2)(-c_{20}c_{31}+c_{22}c_{33}-c_{32}c_{23}
+c_{24}c_{35}-c_{34}c_{25}+c_{26}c_{37}-c_{36}c_{27})
\cr
&+(y_4^2+y_5^2)(-c_{40}c_{51}+c_{50}c_{41}+c_{42}c_{53}-c_{52}c_{43}+c_{44}
c_{55}-c_{54}c_{45}+c_{46}c_{57}-c_{56}c_{67})
\cr
&+(y_6^2+y_7^2)(-c_{60}c_{71}+c_{70}c_{61}+c_{62}c_{73}-c_{72}c_{63}+c_{64}
c_{73}-c_{74}c_{63}+c_{66}c_{77}-c_{76}c_{67})
\cr
{}~&=~(y_0^2+y_1^2)(cos^2 t+sin^2 t)+(y_2^2+y_3^2)(cos^2 t+sin^2 t)
\cr
&+(y_4^2+y_5^2)
     (cos^2 t+sin^2 t)+(y_6^2+y_7^2)(cos^2 t+sin^2 t)~=~1
\cr}
\eqno(50)$$
and $h_{1j}~=~0,~j\ne 1$.
However,
$$\eqalign{h_{22}~&=~(cos^2 t_1~-~sin^2
t_1)(y_0^2+y_1^2+y_2^2+y_3^2+y_4^2+y_5^2+y_6^2
                                      +y_7^2)
        ~=~cos(2t_1)
\cr
h_{23}~&=~-sin(2t_1)(y_0^2+y_1^2+y_2^2+y_3^2-y_4^2-y_5^2-y_6^2-y_7^2)
\cr
h_{24}~&=~-2~sin(2t_1)(y_3y_4-y_2y_5+y_0y_7+y_1y_6)
\cr
&\vdots
\cr}
\eqno(51)$$
Thus, a y-dependent matrix is obtained when $R_g^T$ is given by (49) or any
$e^{t_i X_i}$.  A y-independent gauge transformation rule from the SO(8) action
on an $S^7$ fibre, by analogy with the tranformation of the connection form
resulting from an SO(4) action on $S^3$, therefore cannot be constructed.

Constraints on the connection form transformation rule can also be obtained
for bundles with fibres that are submanifolds of $S^7$.  The structure group
initially can be chosen to be any subgroup of SO(8) which preserves the
submanifold.  However, there also should be less conditions on the allowed
transformations, because there are fewer fibre coordinates to be eliminated.
Therefore, this leaves open the possibility of a residual gauge symmetry
associated with the action of a subgroup of the original structure group
on a submanifold of the standard fibre.

For the $S^7$ bundle, one choice for the submanifold is
$S^3~=~\{y \in S^7~\vert~y_0^2+y_1^2+y_2^2+y_3^2=1,~y_4=y_5=y_6=y_7=0\}$.
Independence of the inhomogeneous term, or equivalently $c_1,~...,~c_7$,
with respect to $y_1,~...,~y_3$ leads to the
15 conditions
$$\eqalign{d_{01}~&=~-d_{23}~~~~~~~d_{02}~=~d_{13}~~~~~~~d_{03}~=~-d_{12}
\cr
d_{04}~&=~d_{15}~=~d_{26}~=~-d_{37}
\cr
d_{05}~&=~-d_{14}~=~-d_{27}~=~-d_{36}
\cr
d_{06}~&=~d_{17}~=~-d_{24}~=~d_{35}
\cr
d_{07}~&=~-d_{16}~=~d_{25}~=~d_{34}
\cr}
\eqno(52)$$
and the generators of the remaining 13 transformations are $J_{01}-J_{23},~
J_{02}+J_{13},~J_{03}-J_{12},~J_{04}+J_{15}+J_{26}-J_{37},~J_{05}-J_{14}
-J_{25}-J_{36},~J_{06}+J_{17}-J_{24}+J_{35},~J_{07}-J_{16}+J_{25}+J_{34},~
J_{45},~J_{46},~J_{47},~J_{56},~J_{57}$ and $J_{67}$.  Independence
of the homogeneous terms can be checked for these generators by establishing
the y-independence of $L_y^T R_g^T L_{(y \cdot g)^{-1}}^T$,
with $L_y^T$ given by (36) after replacing $y_i$ by $-y_i$ when $i=1,2,3$ and
setting $y_j~=~0,~j \ge 4$, and $R_g^T$ equal to the exponential of the
generator.  This property can be verified for the first seven generators and
holds trivially for the last six generators.

It is now necessary to note that the transformations generated by
$J_{04}+J_{15}+J_{26}-J_{37},~J_{05}-J_{14}-J_{25}-J_{36},~J_{06}+J_{17}-J_{24}
+J_{35}$ and $J_{07}-J_{16}+J_{25}+J_{34}$ do not leave $S^3$ invariant
and instead map it into four-dimensional submanifolds of $S^7$, consisting
of a one-parameter family of three-spheres.  The coordinates of these
four-dimensional submanifolds are, respectively,
$$\eqalign{(y_0&~cos~d_{04},~y_1~cos~d_{04},~y_2~cos~d_{04},~y_3~cos~d_{04},
{}~y_0~sin~d_{04},
{}~y_1~sin~d_{04}
\cr
{}~&~~~~~~~~~~~~~~~~~~~~~~~~~~~~~~~~~~~~~~~~~~~~~~~~~~~~~~~~~~~~y_2~sin~d_{04},
{}~-y_3~sin~d_{04})
\cr
 (y_0&~cos~d_{05},~y_1~cos~d_{05},~y_2~cos~d_{05},~y_3~cos~d_{05},
{}~-y_1~sin~d_{05},~y_0~sin~d_{05},
\cr
&~~~~~~~~~~~~~~~~~~~~~~~~~~~~~~~~~~~~~~~~~~~~~~~~~~~~~~~~~~~~-y_3~sin~d_{05},
{}~-y_2~sin~d_{05})
\cr
(y_0&~cos~d_{06},~y_1~cos~d_{06},~y_2~cos~d_{06},~y_3~cos~d_{06},
-y_2~sin~d_{06},~y_3~sin~d_{06},
\cr
&~~~~~~~~~~~~~~~~~~~~~~~~~~~~~~~~~~~~~~~~~~~~~~~~~~
{}~~~~~~~~~~~~~~y_0~sin~d_{06},
{}~y_1~sin~d_{06})
\cr
(y_0&~cos~d_{07},~y_1~cos~d_{07},~y_2~cos~d_{07},~y_3~cos~d_{07},
{}~y_3~sin~d_{07},~y_2~sin~d_{07},
\cr
{}~&~~~~~~~~~~~~~~~~~~~~~~~~~~~~~~~~~~~~~~~~~~~~~~~~~~~~~~~~~~~~
-y_1~sin~d_{07},
{}~y_0~sin~d_{07})
\cr}
\eqno(53)$$
and since there is a bijective, continuous map to the coordinates
$\{((y_0,y_1,y_2,y_3),~\theta)~\vert~y_0^2+y_1^2+y_2^2+y_3^2=1\}$, each
of these submanifolds is topologically $S^3 \times S^1$.  Moreover,
since the maps between the coordinates are also diffeomorphisms, one
might consider constructing an $S^3 \times S^1$ bundle.

However, the enlargement of the fibre from $S^3$ to $S^3 \times S^1$
with non-zero entries in the last four components of the row vectors (53)
implies that the quantities $c_i$ in (37) - (43) will not necessarily be
independent of $y$.  Three extra conditions
$$\eqalign{d_{01}~+~d_{45}~&=~d_{23}~-~d_{67}
\cr
d_{02}~+~d_{46}~&=~-d_{13}~+~d_{57}
\cr
d_{03}~-~d_{47}~&=~d_{12}~+~d_{56}
\cr}
\eqno(54)$$
must be satisfied before  $c_4,~...,~c_7$ are independent of the
coordinates (53).  There would then be 10 remaining generators
$J_{01}-J_{23}+\alpha J_{45}-(\alpha+2) J_{67},~J_{02}~+~J_{13}~+~\beta
J_{46}
\hfil\break
+(\beta+2) J_{57},~J_{03}-J_{12}+\gamma J_{47}+(2 - \gamma) J_{56},~
J_{04}+J_{15}+J_{26}-J_{37},~J_{05}-J_{14}-J_{25}-J_{36},~J_{05}-J_{14}
-J_{25}-J_{36},~J_{06}+J_{17}-J_{24}+J_{35},~J_{07}-J_{16}+J_{25}+J_{34},~
J_{45}-J_{67},~J_{46}+J_{57}$ and $J_{47}-J_{56}$.

The computation of $L_y^T R_g^T L_{(y \cdot g)^{-1}}^T$ must now be
repeated for these 10 generators with the new coordinates (53) to
determine whether the dependence on $y_i$ and $\theta$ can be eliminated
in the homogeneous term in the transformation rule of the connection form.
Denoting the coordinates of $S^3 \times S^1$ by
$y_\theta~=~y \cdot g_{0l}(\theta), l~=~4,5,6,7$ so that
$$\eqalign{g_{04}(\theta)~&=~(cos~\theta,~0,~0,~0,~sin~\theta,~0,~0,~0)
\cr
g_{05}(\theta)~&=~(cos~\theta,~0,~0,~0,~0,~sin~\theta,~0,~0)
\cr
g_{06}(\theta)~&=~(cos~\theta,~0,~0,~0,~0,~0,~sin~\theta,~0)
\cr
g_{07}(\theta)~&=~(cos~\theta,~0,~0,~0,~0,~0,~0,~sin~\theta)
\cr}
\eqno(55)$$
the following identity for alternative algebras [24]
$$L_{y\cdot y^\prime}~=~L_y L_{y^\prime}~+~[L_y, R_{y^\prime}]
\eqno(56)$$
implies that
$$\eqalign{L_{y \cdot g_{0l}(\theta)}^T~&=~(L_y L_{g_{0l}(\theta)}
{}~+~L_y R_{g_{0l}(\theta)}~-~R_{g_{0l}(\theta)} L_y)^T
\cr
{}~&=~L_{g_{0l}(\theta)}^T L_y^T~+~R_{g_{0l}(\theta)}^T L_y^T~-~
L_y^T R_{g_{0l}(\theta)}^T
\cr}
\eqno(57)$$
Since
$$\eqalign{L_{((y\cdot g_{0l}(\theta))\cdot R_g^T)^{-1}} y^\prime~&=~
((y \cdot g_{0l}(\theta))\cdot R_g^T)^{-1} y^\prime
{}~=~(R_g^T)^{-1} (y \cdot g_{0l}(\theta))^{-1} y^\prime
\cr
{}~&=~(R_g^T)^{-1} (g_{0l}(\theta)^{-1} \cdot y^{-1}) \cdot y^\prime
{}~=~(R_g^T)^{-1} L_{g_{0l}(\theta)^{-1} \cdot y^{-1}} \cdot y^\prime
\cr}
\eqno(58)$$
and
$$L_{g_{0l}(\theta)^{-1} \cdot y^{-1}}~=~L_{g_{0l}(\theta)^{-1}} L_{y^{-1}}
{}~+~L_{g_{0l}(\theta)^{-1}} R_{y^{-1}}~-~R_{y^{-1}} L_{g_{0l}(\theta)^{-1}}
\eqno(59)$$
it can be shown that
$$\eqalign{L_{y_\theta}^T~R_g^T~L_{(y_\theta \cdot g)^{-1}}^T~&=~
L_{y \cdot g_{0l}(\theta)}^T~R_g^T~L_{((y\cdot g_{0l}(\theta)) \cdot g)^{-1}}^T
\cr
{}~&=~(L_{g_{0l}(\theta)}^T L_y^T~+~R_{g_{0l}(\theta)}^T L_y^T~-~L_y^T
R_{g_{0l}(\theta)}^T) R_g^T
\cr
&~~~~~~~~~~~~~~~~~~~~ (L_{y^{-1}}^T L_{g_{0l}(\theta)^{-1}}^T
{}~+~R_{y^{-1}}^T L_{g_{0l}(\theta)^{-1}}^T~-~ L_{g_{0l}(\theta)^{-1}}^T
R_{y^{-1}}^T) R_g^{-1}
\cr}
\eqno(60)$$
The next step in determining whether the fibre-coordinate dependence can be
eliminated from this expression would involve moving all of the
$\theta$-dependent matrices to the center.  Although it can be verified that
$L_{g_{0l}(\theta)} L_y~=~L_y L_{g_{0l}(\theta)}$, the commutators
$[L_y, R_{g_{0l}(\theta)}]$ and $[R_y, L_{g_{0l}(\theta)}]$ do not
similarly vanish.  Thus, the product (60) will be dependent on $y$ and
$\theta$ for the general group element obtained by exponentiating the
10 generators listed above.

Consequently, to find any residual gauge symmetry, the coordinates must
be restricted to the $S^3$ submanifold considered initially.  From the above
considerations, it follows that
only 9 generators $J_{01}-J_{23},~J_{02}+J_{13},~J_{03}-J_{12},~J_{45},~
J_{46},~J_{47},~J_{56},~J_{57},~J_{67}$ leave this submanifold invariant,
and the last 6 generators act trivially on this three-sphere.  As the
remaining generators form an su(2) Lie algebra, there remains an SU(2) group
of symmetry transformations acting on an $S^3$ fibre, which represents the
gauge invariance of an SU(2) Yang-Mills theory corresponding to the SU(2)
principal bundle.

Since this study began with an SO(8) structure group, it appeared possible
that an allowed symmetry group larger than the fibre could be obtained
if the fibre was chosen to be an appropriate submanifold of $S^7$.  The
computation of the fibre coordinate dependence of the connection form
transformation rule for the $S^3$ submanifold demonstrates that the
constraints reduce the structure group to SU(2), producing the standard
principal bundle structure.  This result shall be derived for general bundles
in the next section.

\noindent{\bf 6. Independence of the Transformation of the Connection Form
               with respect to the Fibre Coordinate}

Let us recall that for a general fibre bundle with atlas $\{U_\alpha,
\psi_\alpha\}$, the trivializations $\psi_\alpha,~\psi_\beta$ determine two
local sections $\sigma_\alpha(x)=\psi_\alpha^{-1}(x, y_0), \sigma_\beta(x)=
\psi_\beta^{-1}(x, y_0)$ which may be mapped by a diffeomorphism to
$(x, y(x))$ and $(x, y^\prime(x))$.  If $\psi_{\beta\alpha}:U\times F \to F$,
where $U\subset U_\alpha\cap U_\beta$, is defined by $\psi_{\beta\alpha}(x, y)
{}~=~y^\prime$, then the tangent spaces to the two sections are given
by $(\xi_x, V^\alpha)$ and $(\xi_x, T\psi_{\beta\alpha}(\xi_x, V^\alpha))$.
A connection provides a splitting of the tangent bundle into vertical and
horizontal sub-bundles, and the image of the horizontal subspaces associated
with the two sections in $TU\times TF$ are spanned by vectors of the form
$(\xi_x, C^\alpha(y))$ and $(\xi_x, C^\beta(y^\prime))$.  Defining the
connection forms $\Gamma^\alpha$ by $V^\alpha-C^\alpha$, one finds that they
transform under a change of section as
$$\Gamma^\beta(\xi_x, y^\prime)~=~T\psi_{\beta\alpha}^y(\xi_x)~+~T\psi_{\beta
\alpha}^x\cdot \Gamma^\alpha(\xi_x, y)
\eqno(61)$$
where $\psi_{\beta\alpha}^x(y)\equiv \psi_{\beta\alpha}^y(x)=y^\prime$.

The tangent mapping  $T\psi_{\beta\alpha}^x: TF\to TF$ is an isomorphism if the
tangent bundle of the fibre is trivializable.  Defining
$\phi_y$ to be a mapping from a vector space V to the tangent space $T_y(F)$,
so that $\Gamma^\alpha(\xi_x, y)=\phi_y \Gamma^\alpha(\xi_x),~
\Gamma^\alpha(\xi_x)\in V$, equation (52) becomes
$$\phi_{y^\prime}
\Gamma^\beta(\xi_x)~=~T\psi_{\beta\alpha}^y(\xi_x)~+~T\psi_{\beta\alpha}^x
\cdot \phi_y \Gamma^\alpha(\xi_x)
\eqno(62)$$
To interpret the relation between $\Gamma^\alpha(\xi_x)$ and
$\Gamma^\beta(\xi_x)$ as a gauge transformation of potentials taking values on
 the base space, it is necessary to eliminate the fibre coordinate dependence.
This can be achieved if the right-hand side of the equation can be expressed
as $\phi_{y^\prime} X,~X\in V$ and $\phi_y$ is an injective mapping.  Moreover,
V can be extended to have the same dimension as F, so that
$\Gamma^\alpha(\xi_x,
y)$ ranges over all of $T_y(F)$ and $\phi_y$ is a surjective mapping.
Thus $\phi_y$ should be a bijection.  Writing $T\psi_{\beta\alpha}^y(\xi_x)$
as $\phi_{y^\prime} X_1,~X_1\in V$ and $T\psi_{\beta\alpha}^x \cdot \phi_y
\Gamma^\alpha(\xi_x)$ as $\phi_{y^\prime} X_2,~X_2\in V$, a fibre
coordinate-independent gauge transformation rule may be obtained if
$$\phi_{y^\prime}(X_1)~+~\phi_{y^\prime}(X_2)~=~\phi_{y^\prime}(X_1~+~X_2)
 \eqno(63)$$
expressing linearity of $\phi_y$.  It may be noted that a bijective mapping
between the vector spaces that is nonlinear, analogous to the mapping
$x \to x^3$ on $({\Bbb R}, +)$, could also have been considered.   As
$\Gamma^\alpha(\xi_x,\cdot)$ is a $C^r,~r>1$ vector field on F, the mapping
$\phi: V\times F
\to TF,~ \phi(\cdot, y)=\phi_y$ must be a differentiable function of the
fibre coordinate.  Moreover, $\phi_y$ should also be a vector space
isomorphism, since one takes $\Gamma^\alpha(\xi_x, y)~+~\Gamma^\alpha(\eta_x,
y)$ to be
 $\Gamma^\alpha(\xi_x +\eta_x, y)$ as a result of the trivializations
$\{\psi_\alpha\},~ \psi_\alpha:\pi^{-1}(U_\alpha)\to U_\alpha\times F$ being
diffeomorphisms.  The requirement that $\phi_y$ be a vector space isomorphism
would exclude nonlinear mappings and ensure that equation (62) can be used
in reducing the transformation rule of the connection form to a gauge
transformation involving a dependence on the base space coordinates only.

It follows that $\phi_y$ is a parallelism on the fibre F.  There are in
fact several types of fibre parallelisms [25]:
\hfil\break\hfil\break
   i)  fibre parallelism
\hfil\break
There exists a map $C: E\times_M E\to {\cup\atop {(y, y^\prime)\in E\times_M
E}}
{}~Isom (T_y E, T_{y^\prime} E)$ such that $C(y, y^\prime)=\omega_{y^\prime}
\circ \omega_y^{-1} :T_y E \to T_{y^\prime} E$ where $y, y^\prime\in E_x$
and $\omega_y: (T_{red} E)_x \to T_y E$.  The last map gives rise to an
isomorphism $\Omega: T_{red} E\times_M  E\to TE$.
\hfil\break\hfil\break
  ii)  vertical fibre parallelism
\hfil\break
There exists a map ${\tilde C}: E\times_M E \to {\cup\atop {(y, y^\prime)
\in E\times_M E}}~ Isom(T_y E_x, T_{y^\prime} E_x)$ such that
${\tilde C}(y, y^\prime): T_y E_x= V_y E\to T_{y^\prime} E_x = V_{y^\prime} E$.
This parallelism is given by the isomorphism ${\tilde \Omega}: VT_{red}
E\times_M E\to VTE$.
\hfil\break\hfil\break
 iii)  integrable fibre parallelism
\hfil\break
A fibre parallelism is integrable if, for any $(y, y^\prime)\in E\times_M E$,
there exists a translation $\tau_{y^\prime y}: E\to E,~\tau_{y^\prime y}(y)=
y^\prime$ such that $T\tau_{y^\prime y}\in C(E\times_M E)$.  For a vertical
fibre parallelism, there exists a translation $\tau_{y^\prime y}^{vert}: E_x
\to E_x$ such that $T\tau_{y^\prime y}^{vert} \in {\tilde C}(E\times_M E)$.
It may be noted that a surmersion $(E, M, \pi)$ that admits a vertical
fibre parallelism and a connection is a projectable fibre parallelism.
[If $(E, M , \pi)$ is a fibration, then a vertical fibre parallelism
implies the existence of a connection.]  A projectable fibre parallelism
with an integrable parallelism on each fibre, or equivalently such that the
commutator of invariant vertical vector fields is invariant, can also be
regarded as an integrable vertical fibre parallelism.
\hfil\break\hfil\break
  iv)  globally integrable fibre parallelism
\hfil\break
It can be shown that a fibration which admits a globally integrable fibre
parallelism  is a principal bundle.  The translation $\tau_{y^\prime y}: E\to
E$
can be extended to a global diffeomorphism on E.

For a principal bundle,
$$\eqalign{VTP~&=~ T_e(G)\times P~=~VTP/G\times_M P~=~VT_{red} E\times_M P
\cr
VTP/G~&=~\{right-invariant~vector~fields~tangent~to~the~fibres~of~E\}
\cr
VT_{red} E~&\sim~ T_e(G)\times M
\cr}
\eqno(64)$$
$VTP/G$ is clearly a trivial bundle, because $(VTP/G)_x \sim T_e(G)~\forall
{}~x\in M$, and the transition functions are elements of G, leaving invariant
any vector field in $VTP/G$.

The parallelism in a principal bundle can be taken to be the one induced by
left multiplication and the diffeomorphism $\psi_{\beta\alpha}^x$ to be right
multiplication by a group element.  Since
$$\eqalign{R_{g\ast}~L_{y\ast}~&=~L_{(y\cdot g)\ast}~Ad(g^{-1})
\cr
T\psi_{\beta\alpha}^y(\xi_x)~&=~L_{y\ast}\cdot (g_\ast \cdot \xi_x)
{}~=~L_{y\cdot g\ast} [L_{g\ast}^{-1}(g_\ast \cdot \xi_x)]~=~
\phi_{y^\prime} [L_{g\ast}^{-1} (g_\ast\cdot \xi_x)]
\cr}
\eqno(65)$$
the standard gauge transformation
$$\Gamma^\beta(\xi_x)~=~Ad(g^{-1}) \Gamma^\alpha
(\xi_x)~+~L_{g\ast}^{-1}(g_\ast
\cdot \xi_x)
\eqno(66)$$
is a consequence of equation (62).

The problem of determining which bundles allow  the dependence on the
fibre coordinate to be eliminated has been considered in [12].  A necessary
condition is that
$$T\psi_{\beta\alpha}^x \phi_y \Gamma^\alpha(\xi_x)~=~\phi_{y^\prime} A(x)
\Gamma^\alpha(\xi_x)
\eqno(67)$$
or equivalently, that there exists a global diffeomorphism which is a
solution to the differential system
$$T\psi_{\beta\alpha}^x~=~\phi_{y^\prime}\cdot A \cdot \phi_y^{-1}
\eqno(68)$$
By Frobenius' theorem, the integrability conditions for the differential system
require that the commutator of vector fields on the fibre are invariant
with respect to the map $\phi_{y^\prime}\cdot A \cdot \phi_y^{-1}$.
Thus,
$$[\phi_{y^\prime}\cdot A\cdot\phi_y^{-1} X, \phi_{y^\prime}\cdot A\cdot
\phi_y^{-1} Y]_{y_0^\prime}~=~\phi_{y^\prime} \cdot A\phi_y^{-1} [X, Y]_{y_0}
\eqno(69)$$
for any vector fields $X, Y \in TF$, implying
$$c_{ijk}(y_0) A_{mk}~=~c_{klm}(y_0^\prime) A_{ki} A_{lj}
\eqno(70)$$
with the coefficients $c_{ijk}(y)$ given by $[\xi_i(y), \xi_j(y)]~=~c_{ijk}(y)
\xi_k(y)$, with
\hfil\break
 $\xi_i(y)~=~\phi_y\cdot e_i, e_i\in V$
 representing an orthonormal basis for $T_y(F)$.

The most general gauge matrix A that can be allowed is
$A~=~A_1(y^\prime) A_2(x) A_1(y)^{-1}$ as this gives
$$T\psi_{\beta\alpha}^x~=~\phi_{y^\prime}\cdot A_1(y^\prime) A_2(x)
A_1(y)^{-1} \cdot \phi_y^{-1} ~=~\phi_{y^\prime}^\prime \cdot A_2(x) \cdot
\phi_y^{\prime -1}
\eqno(71)$$
if the new parallelism $\phi_y^\prime$ is given by $\phi_y \cdot A_1(y)$.
The integrability condition for the differential system given by equation (71)
is
$$\eqalign{[\phi_{y^\prime} \cdot A_1(y^\prime) A_2(x) A_1(y)^{-1}  \cdot
\phi_y^{-1} \cdot X,&~
\phi_{y^\prime} \cdot A_1(y^\prime) A_2(x) A_1(y)^{-1} \cdot \phi_y^{-1}
\cdot Y]_{y_0^\prime}
\cr
{}~&=~\phi_{y_0^\prime} \cdot A_1(y^\prime) A_2(x)
A_1(y)^{-1}
{}~\cdot~\phi_{y_0}^{-1}\cdot [X, Y]_{y_0}
\cr}
\eqno(72)$$
Letting $X~=~\xi_i(y)~=~\phi_y\cdot e_i$, one finds that
$$c_{ijk} [A_1(y_0^\prime) \cdot A_2(x) \cdot A_1(y_0)^{-1}]_{mk}
{}~=~c_{klm} [A_1(y_0^\prime) A_2(x) A_1(y_0)^{-1}]_{ki}
            [A_1(y_0^\prime) A_2(x) \cdot A_1(y_0)^{-1}]_{lj}
\eqno(73)$$
Defining the basis $\xi_i^\prime(y)$ using the new parallelism
$\phi_y^\prime\cdot e_i~=~\phi_y\cdot A_1(y) e_i$, and the structure constants
$c_{ijk}^\prime (y)$ by the commutation relations
$[\xi_i^\prime, \xi_j^\prime]_y~=~c_{ijk}^\prime(y) \xi_k^\prime (y)$,
it follows that
$$c_{ijk}^\prime(y) (A_1(y))_{nk}~=~c_{lmn} (A_1(y))_{li} (A_1(y))_{mj}
\eqno(74)$$
Using this relation and multiplying equation (73) by $(A_1(y_0))_{ir}
(A_1(y_0))_{j s} (A^{-1}(y_0^\prime))_{pm}$ gives
$$c_{rsn}^\prime (y_0) A_{2pn}~=~c_{lmp}^\prime(y_0^\prime) A_{2lr} A_{2ms}
\eqno(75)$$
which is equivalent to the previous condition (70).

It may be noted that the proof of the existence of globally integrable
parallelism on the standard fibre F depends on the following lemma, with a more
detailed proof than that provided in reference [12] given below:
\hfil\break\hfil\break
{\bf Lemma}.  The set ${\cal C}_{y_1}~=~\{y\in F\vert
c_{ijk}(y)~=~c_{ijk}(y_1)\}$ either
contains all of F or is a set of dimensionality less than one in F.
\hfil\break\hfil\break
{\bf Proof}.  It is shown in [12] that an identity satisfied by the vector
fields $\{\xi_i\}$ induced
by the parallelism $\phi_y$ [26] implies that
$$c_{ijl}(y^\prime)c_{klm}(y^\prime)-\xi_k(c_{ijm})(y^\prime)~=~c_{ijl}(y)
 c_{klm}(y)~-~\xi_k(c_{ijm})(y)
\eqno(76)$$
where $\xi_k(c_{ijm})$ is defined by the action of the vector field on a
function on F.
Assuming initially that ${\cal C}_{y_1}$ is a continuous curve through $y_1$
with tangent
vector $X_{y_1}~=~X_k \xi_k(y_1)$ and
contracting equation (76) by $X_k$,
$$c_{ijl}(y)c_{klm}(y) X_k~-~X_k\xi_k (c_{ijm})(y)~=~c_{ijl}(y_1)c_{klm}(y_1)
X_k~-~X_k\xi_k(c_{ijm})(y_1)
\eqno(77)$$
 which implies, on ${\cal C}_{y_1}$, that $X_k \xi_k(c_{ijm})(y)~=~
X_k \xi_k(c_{ijm})(y_1)~=~0$ when $X_k$ is constant.
Similarly, the vanishing of the derivative of
$c_{ijl}(y)c_{klm}(y)~-~\xi_k(c_{ijm})(y)$ by equation (76) gives
$$\eqalign{(X_n\xi_n(c_{ijl})(y_1))X_kc_{klm}(y_1)~+~X_k c_{ijl}(y_1)(X_n
\xi_n(c_{klm})(y_1))~&-~X_n\xi_n X_k\xi_k(c_{ijm})(y_1)
\cr
{}~=~
-~X_n \xi_n & X_k \xi_k(c_{ijm})(y_1)~=~0
\cr}
\eqno(78)$$
and repeated differentiation leads to $X_{n_1}\xi_{n_1}...X_{n_r}\xi_{n_r}
(c_{ijm})(y_1)~=~0$.  Consequently, $c_{ijm}$ is constant along the integral
curve of X, and using the assumption that the set ${\cal C}_{y_1}$ is at least
one-dimensional,  one may conclude that it contains the integral
curve of a parallel vector field X passing through $y_1$.

A two-dimensional surface $S_Y$ is spanned by the integral curves of another
 parallel vector field Y intersecting ${\cal C}_{y_1}$ and an n-dimensional
neighbourhood of $y_1$ is covered by the surfaces $S_{\sum_i a_i Y_i}$
corresponding to arbitrary linear combinations of the independent vector fields
$Y_1, ..., Y_{n-1}$.
The surfaces $S_Y$ and $S_{Y^\prime}$ will initially
coincide at ${\cal C}_{y_1}$ if
\hfil\break
$Y^\prime~=~aX+bY$.
However, translation of the vector field X a distance t along the integral
curve of Y using the one-parameter family of diffeomorphisms $\{\chi_t\}$ gives
 a vector field
$$\chi_{t\ast} X~=~X~-~t[X, Y]~+~{{t^2}\over 2}[[X, Y], Y]~-~...
\eqno(79)$$
Parallel transport of the vector field X along the integral curve of $Y^\prime$
using the one-parameter family of diffeomorphisms $\{\chi^\prime_{v\ast}\}$
gives the vector field
$$\chi^\prime_{v\ast} X~=~X~-~bv[X,Y]~+~b^2 {{v^2}\over 2} [[X, Y], Y]~-~...
                        ~+~ab{{v^2}\over 2} [[X, Y], X]~-~...
\eqno(80)$$
Even after setting t equal to $\sum_j~b_j~v^j$, with $b_1~=~b$,
higher-order terms such as that containing
$ab{{v^2}\over 2}$ will vitiate the possibility of equating $\chi_{t\ast} X$
with $\chi^\prime_{v \ast} X$.  Thus, the tangent spaces to $S_Y$ and
$S_{Y^\prime}$, spanned by the bases
$\{\chi_{t\ast} X, Y\}$ and $\{\chi^\prime_{v \ast} X, Y^\prime\}~=~
\{aX+bY, \chi^\prime_{v\ast} X\}$ respectively, are equivalent at
${\cal C}_{y_1}$ but differ when $t,~v~\ne~0$.
Since $S_{Y^\prime}$ does not always coincide with $S_Y$, it must intersect
with
a surface $S_{Y^{\prime\prime}}$ (Fig. 1), where $Y^{\prime\prime}$ is another
 linear combination of the vector fields $Y_1,~ ...,~Y_{n-1}$,
$\sum_i ~a_i^{\prime\prime} Y_i$.

\input epsf.tex

\vbox{
\epsfxsize=3.5in
\epsfysize=2.4in
\centerline{\epsfbox{curve.eps}}
\vskip 0.2in
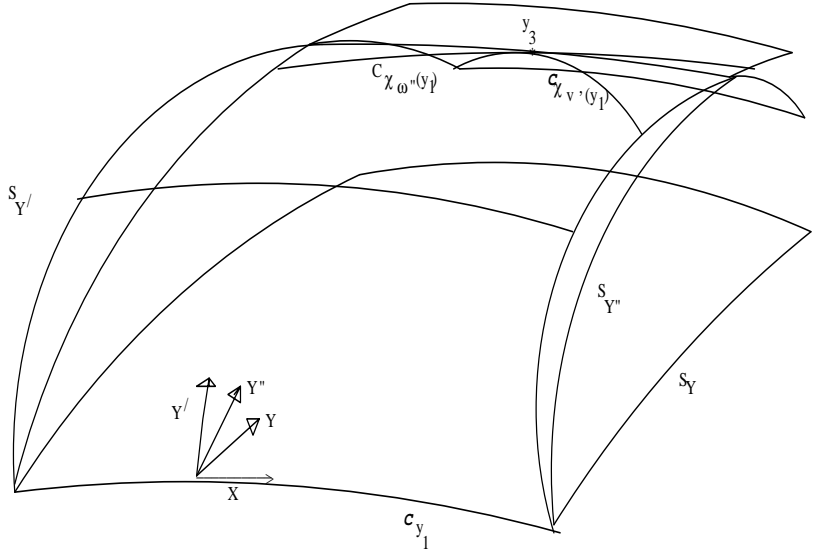
\noindent{\bf Fig. 1  The intersection of the surface $S_{Y^\prime}$
with $S_Y$ and $S_{Y^{\prime\prime}}$ .}}

Translation of ${\cal C}_{y_1}$  a distance v along the integral curves of
$Y^\prime$ results in a curve of constant $c_{ijk}$, ${\cal
C}_{\chi^\prime_v(y_1)}$, because
$$(Y^\prime_n \xi_n c_{ijl}(y)) Y^\prime_k c_{klm}(y)~+~(Y^\prime_k
 c_{ijl}(y))(Y^\prime_n \xi_n c_{klm}(y))~-~Y^\prime_n \xi_n Y^\prime_k \xi_k
(c_{ijm})(y)~=~0
\eqno(81)$$
together with equation (76) implies that
$$Y^\prime_n \xi_n Y^\prime_k \xi_k(c_{ijm})(y)~=~Y^\prime_n \xi_n Y^\prime_k
\xi_k(c_{ijm})(y_1)
\eqno(82)$$
and repeated differentiation gives
$$Y^\prime_{n_1}\xi_{n_1}...Y^\prime_{n_r}\xi_{n_r}(c_{ijm})(y)~=~
Y^\prime_{n_1}\xi_{n_1} ... Y^\prime_{n_r} \xi_{n_r} (c_{ijm})(y_1)
{}~~~~~\forall y\in {\cal C}_{y_1}
\eqno(83)$$
Similarly, translation of ${\cal C}_{y_1}$  a distance w along the integral
curves of $Y^{\prime\prime}$ produces another curve of constant $c_{ijk}$,
${\cal C}_{\chi^{\prime\prime}_w(y_1)}$.

If a point $y_3$ lies in the intersection of $S_{Y^\prime}$ and $S_{Y^{\prime
\prime}}$, then it must be located on an a curve ${\cal
C}_{\chi^\prime_v(y_1)}$
for some v and ${\cal C}_{\chi^{\prime\prime}_w(y_1)}$ for some w.  Parallel
transport of one of these curves along the other will sweep out a
two-dimensional surface of constant $c_{ijk}$.  When this surface is translated
back to $y_1$, one obtains a two-dimensional neighbourhood of constant
$c_{ijk}$ about the point $y_1$.  This procedure can be repeated until an
n-dimensional neighbourhood of $y_1$  of constant $c_{ijk}$ is constructed.
If F is compact, it may be covered by a finite number of these neighbourhoods
and ${\cal C}_{y_1}$ contains all of F.  Otherwise, it has dimensionality
less than one.

This lemma has been used previously [12] to demonstrate that the standard fibre
is $G/D$, where G is a Lie group and D is a discrete subgroup.
When the standard fibre is a group,  and the gauge matrix A represents the
adjoint action of $g^{-1}$, then if $g=exp(t_j X_j)$,
$$\eqalign{exp(-t_j X_j) X_i exp(t_j X_j)~&=~X_m(\delta_{im}-t_jc_{ijm}
+{1\over 2}
                                         t_jt_k c_{ikl} c_{ljm}+...)
\cr
A_{mi}~&=~\delta_{mi}-t_j c_{ijm} +{1\over 2} t_jt_k c_{ikl} c_{ljm}+...
\cr}
\eqno(84)$$
Substituting the formula for $A_{mi}$ into equation (70) and equating the
coefficients at each order in t
gives
$$\eqalign{ O(t^0):~~c_{ijm}~&=~c_{ijm}
\cr
O(t):~~c_{ijk}c_{klm}~&=~c_{ilk}c_{kjm}~+~c_{ikm}c_{kjl}
\cr
O(t^2):~~c_{ijk}c_{klp}c_{lnm}~&=~c_{kjm}c_{ipt}c_{tnk}~+~c_{ilm}c_{jpt}c_{tnl}
                                       ~-~2 c_{klm}c_{ink}c_{jpl}
\cr
&~\vdots
\cr}
\eqno(85)$$
The relations for $O(t^r),~r \ge 1$ all follow from the Jacobi identity for the
Lie algebra structure constants.

One may wish to consider a more general linear transformation A of the form
$$A_{mi}~=~\delta_{mi}~-~t_jd_{1 ijm}~+~{1\over 2}t_jt_k d_{2 ikl}
 d_{3ljm}~+~...
\eqno(86)$$
leading to the relations
$$\eqalignno{c_{ijk}d_{1 klm}~&=~d_{1 ijk} c_{kjm}~+~c_{ikm}d_{1 kjl}~+~...
& (87)
\cr
c_{ijk}d_{2 klp} d_{3 lnm}~&=~c_{kjm}d_{2 ipt}d_{3 tnk}~+~c_{ilm}d_{2 jpt}
d_{3 tnl}~-~2 c_{klm}d_{1 ink} d_{1 jpl}
& (88)
\cr
&~\vdots
\cr}
$$
Since the structure constants $c_{ijk}$ are anti-symmetric, equation (87)
comprises of ${{n(n-1)}\over 2}$ relations for the $n^3$ components of
the tensor $d_{1 ijk}$, so that it does not fully determine $d_1$.
However, equation (78) consists of ${{n(n-1)}\over 2}\cdot n^3$ relations for
the $3n^3$ components of $d_{1 ijk},~d_{2 ijk}$ and $d_{3 ijk}$.  Similarly,
higher orders of t give rise to equations which will include a larger number
of conditions on the coefficients $d_{M ijk}$.  Some of these constraints may
be redundant, but they should eliminate the arbitrariness in the choice of
$d_{M ijk}$ and leave coefficients that are equal to the structure constants
$c_{ijk}$.

\noindent {\bf 7. Gauge Theories and the Seven-Sphere}

The success of the standard model, based on the symmetry group $SU(3) \times
SU(2) \times U(1)$, immediately suggests the problem of finding an explanation
for the gauge groups and the number of forces.  One of the simplest and yet
potentially fundamental ways of deriving the standard model from theoretical
principles would be through an identification of the four forces with the
existence of only four normed real division algebras.  By a well-known theorem
[17][18][19],
this sequence of division algebras corresponds to the sequence of
parallelizable spheres, with the first two spheres
coinciding with the gauge groups U(1) and SU(2).  This correspondence
leads one to conjecture the existence
of a gauge theory based on $S^7$, which nearly possesses a Lie group structure.
In earlier attempts to construct an octonionic gauge theory [13][14][27], there
was no invariance under gauge transformations of the potential taking values
only in the base space.  The reasons for the lack of a gauge principle have
been illuminated in the previous two sections in the study of the
fibre-coordinate dependence of the connection form transformation rule for
$S^7$ bundles.

The extension of the fibre bundle description of gauge theories to a
Kaluza-Klein unification of gravity with the elementary particle interactions
through eleven-dimensional N=1 supergravity also involves the seven-sphere,
as the ground state solution for the metric field, $(AdS)_4 \times S^7$
provides a mechanism for the spontaneous compactification of seven of the
eleven
dimensions.  Because $S^7$ admits 8 Killing spinors, generating 8
supersymmetries, dimensional reduction of the d=11 theory gives N=8 gauged
supergravity in four dimensions [28], and although spontaneous compactification
over $S^7$ is initially promising for the identification of the elementary
particle interactions and the sequence of parallelizable spheres, the gauge
group SO(8) represents the invariances of a theory described by a principal
bundle with both standard fibre and structure
group SO(8), without reference to $S^7$.  Moreover, the number of gauge bosons
is considerably larger
than that associated with any of the known elementary particle interactions and
specifically the strong nuclear force.  It therefore represents a considerable
departure from the original scheme of using the sequence of parallelizable
spheres as the basis for a description of the electromagnetic, weak and strong
interactions.

Since the construction of the N=1, d=11 supergravity Lagrangian, several other
investigations have been undertaken regarding the possibility of finding a
gauge theory based on $S^7$.  Indeed, it has been shown in [12] and the
previous section that it is not feasible to build a purely octonionic
generalization of Yang-Mills theory.  Nevertheless, an octonionic gauge
theory has been proposed in a recent study [24], using a bimodule
representation of the octonion algebra.

Recalling that the octonion algebra is defined by the multiplication relations
$$\eqalign{e_0^2~&=~e_0~~~~~~~~e_0e_i~=~e_ie_0~=~e_i
\cr
e_ie_j~&=~-\delta_{ij} e_0~+~\epsilon_{ijk} e_k
\cr
\epsilon_{ijk}~&=~1~for~(ijk)~=~(123), (145), (167), (264), (257), (347), (356)
 \cr}
\eqno(89)$$
using the conventions of [29], which can be summarized as $e_\mu
e_\nu~=~C^\lambda_{\mu\nu} e_\lambda$,
\hfil\break
$\mu,~\nu,~\lambda~=~0,1,...,7$.  The left representation is
$(L_\mu)_{\lambda\nu}~=~(C_\mu)^\lambda_\nu~\equiv~C^\lambda_{\mu\nu}$
and the right representation is $(R_\mu)_{\lambda\nu}~=~
(\tilde C_\mu)^\lambda_\nu~\equiv~C^\lambda_{\nu\mu}$ with $L_0~=~R_0~=~Id$.
Then $\lambda_i~=~i{{L_i}\over 4}$ and $\rho_j~=~i{{R_j}\over 4}$ satisfy
the commutation relations
$$\eqalign{[\lambda_i,~\lambda_j]~&=~{i\over 2} \epsilon_{ijk}~\lambda_k~+~
                                 2[\rho_j, \lambda_i]
        \cr
[\rho_i,~\rho_j]~&=~-{i\over 2} \epsilon_{ijk}~\rho_k~+~
                                 2 [\lambda_j, \rho_i]
\cr}
\eqno(90)$$
and the trace relations
$$Tr(\lambda_i \lambda_j)~=~Tr(\rho_i \rho_j)~=~{1\over 2}~\delta_{ij}
\eqno(91)$$
Suppose a gauge potential is defined to be $A_\mu (x)~=~A_\mu^i (x) \lambda_i$,
leading to the covariant derivative ${\cal D}_\mu~=~\partial_\mu~+~ig~A_\mu$.
Under a gauge transformation corresponding to the element
$\Omega(x)~=~e^{i\alpha^i (x) \lambda_i}$, ${\cal D}_\mu \psi(x) \to
e^{i \alpha^i \lambda_i}
{}~{\cal D}_\mu \psi(x)$ for any fermion field $\psi(x)$, and
$[{\cal D}_\mu, {\cal D}_\nu]$ also transforms covariantly, allowing one
to immediately find an invariant $-{1\over 2} Tr(F_{\mu\nu} F^{\mu\nu})$.

However, the gauge tranformation which leaves this action invariant
$$\eqalign{A_\mu(x)~\to~A_\mu^\prime (x)~&=~\Omega(x) A_\mu (x) \Omega^{-1}(x)
{}~+~{i\over g} (\partial_\mu \Omega(x)) \Omega^{-1}(x)
\cr
{}~&=~e^{i \alpha \cdot \lambda} A_\mu e^{-i \alpha \cdot \lambda}~-~
                                         {1\over g} \partial_\mu \alpha^i
                                                                \lambda_i
\cr
{}~&=~A_\mu~+~[i \alpha \cdot \lambda, A_\mu]~-~{1\over 2}
                                                  [\alpha \cdot \lambda,
                                                      [\alpha \cdot \lambda,
                                                                 A_\mu]]
{}~+~...-{1\over g} \partial_\mu \alpha^i \lambda_i
\cr}
\eqno(92)$$
contains not only the generators $\lambda_i$ associated with the left
representation of the octonion algebra but also generators arising from
multiple commutators of $\lambda_i$ and $\rho_j$.
To linear order in $\alpha$,
$$\eqalign{A_\mu^\prime~&=~A_\mu^{\prime i} \lambda_i~+~2i \alpha^j~A_\mu^k~
                                                           [\rho_k, \lambda_j]
\cr
A_\mu^{i\prime}~&=~A_\mu^i~-~{1\over 2} \epsilon_{ijk} \alpha^j A_\mu^k
                    ~-~{1\over g}  \partial_\mu \alpha^i
\cr}
\eqno(93)$$
and
$$\eqalign{F_{\mu\nu}^{\prime}~&=~F_{\mu\nu}^{\prime i}~\lambda_i
              ~-~2g \alpha^k (A_\mu^{\prime i} A_\nu^{\prime j}~-~
                                A_\nu^{\prime i} A_\mu^{\prime j})
                              [\lambda_i, [\rho_k, \lambda_j]]
\cr
{}~&+~2i [\partial_\mu(\alpha^j A_\nu^{\prime k})~-~\partial_\nu (\alpha^j
A_\nu^{\prime k})] [\rho_k, \lambda_j]
\cr
F_{\mu\nu}^{\prime i}~&=~\partial_\mu A_\nu^{\prime i}~-~\partial_\nu
A_\mu^{\prime i}~-~{1\over 2} g \epsilon_{ijk} A_\mu^{\prime j} A_\nu^{\prime
k}
\cr}
\eqno(94)$$
The trace relations
$$\eqalign{Tr(\lambda_i [\rho_j, \lambda_k])~&=~{i\over 8} \epsilon_{ijk}
\cr
Tr([\lambda_i, \rho_j][\lambda_k, \rho_l])~&=~
{{-1}\over {32}} [\epsilon_{ijkl}~+~2(\delta_{ik} \delta_{jl}~-~\delta_{il}
                                                     ~\delta_{jk}) ]
\cr}
\eqno(95)$$
can be used to express the Lagrangian in component form
$${\cal L}~=~-{1\over 4} (\partial_\mu A_\nu^i~-~\partial_\nu A_\mu^i)
                                   (\partial^\mu A^{\nu i}~-~\partial^\nu
                                                               A^{\mu i})
              -{1\over {16}} g^2 A_\mu^j A_\nu^k (A^{j \mu} A^{k \nu}~-~
                                                      A^{k \mu} A^{j\nu})
\eqno(96)$$
but the action is not invariant under the substitution $A_\mu^i \to
 A_\mu^{\prime i}$ as (96) becomes
$$\eqalign{&-{1\over 4} (\partial_\mu A_\nu^i~-~\partial_\nu A_\mu^i)
                (\partial^\mu A^{\nu i}~-~\partial^\nu A^{\mu i})
{}~-~{1\over {16}} g^2 A_\mu^j A_\nu^k (A^{j\mu} A^{k \nu}~-~A^{k \mu} A^{j
\nu})
\cr
{}~&+~{1\over 2} [\partial_\mu(\epsilon_{ijk} \alpha^j A_\nu^k~+~{1\over g}
                             \partial_\nu \alpha^i)
                ~-~\partial_\nu (\epsilon_{ijk} \alpha^j A_\mu^k~+~
                                      {1\over g} \partial_\mu \alpha^i)]
                 \cdot \partial^\mu A^{\nu i}
\cr
{}~&+~{1\over 8} g^2 (A_\mu^j A_\nu^k~-~A_\nu^j A_\mu^k)
                       ({1\over 2}\epsilon^j_{pq} \alpha^p A_\mu^q~+~
                                 {1\over g} \partial^\mu \alpha^j)
                          \cdot A_k^\nu
\cr}
\eqno(97)$$
and the extra terms in (97) are not total derivatives.  Consequently, the
transformation of the potential $A_\mu~\to~A_\mu^\prime$ is required and
this involves generators other than those corresponding to the left
representation of the octonion algebra.

The appearance of the non-closed algebraic part of the variation involving
the generators $\rho_k$ is a manifestation of the fibre-coordinate dependence
in the transformation rule of the connection form (61).  One may recall that
$T \psi_{\beta\alpha}^y (\xi_x)$ is $\phi_{y \prime} X_1$ for some $X_1 \in V$
when the standard fibre admits a parallelism, and thus the y-dependence of this
term can be eliminated for an $S^7$ bundle.  From equation (67) and proof of
the subsequent lemma, it follows that the y-dependence of
$T \psi_{\beta\alpha}^x \cdot \phi_y \Gamma^\alpha(\xi_x)$ can be eliminated
only when the standard fibre is a group manifold.  For an $S^7$ bundle, the
remaining y-dependence should be associated with this term, and this
has been confirmed in $\S 5$ and equation (93).  Moreover, the analysis of
$\S 5$ leads one to conjecture that the Lagrangian is part of a larger theory
possessing $G_2$ invariance.  Writing $A_\mu^\prime~=~A_\mu^{\prime AB}
J_{AB}$,
a transformation
$A_\mu^\prime \to A_\mu~=~\Omega^{-1} A_\mu \Omega~-~{i\over g}
                                   \Omega^{-1} (\partial_\mu \Omega),~\Omega
  \in G_2$
can be found such that
$A_\mu$ has only non-zero components $A_\mu^i$ multiplying the generators
$\lambda_i$.  From (94), $F_{\mu\nu}~=~F_{\mu\nu}^{AB} J_{AB}$ has non-zero
components corresponding to the generators of $G_2$ and the
Lagrangian $-{1\over 2} Tr(F_{\mu\nu} F^{\mu\nu})$ actually should be invariant
under the entire group of $G_2$ transformations.  This symmetry is broken only
when one specializes to a particular choice for the vanishing components of the
gauge potential.

This result may also be understood from the context of non-associative
deformations of gauge theories [30].  These generalizations are based on
an algebraic structure consisting of M set of generators $\{T_{pi}\},~
p~=~1,...,M$.  Together, the entire set of generators form an
associative Lie algebra structure.  Restriction to one set of
generators, $T_i^p$, p fixed, leads to a problem with closure of the
algebra.
$$[T_i^p, T_j^p]~=~f_{ijk}^p~T_k^p~+~\sum_{n=1}^M~\sigma_n^p~[T_j^n, T_i^p]
\eqno(98)$$
The deviation from associativity can be measured by the associator
$$\eqalign{J(T_i^p,~T_j^p,~T_k^p)~&=~\epsilon^{ijk} (T_i^p,~T_j^p,~T_k^p)
{}~=~\epsilon^{ijk}~[~(T_i^p~T_j^p)~T_k^p~-~T_i^p(T_j^p~T_k^p)~]
\cr
{}~&=~ \sigma_n^p([T_i^p, [T_k^n, T_j^n]]~+~[T_j^p,[T_i^n, T_k^p]]
                 ~+~[T_k^p,[T_j^n, T_i^p]])
\cr}
\eqno(99)$$
The lack of closure of the algebra associated with a single set of
generators, or equivalently, the coupling of the gauge potentials
corresponding to distinct sets of generators, leads to a Lagrangian, based on
only the field strengths $F_{\mu\nu}^p$, which contains extra nonlinear terms.
Therefore, when the algebra $T_i^p,~p~fixed$ represents the octonions,
the entire algebra is given by the matrices associated with left and right
multiplication introduced earlier.  The theory obtained is therefore
a special case of this general procedure of deforming gauge theories, where
the Lagrangian is part of an action with the larger $G_2$ symmetry generated
by the combined set of fourteen generators.

It is also of interest to note that a general procedure for
constructing a non-associative gauge theory has been developed in [31], where
the potential takes values in a non-associative algebra A.  The gauge symmetry
of the action, however, is the automorphism group of A, $G_A$, so that given a
symmetric, bi-linear non-degenerate form $\langle~u~\vert~v~\rangle$,
\hfil\break
$~u,~v~\in~A$, satisfying the invariance condition $\langle~
gu~\vert~gv~\rangle$, a
Lagrangian
$$L_0~=~{1\over 4}~\langle~F_{\mu\nu}~\vert~F_{\mu\nu}~\rangle
\eqno(100)$$
may be constructed.  If $\Lambda_p$ is the set of A-valued p-forms,  then it is
necessary to assume that $g \in gl(\Lambda_1),~ \xi~\in~\Lambda_1$ and
$$\eqalign{&(i)~~~~g(\omega\omega)~=~(g\omega)(g\omega)
\cr
&(ii)~~~d \xi~+~\xi~\xi~=~0
\cr
&(iii)~~~~d(g \omega)~=~g(d\omega)~-\xi(g \omega)~-~(g\omega) \xi
\cr}
\eqno(101)$$
Although it is not known if there is a solution for g and $ \xi$  for
any given non-associative algebra, it can be assumed that one exists when
A is the octonion algebra.  The symmetry group $G_A$ is then $G_2$, and the
Lagrangian resembles the one given in equation (96).

A common property of all of these theories is that the action possesses
a Lie group symmetry even though it has been constructed so that it
seems to include only components in the non-associative octonion algebra.
This provides further support for the assertion proven in [12] and $\S 6$
 that a pure gauge theory with a symmetry defined only by the non-associative
algebra does not exist.  In addition, it may be noted that although $(g^{-1}
A_\mu) g~=~g^{-1}(A_\mu g)$ for alternative algebras by Artin's theorem,
$$\{g^{-1}(A_\mu g)\}\{(g^{-1}A_\nu) g\}~\ne~
      g^{-1}~(A_\mu~A_\nu)~g
\eqno(102)$$
in general for non-associative algebras, so that the validity of invariance
under finite gauge transformations as a consequence of invariance under
infinitesimal transformations cannot be proven for non-associative algebras
[31]. Since an extensive investigation of the possibility of constructing a
pure gauge theory based only on the non-associative algebra has revealed that a
Lie
group structure is essential, this property will be assumed in the following
sections, although the seven-sphere shall continue to used in this
geometrical approach to the internal symmetry spaces.

\noindent{\bf 8. Division Algebras and the Standard Model}

A connection between the gauge groups in the standard model and the division
algebras has been established recently using Clifford algebras.  In this
approach, the division algebras are the spinor spaces, while the Clifford
algebras, which may be known as adjoint division algebras, may be used to
derive the gauge symmetries of the action [32].
Denoting this Clifford algebra of $R^{p,q}$ by $R_{p,q}$, generated by the
elements $\Upsilon_\alpha$ satsifying
$$\eqalign{\Upsilon_\alpha \Upsilon_\beta~+~\Upsilon_\beta \Upsilon_\alpha~&=~
2\eta_{\alpha\beta} \epsilon
\cr
 \eta_{\alpha\beta}~&=~diag(1,...,1,-1,...,-1),~\epsilon~=~Id_{p+q}
\cr}
\eqno(103)$$
the Pauli algebra is $P~\sim~R_{3,0}~\sim~{\Bbb C} \otimes {\Bbb H}$.
Given elements $u_1, ..., u_n,~x$ in an algebra A, the left adjoint map
$x \to u_n( ... (u_2(u_1 x)) ... )$ defines an associative algebra of left
actions.  Using the basis $\{e_a\}$ of A, $A_L$ consists of elements of the
form $1_L,~e_{La},~e_{Lab}~=~e_{La}\cdot e_{Lb},~...$.  For the division
algebras ${\Bbb C},~{\Bbb H}$ and ${\Bbb O}$, it can be shown that
${\Bbb C}_L~=~{\Bbb C}_R~=~{\Bbb C}$, ${\Bbb H}_L~\sim~{\Bbb H}_R~\sim~{\Bbb
H}$, and ${\Bbb O}_L~=~{\Bbb O}_R~\sim~R(8)$.  While the adjoint left algebra
$P_L$ acts on the space of $2 \times 1$ complex Pauli spinors,
$P_L(2)~=~R(2)~\otimes~P_L~\sim~C(4)~\sim~{\Bbb C}~\otimes~R_{1,3}$ is the
Dirac algebra, which is the complexification of the Clifford algebra
$R_{1,3}~\sim~{\Bbb H}(2)$.  The algebras of left actions and right actions of
${\Bbb H}(2)$ on the space of $2 \times 1$ matrices over ${\Bbb H}$ commute,
so that the algebra of left actions represent the Clifford algebra of
4-dimensional Minkowski space-time and the algebra of right actions generates
a $SU(2)~\times~U(1)$ internal symmetry.
Now considering the tensor product $T~=~{\Bbb R} \otimes {\Bbb C} \otimes
{\Bbb H} \otimes {\Bbb O}$, the adjoint $T_L~\sim~R_{0,9}$ corresponds to the
Pauli algebra,
whereas $T_L(2)~\sim~{\Bbb C}(32)$, the complexification of $R_{1,9}$, the
 equivalent of the Dirac algebra in the 10-dimensional Minkowski space-time.
As the spinor space T is 64-dimensional, it is just large enough to describe a
 family consisting of a lepton doublet and a quark doublet with three distinct
colours and the corresponding anti-family.

The subspace of 2-vectors of $R_{p,q}$ closes under commutation and it is
isomorphic to the Lie algebra so(p,q).  Thus, the two-vector basis
$\{ e_{Lpq},~p, q = 1,...,6, p \ne q\}$ of ${\Bbb O}_L~\sim~R_{0,6}$ as
15-dimensional and isomorphic to $so(6)~\sim~su(4)$.  The intersection of
su(4) with $LG_2~=~\{e_{Lab}~-~e_{Lcd}:e_ae_b~=~e_ce_d\}$ is
\hfil\break
$su(3)~=~\{e_{Lpq}~-~e_{Lrs}:~e_pe_q~=~e_re_s, p, q, r, s~ \ne~7\}$.
The SU(3) gauge symmetry of the strong interactions therefore arises as part of
the SO(1,9) Lorentz transformations and not as an internal symmetry in
10 dimensions, although it can be regarded as an internal symmetry in
4 dimensions.  This is consistent with the use of string
theory to describe the strong interactions and gravity.

These considerations suggest that one may begin with a ten-dimensional
Lagrangian
$$\eqalign {{\cal L}_{1,9}~&=~{\cal L}_{gauge}~+~{\cal L}_\phi~+~
                            {\cal L}_{1,9}^{ferm}
\cr
{\cal L}_{1,9}^{ferm}~&=~\langle \Psi\vert {\crosspart\partial}_{1,9} \Psi
\rangle
\cr
{\cal L}_\phi~&=~ \sum_{a=0}^9~\langle \partial_a \phi\vert \partial^a \phi
\rangle~-~\mu^2 \langle \phi\vert \phi \rangle~-~\lambda \langle
\phi \vert \phi \rangle^2
\cr
\sum_{p=4}^9~(\partial_p \partial^p)\phi_i~&=~0
\cr}
\eqno(104)$$
and
${\cal L}_{gauge}$ is a ten-dimensional action for the spin-one gauge field.
After using the projector distinguishing between matter and anti-matter
multiplets, $R_{1,9}$ is projected to $R_{1,3} \otimes SO(6)$, the bosonic
part of the action is based on the covariant derivative appropriate for
$R_{1,3} \otimes SU(4)$ .  The SU(4) symmetry must then be
broken to SU(3) to reproduce the QCD action, while the $SU(2) \times U(1)$
symmetry, arising from $R_{1,3}$ leads to the Weinberg-Salam model.
While the scalar Lagrangian is used for spontaneous symmetry breaking, the
fermion Lagrangian may be reduced to the standard lepton-quark Lagrangian
in four dimensions.  Using the projector $\rho_{\pm}~=~{{(1~\pm i~e_{L7})}
\over 2}$, one may re-express the fermion term as
$$\eqalign{\langle \rho_+ \Psi\vert {\crosspart\partial}_{1,3} (\rho_+ \Psi)
\rangle~&+~
                               \langle \rho_-\Psi\vert
 {\crosspart\partial}_{1,3}(\rho_- \Psi)
                                               \rangle
\cr
{}~&+~\langle\rho_+ \Psi \vert {\crosspart\partial}_{0,6} (\rho_- \Psi) \rangle
{}~+~ \langle \rho_-\Psi \vert {\crosspart\partial}_{0,6}(\rho_+ \Psi) \rangle
\cr}
\eqno(105)$$
The last two terms represent matter/anti-matter transitions that are not
observed, and they vanish upon imposing the conditions
$$ {\crosspart\partial}_{0,6} (\rho_{\pm} \Psi)~=~0
\eqno(106)$$
The solutions to these constraints have a dependency on the coordinates
in the extra six dimensions which gives rise to the SU(2) and SU(3) symmetries
of the standard model.

\noindent{\bf 9.  Gauge Transformation Constraints for an SU(4) Action on
an $S^7$ Bundle}

The special place for an SU(4) gauge symmetry in the formulation of the
standard model given in the previous section suggests that it may be of
interest to study the SU(4) action on the seven-sphere, $S^7~=~SU(4)/SU(3)$.
Although it has been demonstrated that the generalized gauge transformations
are still restricted to Lie groups, the procedure followed in sections
5-7 can be applied, in principle, to the bundle with $S^7$ fibre with an
SU(4) structure group.  This allows for the possibility of obtaining an action
with the appropriate SU(3) gauge symmetry, based on an  action with
a larger symmetry group SU(4) more directly connected with the Clifford algebra
   $R_{1,9}$.

The action of SU(4) on the seven-sphere follows from the invariance of
the bilinear form ${\bar z}_0^\prime z_0~+~{\bar z_1}^\prime z_1~+~
{\bar z}_2^\prime z_2~+~{\bar z}_3^\prime z_3$, where $z~=~(z_0~z_1~z_2~z_3)
\in {\Bbb C}^4$.  To define the equivalent of the inhomogeneous term in the
transformation rule of the connection form, one needs the embedding of
   $S^7$ into SU(4).  Left multiplication by a unit octonion maps any point
$y^\prime \in S^7$ to a point $y^{\prime\prime}~=~y y^{\prime}$, and
since a transitive group action on the sphere would take any pair of
points into each other, left multiplication by the octonion y
can be represented as $\iota_L(y) \in SU(4)$, where $\iota_L: S^7 \to SU(4)$.

Quaternions can be included in the group SO(4), as noted earlier, through
$$(y_0~y_1~y_2~y_3)~\to~\left(\matrix{y_0&-y_1&-y_2&-y_3
                                         \cr
                                        y_1& y_0&-y_3&y_2
                                          \cr
                                         y_2 & y_3 & y_0 & -y_1
                                          \cr
                                         y_3& -y_2& y_1 & y_0
                                           \cr}
                                              \right)
\eqno(107)$$
and they can also be represented as SU(2) matrices
$$\left(\matrix{y_0~+~iy_2 & y_1~+i y_3
                  \cr
                  -y_1~+~iy_3 &  y_0~-~iy_2
                           \cr}\right)
\eqno(108)$$
acting on the column vectors
$$\left(\matrix{&y_0~+~i y_2
                        \cr
                 &y_1~+~i y_3
                           \cr}
                            \right)
\eqno(109)$$
Similar considerations apply to the embedding of $S^7 \to SU(4)$.  The problem
of representing left multiplication by octonions therefore reduces to
the problem of determining whether there is a correct choice of coordinate
axes.

\noindent{\bf Proposition.}  There are no choices of octonion multiplication
     rules and coordinate axes in ${\Bbb C}^4$ such that left multiplication
by unit octonions can be represented as a $4 \times 4$ matrix with entries
of the form $\pm y_{r_i}\pm
 i y_{r_j}$.

\noindent{\bf Proof.}  Consider the transformation

$$ y^{\prime\prime}~=~Z~ y^\prime
\eqno(110)$$
 where
$Z~=~(z_{ij})$ and $y^\prime$
and $y^{\prime\prime}$ are column
 vectors
with the coordinate axes
\hfil\break
$(0~r_1),~(r_2~r_3),~(r_4~r_5)$ and $(r_6~r_7)$.
Since

$$\eqalign
{ y_0^{\prime\prime}~+~i y_{r_1}^{\prime\prime}~=&~
[Re~z_{00}~y_0^\prime~-~Im~z_{00}~y_{r_1}^\prime
{}~+~Re~z_{01}~y_{r_2}^\prime ~-~Im~z_{01}~y_{r_3}^\prime ]
\cr
&~-~ [Re~z_{02}~y_{r_4}^\prime
         ~-~ Im~z_{02}~y_{r_5}^\prime~+~ Re~z_{03}~y_{r_6}^\prime
                                             ~-~ Im~z_{03}~y_{r_7}^\prime]
\cr
{}~&~+~i~[ Im~z_{00}~y_0^\prime~+~Re~z_{00}~y_{r_1}^\prime
 ~+~  Im~z_{01}~y_{r_3}^\prime
{}~+~Re~z_{01}~y_{r_3}^\prime]
       \cr
&~+~i~[Im~z_{02}~y_{r_4}^\prime~+~ Re~z_{02}~y_{r_5}^\prime
{}~+~Im~z_{03}~y_{r_6}^\prime~+~ Re~z_{03}~y_{r_7}^\prime]
                             \cr}
\eqno(111)$$

$$\eqalign{Re~z_{00}~=&~ y_0~~~~~~~~~~Re~z_{01}~=~-y_{r_2}
\cr
Re~z_{02}~=&~-y_{r_4} ~~~~~ Re~z_{03}~=~-y_{r_6}
\cr
Im~z_{00}~=&~ y_{r_1}~~~~~~~~Im~z_{01}~=~y_{r_3}
\cr
Im~z_{02}~=&~ y_{r_5}~~~~~~~~Im~z_{03}~=~y_{r_7}
\cr}
\eqno(112)$$
so that
$$\eqalign{y_{r_1}^{\prime\prime}~&=~y_{r_1} y_0^\prime~+~y_0
y_{r_1}^\prime~-~y_{r_2}
                                  y_{r_3}^\prime~+~y_{r_3} y_{r_2}^\prime
\cr
                            &~-~ y_{r_4} y_{r_5}^\prime~+~y_{r_5}
y_{r_4}^\prime
                                ~-~y_{r_6} y_{r_7}^\prime
                                  ~+~y_{r_7} y_{r_6}^\prime
\cr}
\eqno(113)$$
which is consistent with the triples $(r_1 r_3 r_2)$, $(r_1 r_5 r_4)$
and
\hfil\break
$(r_1 r_7 r_6)$.
The relation
$$\eqalign{y_{r_2}^{\prime\prime}~+~i y_{r_3}^{\prime\prime}~=&~[Re~z_{10}
{}~y_0^\prime
{}~-~Im~z_{10}~y_{r_1}^\prime
{}~+~Re~z_{11}~y_{r_2}^\prime~-~ Im~z_{11}~y_{r_3}^\prime ]
\cr
&~+~ [ Re~z_{12}~y_{r_4}^\prime~-~Im~z_{12}~y_{r_5}^\prime
{}~+~Re~z_{13}~
y_{r_6}^\prime~-~Im~z_{13}~y_{r_7}^\prime]
\cr
&~+~i~[ Im~z_{10}~y_0^\prime~+~Re~z_{10}~y_{r_1}^\prime
{}~+~Im~z_{11}~
y_{r_2}^\prime~+~Re~z_{11}~y_{r_3}^\prime]
\cr
&~+~i~[ Im~z_{12}~y_{r_4}^\prime~+~Re~z_{12}~
 y_{r_5}^\prime
{}~+~Im~z_{13}~y_{r_6}^\prime
 ~+~Re~z_{13}~y_{r_7}^\prime]
\cr}
\eqno(114)$$
implies that
$$\eqalign{Re~z_{10}~=&~y_{r_2}~~~~~~~~~~~~~~~~Re~z_{11}~=~y_0
\cr
Re~z_{12}~=&~-y_{r_{T(r_2,r_4)}}
{}~~~~~Re~z_{13}~=~-y_{r_{T(r_2, r_6)}}
\cr
Im~z_{10}~=&~y_{r_3}~~~~~~~~~~~~~~~Im~z_{11}~=~-y_{r_1}
\cr
Im~z_{12}~=&~y_{r_{T(r_2, r_5)}}~~~~~~~~Im~z_{13}~=~ y_{r_{T(r_2, r_7)}}
\cr}
\eqno(115)$$
where $e_{r_{T(r_i, r_j)}}~\equiv~e_{r_i}e_{r_j}$.  Since
$$\eqalign{y_{r_3}^{\prime\prime}~=&~y_{r_3} y_0^\prime~+~y_{r_2}
y_{r_1}^\prime~+~
                            y_0 y_{r_3}^\prime~-~y_{r_1} y_{r_2}^\prime
\cr
                           &~+~y_{r_{T(r_2, r_5)}} y_{r_1}^\prime
                            ~-~y_{r_{T(r_2,r_4)}}y_{r_5}^\prime
                              ~+~y_{r_{T(r_2, r_7)}} y_{r_6}^\prime
                               ~-~ y_{r_{T(r_2, r_6)}} y_{r_7}^\prime
\cr}
\eqno(116)$$
the following triples, $(r_3~r_{T(r_2, r_5)}~r_4)$, $(r_3~r_5~r_{T(r_2,
r_4)})$,
$(r_3~r_{T(r_2, r_7)}~r_6)$ and
\hfil\break
$(r_3~r_7~r_{T(r_2, r_6)})$.
The relation
$$\eqalign{y_{r_4}^{\prime\prime}~+~i y_{r_5}^{\prime\prime}~=&~
[Re~z_{20}~y_0^\prime~-~Im~z_{20}~y_{r_1}^\prime
{}~-~Re~z_{21}~y_{r_2}^\prime~-~
Im~z_{21}~y_{r_3}^\prime]
\cr
&~+~[ Re~z_{22}~y_{r_4}^\prime~-~Im~z_{22}~y_{r_5}^\prime~+~ Re~z_{23}
{}~y_{r_6}^\prime~-~Im~z_{23}~y_{r_7}^\prime]
\cr
&~+~i~[ Im~z_{20}~y_0^\prime~+~Re~z_{20}~y_{r_1}^\prime
{}~+~Im~z_{21}~y_{r_2}^\prime~+~Re~z_{21}~y_{r_3}^\prime]
\cr
&~+~i~[ Im~z_{22}~y_{r_4}^\prime~+~Re~z_{22}~y_{r_5}^\prime
{}~+~Im z_{23}~y_{r_6}^\prime~+~ Re~z_{23}~y_{r_7}^\prime]
\cr}
\eqno(117)$$
implies
$$\eqalign{Re~z_{20}~=&~y_{r_4}~~~~~~~Re~z_{21}~=~y_{r_{T(r_2, r_4)}}
\cr
Re~z_{22}~=&~y_0~~~~~~~~Re~z_{23}~=~-y_{r_{T(r_4,r_6)}}
\cr
Im~z_{20}~=&~y_{r_5}~~~~~~Im~z_{21}~=~-y_{r_{T(r_3, r_4)}}
\cr
Im~z_{22}~=&~-y_{r_1}~~~Im~z_{23}~=~ y_{r_{T(r_4, r_7)}}
\cr}
\eqno(118)$$
and
$$\eqalign{y_{r_5}^{\prime\prime}~=&~y_{r_5} y_0^\prime~+~y_{r_4}
y_{r_1}^\prime
{}~+~y_{r_{T(r_2, r_4)}} y_{r_3}^\prime~-~y_{r_{T(r_3, r_4)}} y_{r_2}^\prime
\cr
&~~~-~y_{r_1} y_{r_4}^\prime~+~y_0 y_{r_5}^\prime~+~y_{r_{T(r_4, r_7)}}
y_{r_6}^\prime~-~y_{r_{T(r_4, r_6)}} y_{r_7}^\prime
\cr
{}~&=~y_{r_5} y_0^\prime~+~y_{r_4} y_{r_1}^\prime~+~y_{r_{T(r_3, r_5)}}
y_{r_3}^\prime~+~y_{r_{T(r_2, r_5)}} y_{r_2}^\prime
\cr
&~~~-~y_{r_1} y_{r_4}^\prime~+~y_0 y_{r_5}^\prime~+~y_{r_{T(r_4, r_7)}}
y_{r_6}^\prime~-~y_{r_{T(r_4, r_6)}} y_{r_7}^\prime
\cr}
\eqno(119)$$
consistent with the triples $(r_{T(r_4, r_7)}~r_6~r_5)$ and
$(r_{T(r_4~r_6)}~r_5~r_7)$.
Consequently, as
$$\eqalign{y_{r_6}^{\prime\prime}~+~i y_{r_7}^{\prime\prime}~=~&
[Re~z_{30}~y_0^\prime~-~Im~z_{30}~y_{r_1}^\prime
{}~+~Re~z_{31}~y_{r_2}^\prime
{}~-~Im~z_{31}~y_{r_3}^\prime ]
\cr
&~+~[Re~z_{32}~y_{r_4}^\prime~-~Im~z_{32}~y_{r_5}^\prime~+~
{}~Re z_{33}
{}~y_{r_6}^\prime~-~Im~z_{33}~y_{r_7}^\prime]
\cr
&~+~i~[Im~z_{30}~y_0^\prime~+~Re~z_{30}~y_{r_1}^\prime
{}~+~Im~z_{31}~y_{r_2}^\prime
+~Re~z_{31}~y_{r_2}^\prime]
\cr
&~+~i[Re~z_{32}~y_{r_4}^\prime~+~Im~z_{32}~y_{r_4}
{}~+~Re~z_{33}~y_{r_7}^\prime
{}~+~Im~z_{33}~y_{r_6}^\prime]
\cr}
\eqno(120)$$
the matrix elements are
$$\eqalign{Re~z_{30}~=&~y_{r_6}~~~~~~~~~~~~~~~~Re~z_{31}~=~y_{r_{T(r_2, r_6)}}
\cr
Re~z_{32}~=&~y_{r_{T(r_4, r_6)}}~~~~~~~~Re~z_{33}~=~y_0
\cr
Im~z_{30}~=&~y_{r_7}~~~~~~~~~~~~~~~Im~z_{31}~=~-y_{r_{T(r_3, r_6)}}
\cr
Im~z_{32}~=&~-y_{r_{T(r_5, r_6)}}~~~~Im~z_{33}~=~-y_{r_1}
\cr}
\eqno(121)$$
and
$$\eqalign{y_{r_7}^{\prime\prime}~&=~y_{r_7} y_0^\prime~+~y_{r_{T(r_2, r_6)}}
y_{r_3}^\prime~-~y_{r_{T(r_3, r_6)}} y_{r_2}^\prime
\cr
&~~~~+~y_{r_{T(r_4, r_6)}}y_{r_5}^\prime~-~y_{r_{T(r_5, r_6)}} y_{r_4}^\prime
{}~+~y_0 y_{r_7}^\prime~-~ y_{r_1} y_{r_6}^\prime
\cr
{}~&=~y_{r_7} y_0^\prime~+~y_{r_6} y_{r_1}^\prime~+~ y_{r_{T(r_3, r_7)}}
y_{r_3}^\prime~+~y_{r_{T(r_2, r_7)}} y_{r_2}^\prime
\cr
&~~~~+~y_{r_{T(r_5, r_7)}} y_{r_5}^\prime~+~y_{r_{T(r_4, r_7)}} y_{r_4}^\prime
{}~+~ y_0 y_{r_7}^\prime~-~y_{r_1} y_{r_6}^\prime
\cr}
\eqno(122)$$
The four possible choices for the pair $\{r_{T(r_2, r_4)}, r_{T(r_2, r_5)}\}$
are $\{r_6, r_7\}$, $\{r_7, r_6\}$ and $\{r_6, -r_7\}$ and $\{-r_7, r_6\}$,
where $e_{-r_i}~\equiv -e_{r_i}$, there are no sets of triples consistent
with any of the octonion multiplication tables.
The same conclusion could be reached for right multiplication using the
transpose, dropping the primes in the row vector and placing primes on the
elements of the $4 \times 4$ matrix.
$$\eqalign{(y_0^{\prime\prime}~+~iy_{r_7}^{\prime\prime}~~&
y_{r_2}^{\prime\prime}~+~i
y_{r_3}^{\prime\prime}~~y_{r_4}^{\prime\prime}~+~i y_{r_5}^{\prime\prime}
{}~~y_{r_6}^{\prime\prime}~+~iy_{r_7}^{\prime\prime})
\cr
{}~=&~(y_0~+~i y_{r_1}~~y_{r_2}~+~i y_{r_3}~~y_{r_4}~+~i y_{r_5}
{}~~y_{r_6}~+~iy_{r_7})
{}~\left(\matrix{z_{00}^\prime&z_{10}^\prime&
z_{20}^\prime&z_{30}^\prime
\cr
z_{01}^\prime& z_{11}^\prime & z_{21}^\prime & z_{31}^\prime
\cr
z_{02}^\prime & z_{12}^\prime & z_{22}^\prime & z_{32}^\prime
\cr
z_{03}^\prime & z_{13}^\prime & z_{23}^\prime & z_{33}^\prime
\cr}
\right)
\cr}
\eqno(123)$$
This result also can be verified by noting that the first column of the $4
\times 4$ matrix mapping the origin $o$ to y must have entries
$z_{00}~=~y_0~+~iy_1,~z_{10}~=~y_2~+~iy_3,~ z_{20}~=~y_4~+~iy_5$
\hfil\break
and $z_{30}~=~y_6~+~iy_7$ and the unitarity relations
for this matrix cannot be satisfied when the other entries are of the
form $\sigma_{r_i r_{i+1}}(y_{r_i}~
\pm~i y_{r_{i+1}})$, where $\sigma_{r_i r_{i+1}}~=~\pm 1$.

Another way of representing the embedding of $S^7$ into SU(4) might be achieved
through the identification of the unit octonion
$y~=~y_0~+~y_1 e_1~+~y_2 e_2~+~y_3 e_3~+~y_4 e_4~+~y_5 e_5$
\hfil\break
$~+~y_6 e_6~+~
y_7 e_7
{}~=~(y_0~+~iy_1)~+~(y_2~+~i y_3)e_2~+~(y_4~+~i y_5) e_4~+~(y_6~+~i y_7) e_6$
with a  $4 \times 4$ matrix based on the quaternionic subalgebra $(1, e_2,
e_4, e_6)$
$$ \iota_L(y)~=~\left(\matrix {y_0+iy_1 & (y_2+iy_3)e_2 & (y_4+iy_5)e_4 &
(y_6+iy_7)e_6
\cr
(y_2+iy_3) e_2 & y_0+i y_1 &(y_6+iy_7) e_6 & (y_4+iy_5)
e_4
\cr
(y_4+iy_5) e_4 & (y_6+iy_7) e_6 & y_0+ i y_1 &   (y_2+iy_3) e_2
\cr
(y_6+iy_7) e_6 & (y_4 +i y_5) e_4 & (y_2+iy_3) e_2 & y_0 + iy_1
\cr}
\right)
\eqno(124)$$
so that
$$ y^{\prime\prime}~=~y y^\prime~=~\iota_L(y)\left(\matrix{&~y_0^\prime~+~i
y_1^\prime
\cr
&(y_2^\prime~+~i y_3^\prime) e_2
\cr
&(y_4^\prime~+~ i y_5^\prime) e_4
\cr
&(y_6^\prime~+~ i y_7^\prime) e_6
\cr}
\right)
\eqno(125)$$

It is also of interest to note that since
${{\bar y}^{\prime\prime}y^{\prime\prime}}
{}~=~{{\bar y}^\prime y^\prime}~=~ 1$,
$\iota_L(y)$ satisfies the identity ${{\iota_L(y)}^{\dag}} \iota_L(y)~=~1$, the
defining relation for unitary matrices.  However, the
determinant of this matrix contains non-trivial expressions involving
quaternions and non-associativity properties of the octonions.
It is nevertheless possible to use the matrix $\iota_L(y)$
in the computation of the inhomogeneous term in the gauge transformation
rule.

In the identical manner to the calculation of the expression
$y \cdot (d_{AB} J_{AB}) \iota_L(y^{-1})^T$ in section 5, one may use the
$4\times 4$ matrices  $J_A$ representing the generators of SU(4).
Since the generators $J_A$ are anti-hermitian and traceless,

$$ d_A J_A~=~\left( \matrix{ id_7 & d_1+id_4 & d_2 +id_5 & d_3+id_6
                             \cr
                           -d_1+id_4 & -id_7+id_{14} & d_8 +id_9 &
                                                               d_{10}+id_{11}
                               \cr
                             -d_2+id_5 & -d_8+id_9 & -i d_{14} +id_{15} &
                                                           d_{10}+id_{11}
                                \cr
                             -d_3+id_6 & -d_{10} +id_{11} & -d_{12} +id_{13} &
                                                                     -id_{15}
                                         \cr}
                                                \right)
\eqno(126)$$
Using the row vector $y~=~(y_0+iy_1~(y_2+iy_3)
 e_2~(y_4+iy_5)e_4~(y_6+iy_7)e_6)$, it follows that
$y (d_A J_A) \iota_L(y^{-1})^T~=~(c_0~c_1~c_2~c_3)$
where

$$\eqalign{c_0~=&~[~id_7~(y_0^2+y_1^2)~+~i(d_7-d_{14})(y_2^2+y_3^2)
                  ~+~i(d_{14}-d_{15})
            (y_4^2+y_5^2)~+~id_{15} (y_6^2+y_7^2)~]
\cr
+~&[~-2(y_0+iy_1)(y_2+iy_3)(d_1+id_4)~+~2d_{12}(y_4 y_6+y_5 y_7)~-~
2 d_{13}(y_4 y_7 - y_5 y_6)~]~e_2
\cr
+~&[~-2(y_0+iy_1)(y_4+iy_5)(d_2+id_5)~+~2d_{10}(y_2 y_6+y_3 y_7)~-~
2 d_{11}(y_2 y_7 - y_3 y_6)~]~e_4
\cr
+~&[~-2(y_0+iy_1)(y_6+iy_7)(d_3+id_6)~+~2d_8 (y_2 y_4+y_3 y_5)~-~
2 d_9 (y_2 y_5+y_3 y_4)~]~e_6
\cr}$$
$$\eqalign{c_1~=&~[~(y_0+iy_1) id_7~+~(y_2+iy_3) e_2 (-d_1+id_4)~+~ (y_4+iy_5)
e_4 (-d_2+id_5)
\cr
&~~~~~~+~ (y_6+iy_7) e_6 (-d_3+id_6)~]~ (-y_2-iy_3) e_2
\cr
{}~+&~[~(y_0+iy_1)(d_1+id_4)~+~(y_2+iy_3) e_2(-id_7+id_{14})~+~(y_4+iy_5) e_4
(-d_8+id_9)
\cr
&~~~~~~+~(y_6+iy_7) e_6 (-d_{10}+id_{11})~]~(y_0-iy_1)
\cr
{}~+&~[~(y_0+iy_1)(d_2+id_5)~+~(y_2+iy_3) e_2 (d_8+id_9)~+~(y_4+iy_5) e_4
(-id_{14}+id_{15})
\cr
&~~~~~~+~(y_6+iy_7) e_6 (-d_{12}+id_{13})~]~(-y_6-iy_7) e_6
\cr
{}~+&~[~(y_0+iy_1)(d_3+id_6)~+~(y_2+iy_3)e_2(d_{10}+id_{11})~+~(y_4+iy_5) e_4
(d_{12}+id_{13})
\cr
&~~~~~~+~(y_6+iy_7) e_6 (-id_{15})~]~(-y_4-iy_5) e_4
\cr
\vdots &
\cr}
\eqno(127)$$
$c_0$ can only be independent of y if
$$\eqalign{d_7~&=~d_7~-~d_{14}~=~d_{14}~-~d_{15}~=~d_{15}
\cr
d_1~+~id_4~&=~0~~~~~~~~~~~~d_8~=~d_9~=~0
\cr
d_2~+~id_5~&=~0~~~~~~~~~~~d_{10}~=~d_{11}~=~0
\cr
d_3~+~id_6~&=~0~~~~~~~~~~~d_{12}~=~d_{13}~=~0
\cr}
\eqno(128)$$
and then
$$c_1~=~-2~[~id_4(y_2+iy_3) e_2~+~id_5 (y_4+iy_5) e_4~+~ id_6 (y_6+iy_7) e_6~]
             ~(y_2+iy_3) e_2
\eqno(129)$$
which is only independent of y if $d_4~=~d_5~=~d_6~=~0$.  Independence
of the transformation rule of the connection form with respect to the
fibre coordinate y leads to enough constraints on the gauge matrix so that
the coefficients $d_A$ must all vanish.  Thus, there appear to be no
non-trivial
combinations of the generators such that the inhomogeneous term is independent
of the fibre coordinate.

Another embedding
of $S^7$ into the space of matrices satisfying the unitary relation and
its effect on the terms in the
connection form transformation rule will therefore be determined.
The general $4 \times 4$ matrix is given by
$g~=~D(\delta_1, \delta_2, \delta_3, -\delta_1-\delta_2-\delta_3)
U_{34}(\phi_3, \sigma_6) U_{23}(\theta_3, \sigma_5) U_{24}(\phi_2, \sigma_4)
\cdot~U_{12}(\theta_2, \sigma_3) U_{13}(\theta_1, \sigma_2)
U_{14}(\phi_1, \sigma_1)$, where
\hfil\break
$D(\alpha_1, \alpha_2, \alpha_3, \alpha_4)$
is the diagonal matrix with elements $e^{i \alpha_1},~e^{i \alpha_2},~
e^{i \alpha_3},~ e^{i \alpha_4}$ and $U_{pq}(\phi, \sigma)$, which has all
diagonal
elements equal to 1, except for $u_{pp}$ and $u_{qq}$, which should be
$cos~\phi$, and non-zero off-diagonal entries
$u_{pq}~=~sin~\phi~
e^{-i \sigma}$, $u_{qp}~=~sin~\phi~e^{i \sigma}$,
\hfil\break
represents a unitary
transformation in the (p,q)-plane.  Consequently, the elements of g are

$$\eqalign{g_{00}~&=~cos~\theta_2~ cos~\theta_1~cos~\phi_1~e^{i \delta_1}
\cr
g_{01}~&=~-sin~\theta_2~e^{i(\delta_1-\sigma_3)}
\cr
g_{02}~&=~-cos~\theta_2~sin~\theta_1~e^{(\delta_1-\sigma_2)}
\cr
g_{03}~&=~-cos~\theta_2~cos~\theta_1~sin~\phi_1~e^{i(\delta_1-\sigma_1)}
\cr
g_{10}~&=~cos~\theta_3~cos~\theta_2~sin~\theta_2~cos~\theta_1~cos~\phi_1
           e^{i(\delta_2+\sigma_3)}
\cr
&~~~~~~~-~sin~\theta_3~sin~\theta_1~cos~\phi_1
            e^{i(\sigma_2+\delta_2-\sigma_5)}
{}~ - cos~\theta_3~\sin~\phi_2~sin~\phi_1
e^{i(\sigma_1+\delta_2-\sigma_4)}
\cr
g_{11}~&=~ cos~\theta_2~cos~\theta_3~cos~\phi_2~e^{i \delta_2}
\cr
g_{12}~&=~ -cos~\theta_3~cos~\phi_2~sin~\theta_2~sin~\theta_1~e^{i(\delta_2
+\sigma_3-\sigma_2)}~-~sin~\theta_3~cos~\theta_1~e^{i(\delta_2-\sigma_5)}
\cr
g_{13}~&=~ -cos~\theta_3~cos~\phi_2~sin~\theta_2~sin~\phi_1~cos~\theta_1
e^{i(\delta_2+\sigma_3-\sigma_1)}
\cr
&~~~~~~~+~sin~\theta_3~sin~\theta_1~sin~\phi_1
{}~e^{i(\delta_2+\sigma_2-\sigma_1-\sigma_5)}
\cr
&~~~~~~~ -cos~\theta_3~sin~\phi_2~cos~\phi_1~e^{i(\delta_2-\sigma_4)}
\cr
g_{20}~&=~ cos~\phi_3~sin~\theta_3~cos~\phi_2~sin~\theta_2~
cos~\theta_1~cos~\phi_1~e^{i(\delta_3+\sigma_3+\sigma_5)}
\cr
&~~~~~~~+~cos~\theta_3~
cos~\phi_3~sin~\theta_1~cos~\phi_1~e^{i(\delta_3+\sigma_2)}
\cr
&~~~~~~~ -cos~\phi_3~sin~\theta_3~sin~\phi_2~sin~\phi_1~e^{i(\sigma_1+\delta_3
+\sigma_5-\sigma_4)}
\cr
g_{21}~&=~ cos~\phi_3~sin~\theta_3~cos~\phi_2~cos~\theta_2~e^{i(\delta_3+
\sigma_5)}
\cr
g_{22}~&=~-cos~\phi_3~sin~\theta_3~cos~\phi_2~sin~\theta_2~sin~\theta_1~
e^{i(\delta_3+\sigma_3+\sigma_5-\sigma_2)}
\cr
&~~~~~~~~~~~~~+~cos~\phi_3~cos~\theta_3~
cos~\theta_1~e^{i\delta_3}
\cr}
$$
$$\eqalign{
g_{23}~&=~ -cos~\phi_3~sin~\theta_3~cos~\phi_2~sin~\theta_2~cos~\theta_1~
sin~\phi_1~e^{i(\delta_3+\sigma_3-\sigma_1)}
\cr
&~~~~~~~~~~~~~~~-~cos~\phi_3~cos~\theta_3~
sin~\theta_1~sin~\phi_1~e^{i(\delta_3+\sigma_2-\sigma_1)}
\cr
&~~~~~~~~~~~~~~~~ -
cos~\phi_3~sin~\theta_3~sin~\phi_2~cos~\phi_1~e^{i(\delta_3+\sigma_5
-\sigma_4)}
\cr
g_{30}~&=~ sin~\phi_3~sin~\theta_3~cos~\phi_2~sin~\theta_2~cos~\theta_1~
cos~\phi_1~e^{i(\sigma_3+\sigma_5+\sigma_6-\delta_1-\delta_2-\delta_3)}
\cr
&~~~~~~~~~~~~~~+~cos~\phi_3~sin~\phi_2~sin~\theta_2~cos~\theta_1~cos~\phi_1~
e^{i(\sigma_3+\sigma_4-\delta_1-\delta_2-\delta_3)}
\cr
&~~~~~~~~~~~~~+
sin~\phi_3~cos~\theta_3~sin~\theta_1~cos~\phi_1~e^{i(\sigma_2+\sigma_6
-\delta_1-\delta_2-\delta_3)}
\cr
&~~~~~~~~~~~~+~cos~\phi_3~cos~\phi_2~sin~\phi_1~
e^{i(\sigma_1-\delta_1-\delta_2-\delta_3)}
\cr
&~~~~~~~~~~~~-~sin~\phi_3~sin~\theta_3~sin~\theta_2~sin~\phi_1~
e^{i(\sigma_1+\sigma_5+\sigma_6-\sigma_4-\delta_1-\delta_2-\delta_3)}
\cr
g_{31}~&=~ sin~\phi_3~sin~\theta_3~cos~\phi_2~cos~\theta_2
e^{i(\sigma_5+\sigma_6-\delta_1-\delta_2-\delta_3)}
\cr
&~~~~~~~~~~~~~+~cos~\phi_3~sin~\phi_2~cos~\theta_2~
e^{i(\sigma_4-\delta_1-\delta_2-\delta_3)}
\cr
g_{32}~&=~ -sin~\phi_3~sin~\theta_3~cos~\phi_2~sin~\theta_2~sin~\theta_1
e^{i(\sigma_3+\sigma_5-\sigma_2-\sigma_6-\delta_1-\delta_2-\delta_3)}
\cr
&~~~~~~~~~-~cos~\phi_3~sin~\theta_2~sin~\phi_2~e^{i(\sigma_3+\sigma_4-\sigma_2
-\delta_1-\delta_2-\delta_3)}
\cr
&~~~~~~~~-~
sin~\phi_3~cos~\theta_3~cos~\theta_1~e^{i(\sigma_6-\delta_1-\delta_2
-\delta_3)}
\cr
g_{33}~&=~-sin~\phi_3~cos~\theta_3~sin~\theta_1~sin~\phi_1
{}~e^{i(\sigma_2+\sigma_6-\sigma_1-\delta_1-\delta_2-\delta_3)}
\cr
&~~~~~~~-~sin~\phi_3~sin~\theta_3~cos~\phi_2~sin~\theta_2~
cos~\theta_1~sin~\phi_1
e^{(\sigma_3+\sigma_5+\sigma_6-\sigma_1-\delta_1-\delta_2-\delta_3)}
\cr
&~~~~~~~~-cos~\phi_3~sin~\phi_2~sin~\theta_2~cos~\theta_1~cos~\phi_1
{}~e^{i(\sigma_3+\sigma_4-\sigma_1-\delta_1-\delta_2-\delta_3)}
\cr
&~~~~~~~-~sin~\phi_3~sin~\theta_3~sin~\phi_2~cos~\phi_1
{}~e^{i(\sigma_5+\sigma_6-\sigma_4-\delta_1-\delta_2-\delta_3)}
\cr
&~~~~~~~~+cos~\phi_3~cos~\phi_2~cos~\phi_1~e^{-i(\delta_1+\delta_2+\delta_3)}
\cr}
\eqno(130)$$
The embedding of $\iota_L: S^7~\to~SU(4)$ will be defined so that
$$ \eqalign{\iota_L(y)~\left(\matrix{~&1
                       \cr
                             &0
                        \cr
                             &0
                         \cr
                              &0
                          \cr}\right)
{}~&=~\left(\matrix{~& y_0+iy_1
                    \cr
                   & y_2+iy_3
                    \cr
                   & y_4+iy_5
                     \cr
                   & y_6+iy_7
                     \cr}\right)
\cr
(1~0~0~0)~\iota_L(y)^T~&=~(y_0+iy_1~y_2+iy_3~y_4+iy_5~y_6+iy_7)
\cr}
\eqno(131)$$
For simplicity, one may choose g to be $\iota_L(y)^T$ so that
$$\eqalign{y_0+iy_1~&=~g_{00}~=~ cos~\theta_2~cos~\theta_1~cos~\phi_1
{}~e^{i \delta_1}
\cr
y_2+iy_3~&=~g_{01}~=~ -sin~\theta_2~e^{i(\delta_1-\sigma_3)}
\cr
y_4+iy_5~&=~g_{02}~=~ -cos~\theta_2~sin~\theta_1~e^{i(\delta_1-\sigma_2)}
\cr
y_6+iy_7~&=~g_{03}~=~-cos~\theta_2~cos~\theta_1~sin~\phi_1
{}~e^{i(\delta_1-\sigma_1)}
\cr}
\eqno(132)$$
which clearly satisfies ${\bar y} y~=~1$.  The theorem above concerning the
existence of SU(4) matrices mapping $o$ to y, or $y^\prime$ to
$y^{\prime\prime}~=~y y^\prime$ is circumvented because the other entries
of $g~=~\iota_L(y)^T$ are related nonlinearly to $g_{00},~g_{01},~g_{02}$
and $g_{03}$.  Any SU(4) matrix with the the first row given by (122) will map
$o$ to y.  To simplify the calculations involving the entire matrix, the other
eight parameters, $\delta_2,~\delta_3,~\phi_2,~\phi_3,~\theta_3,~\sigma_4,~
\sigma_5,~\sigma_6$ shall be set to zero.
As $\iota_L(y^{-1})^T~=~g^{\dag}$,
$$\eqalign{y&\cdot (d_A J_A) (\iota_L(y^{-1})^T)~\equiv~(c_0~c_1~c_2~c_3)
\cr
&=~(cos~\theta_2~cos~\theta_1~cos~\phi_1~e^{i \delta_1}~
-sin~\theta_2~e^{i(\delta_1-\sigma_3)}~-~cos~\theta_2~sin~\theta_1~
e^{i(\delta_1
-\sigma_2)}
\cr
&~~~~~~~~~~~~~~~~~~~~~~~~~~~~~~~~~~~~~~~~~~~~~~~~~~~~~~~~~-cos~\theta_2~cos~
\theta_1~sin~\phi_1~e^{i(\delta_1-\sigma_1)})
\cr
&~\left(\matrix{id_7&d_1+id_4&d_2+id_5&d_3+id_6
                \cr
                -d_1+id_4&-id_7+id_{14}&d_8+id_9&d_{10}+id_{11}
                \cr
                 -d_2+id_5&-d_8+id_9&-id_{14}+id_{15} & d_{12}+id_{13}
                 \cr
                 -d_3+id_6&-d_{10}+id_{11} &-d_{12}+id_{13} & -id_{15}
                 \cr}
                      \right)
\cr
&\left(\matrix{cos~\theta_2~cos~\theta_1~cos~\phi_1~e^{-i\delta_1}&
sin~\theta_2~cos~\theta_1~cos~\phi_1~e^{-i \sigma_3}
\cr
{}~-sin~\theta_2~e^{-i(\delta_1-\sigma_3)}&cos~\theta_2
\cr
{}~cos~\theta_2~sin~\theta_1~e^{-i(\delta_1-\sigma_2)}&-sin~\theta_2~
sin~\theta_1
{}~e^{i(\sigma_2-\sigma_3)}
\cr
{}~-cos~\theta_2~cos~\theta_1~sin~\phi_1~e^{-i(\delta_1-\sigma_1)}
&~-sin~\theta_2~cos~\theta_1~sin~\phi_1~e^{i(\sigma_1-\sigma_3)}
\cr
\cr
&sin~\theta_1~cos~\phi_1~e^{-i\sigma_2}
{}~~~~~~~~~~~sin~\phi_1~e^{i(\delta_1-\sigma_1)}
\cr
&0~~~~~~~~~~~~~~~~~~~~~~~~0
\cr
&cos~\theta_1~~~~~~~~~~~~~~~~~~0
\cr
&-sin~\theta_1sin~\phi_1~e^{i(\sigma_1-\sigma_2)}~~~~~~
                      cos~\phi_1~e^{i\delta_1}
\cr}
\right)
\cr}
\eqno(133)$$
so that
$$\eqalign{c_0~&=~(cos^2\theta_2~cos^2\theta_1~cos^2\phi_1)(id_7)~+~
sin^2 \theta_2(-id_7+id_{14})
\cr
&~+~(cos^2\theta_2~sin^2 \theta_1)(-id_{14}+id_{15})
{}~+~(cos^2 \theta_2~cos^2\theta_1~sin^2\phi_1)(-id_{15})
\cr
&~+~2i(-sin~\theta_2~cos~\theta_2~cos~\theta_1~cos~\phi_1)~[d_1~sin~\sigma_3~+~
d_4~cos~\sigma_3]
\cr
&~+~2i(-cos^2\theta_2~sin~\theta_1~\cos~\theta_1~cos~\phi_1)~
[d_2~sin~\sigma_2~+~
d_5~cos~\sigma_2]
\cr
&~+~2i(-cos^2\theta_2~cos^2\theta_1~sin~\phi_1~sin~\phi_1)~[d_3~sin~\sigma_1~+~
d_6~cos~\sigma_1]
\cr
&~+~2i(sin~\theta_2~cos~\theta_2~cos~\theta_1~sin~\phi_1)~[d_8~sin~(\sigma_2-
\sigma_3)~+~d_9~cos~(\sigma_2-\sigma_3)]
\cr
&~+~2i(sin~\theta_2~cos~\theta_2~cos~\theta_1~sin~\phi_1)~[d_{10}~sin~(\sigma_1
-\sigma_3)~+~d_{11}~cos~(\sigma_1-\sigma_3)]
\cr
&~+~2i(cos^2\theta_2~sin~\theta_1~cos~\theta_1~sin~\phi_1)~
[d_{12}~sin~(\sigma_1
-\sigma_2)~+~d_{13}~cos~(\sigma_1-\sigma_2)]
\cr}
\eqno(134)$$
Independence of $c_0$ with respect to the angular coordinates requires
$$\eqalign{id_7~&=~-id_7~+~id_{14}~=~-id_{14}~+~id_{15}~=~-id_{15}
\cr
\sigma_1~&=~\sigma_2~=~\sigma_3~=~0
\cr
d_5~&=~d_6~=~d_9~=~d_{11}~=~d_{13}~=~0
\cr}
\eqno(135)$$
The remaining coefficients are $d_1,~d_2,~d_3,~d_8,~d_{10}$ and $d_{12}$
and the fibre is restricted to a four-dimensional submanifold of $S^7$,
$SU(2) \times U(1)$.  Similarly,
$$c_1~=~d_1~cos~\theta_1~cos~\phi_1~e^{i\delta_1}~+~d_8~sin~\theta_1~
e^{i\delta_1}~+~d_{10}~cos~\theta_1~sin~\phi_1~e^{i\delta_1}
\eqno(136)$$
which implies the vanishing of $d_1,~d_8$ and $d_{10}$, and
$$c_2~=~d_2~cos~\theta_2~cos~\phi_1~e^{i\delta_1}~+~d_{12}~cos~\theta_2
{}~sin~\phi_1~e^{i\delta_1}
\eqno(137)$$
which is independent of the angles if $d_2~=~d_{12}~=~0$.  Finally,
$$c_3~=~d_3~cos~\theta_2~cos~\theta_1~e^{i\delta_1}
\eqno(138)$$
so that independence with respect to the fibre coordinates can be
achieved either by setting $d_3~=~0$ or $\theta_1~=~\theta_2~=~\delta_1~=~0$.
The latter choice is obviously preferable as it still leaves a non-trivial
action on the submanifold of the fibre parametrized by $\phi_1$, namely the
action of U(1) on $S^1$.

Independence of the homogeneous part of the gauge transformation with respect
to the fibre coordinates follows from the relation
$$ L_y^T~R_g^T~L_{(y\cdot g)^{-1}}^T~=~\iota_L(y)^T~exp(d_3~J_3)~
[\iota_L(y\cdot g)^{-1})]^T~=~Id_4
\eqno(139)$$
The remaining gauge symmetry is therefore associated with a U(1) gauge
potential transforming as
$$\eqalign{A_\mu~\to~A_\mu~+~(\partial_\mu g) g^{-1}
\cr
A_\mu^3~\to~A_\mu^3~+~\partial_\mu~d_3
\cr}
\eqno(140)$$

\noindent{\bf 10. Dimensional Reduction over Coset Manifolds and Residual
Gauge Symmetry}

The action of generalized gauge transformations on bundles with an $S^7$ fibre,
and the residual gauge symmetries, have been studied in sections 5 and 9.
These results can be compared with the dimensional reduction of 11-dimensional
supergravity over $M_4 \times S^7$ [33] and dimensional reduction of
superstring theory from ten to four dimensions [34].

Given a coset manifold
$S/R$, dimensional reduction of the theory can be achieved automatically
when all of the fields are required to be invariant under the group S.
For a tensor field , with $X^\mu~=~(x, y)$, this implies that

$$T^{\mu_1..\mu_n}(g(x,y))~=~{{\partial g(x, y)^{\mu_1}}\over
                                   {\partial X^{\rho_1}}}...
                                  {{\partial g(x,y)^{\mu_n}}\over
                                   {\partial X^{\rho_n}}}
                                        T^{\rho_1...\rho_n}(x, y)
\eqno(141)$$

It has been shown that S-invariance of a field on $M_4 \times S/R$
follows from R-invariance of the field at a designated base point [33] of
the homogeneous coset manifold.
When the coset manifold is $S^7$, and S is the group SU(4), then
SU(4) invariance of a vector field $V^\alpha(x,y)$,
$$V^\alpha(x, g(y))~=~{{\partial (g(y))^\alpha}\over
                            {\partial y^\gamma}} V^\gamma(x, y)
\eqno(142)$$
is equivalent to SU(3) invariance at the
base point
$$V^\alpha(x, o)~=~{{\partial (h(y))^\alpha}\over
                               {\partial y^\gamma}}\vert_{y=o} V^\gamma(x, o)
\eqno(143)$$
where $h\in R$, the stabilizing group of the origin o.
Consequently, the only unconstrained fields over the base space $M_4$
are the SU(3) singlets.

In this case, the decomposition of the adjoint representation of S
into irreducible representations of R provides the gauge groups of the
dimensionally reduced field theory.  Denoting the isotropy representation
of R by $I_R$, defined by an isomorphism of R into SO(N), where N is the
dimension of S/R, the adjoint representation of S decomposes as
$ad~S~\to ~ad~R~\oplus I_R$.  When $S~=~SO(8)$, ${\underline {28}}
\to ~{\underline{21}}~+~{\underline{7}}$ and when $S~=~SU(4)$,
${\underline{15}}~\to~{\underline{8}}
{}~+~{\underline{3}}~+~{\underline{\bar 3}}~+~{\underline{1}}$.
The number of unconstrained gauge potentials is given by the number
of R-singlets in $I_R$, or equivalently, the dimension of C(R), the
centralizer of R in S.  Consequently, dimensional reduction of
an S-invariant gauge theory in $M_4 \times S/R$ should give a theory
with unconstrained gauge potentials on $M_4$ transforming under the
symmetry group $C(R)$.  When $S~=~SO(8)$, this residual symmetry group only
consists of the identity element, whereas, when $S~=~SU(4)$, the
symmetry group is U(1).  These results are consistent with those obtained
through the calculation of the transformation of the connection form in
sections 5 and 9 respectively.

In a modification of this technique,  S-invariance can be extended to symmetric
gauge fields which satisfy a generalized gauge invariance law [33]
$$\eqalign{g(s,x)~A_\rho~g^{-1}(s,x)~+~\partial_\rho g(s,x) g^{-1}(s,x)~=~
A_\mu(s(x))&~J_\rho^\mu(s,x)
\cr
&~s~\in~ S,~g(s,x)~\in~G
\cr}
\eqno(144)$$
indicating that the potential is invariant up to a gauge transformation.
If the initial gauge group is G, and K(R) is the homomorphic image of R in G,
then after dimensional reduction, the gauge group is the centralizer in G
of K(R) [33].  For different choices of G, the use of symmetric gauge fields
leads to a wider variety of dimensional reduction schemes [33][34] and
therefore
might be used to obtain larger residual gauge groups beginning with an
invariance under $S~=~SU(4)$.  However, the choice of G would be somewhat
arbitrary in general, and this represents a theoretical obstacle to the
implementation of this procedure [35].

The necessity of dimensionally reducing a ten-dimensional SU(4)-invariant
theory to a four-dimensional SU(3)-invariant theory does arise naturally in the
unification scheme described in $\S 8$.  Moreover, it has recently been
demonstrated that the strong-coupling limits of Type IIA theory and the
$E_8 \times E_8$ heterotic string are related to 11-dimensional supergravity
compactified over $S_1$ [36] and ${{S_1}\over {{\Bbb Z}_2}}$ [37] respectively.
Amongst the solutions to the $d=11$ supergravity equations of motion
are those with SU(4) symmetry, in which the metric on $S^7$ is obtained by
stretching U(1) fibres over ${\Bbb {CP}}^3$ [38][39].  Thus, this symmetry
is specially selected in the higher-dimensional unified theories, and it
only remains to dimensionally reduce these theories to four dimensions.
While S-invariance of the fields, according to (141), is required for
independence of the terms in the higher-dimensional action,
automatically allowing for integration over the fibre coordinates, this only
implies that the fields satisfy the condition of R-invariance at the origin o.
R-singlets immediately correspond to unconstrained fields on $M_4$, but
the fields transforming under other irreducible representations of
R may also be used to construct terms in the higher-dimensional
action which may be dimensionally reduced,  as integration over the fibre
coordinates can be performed giving rise to an action on $M_4$.  Specifically,
when $S~=~SU(4)$ and $R~=~SU(3)$, amongst the non-trivial representations in
the
decomposition of $ad~SU(4)$ are those corresponding to the SU(3) gauge field,
which could also be included in the dimensionally reduced action.  This would
therefore  provide a method for obtaining the QCD gauge theory for strong
interactions in four dimensions from a higher-dimensional theory through the
geometrical procedure of reduction over a coset space.

In addition to this method for obtaining an SU(3) gauge theory, the
results of a systematic study of Lagrangians containing fields transforming
linearly under a group R but nonlinearly under a larger group S [40] could be
used in this case.   Since it is known that a theory with fields transforming
linearly under R can be shown to be equivalent to a theory with fields
transforming nonlinearly under S, pure gauge fields on S/R may be added
to gauge fields on R to obtain a Lagrangian with local gauge invariance
under S [41].  Given an element of the coset space $\phi_0(x) \in S/R$,
and gauge fields in the Lie algebra of R, $A_\mu \in {\cal R}$, the pure gauge
fields [41] are defined to be
$$B_\mu~=~\phi_0(x) (\partial_\mu~+~A_\mu) \phi_0^{-1}(x)
\eqno(145)$$
and the set $\{A_\mu,~B_\mu\}$ forms a nonlinear representation
of S and the field content of an S-invariant Lagrangian.  It may be noted
that the extra fields are derived from scalar quantities and therefore
resemble the coordinate fields of a higher-dimensional
theory.  Within the context of the nonlinear realization approach, the physical
equivalence of the R-invariant and S-invariant theories follows from the
elimination of the pure gauge fields by gauge transformations, whereas
the dimensional reduction of a higher-dimensional theory produces a
closely related but nevertheless distinct theory.  The usefulness of
pure gauge fields on coset spaces depends on whether the symmetry of the
spin-one gauge field part of the ten-dimensional action in (104) is fully
SU(4) or an SU(3) symmetry that is being viewed as SU(4) through the
method of induced representations.

In $\S 8$, the standard model was derived through an algebraic procedure
for obtaining the gauge symmetry from the left action algebra acting on the
tensor product ${\Bbb R} \otimes {\Bbb C} \otimes {\Bbb H} \otimes
{\Bbb O}$, which, in turn, is equivalent to the Clifford algebra associated
with the Lorentz metric in $R^{1,9}$.  It was demonstrated in that
section that this formulation of the model naturally led to the introduction
of the group SU(4) as that part of the Lorentz group SO(1,9), the exponential
of two-vector generators $e_{Lab}$ in $R_{1,9}$, corresponding to
${\Bbb O}_L$.  The restriction to SU(3), however, required that extra
conditions be imposed on the Clifford algebra, and these arose as a consequence
of certain properties that necessarily had to satisfied by the fermion terms.
While derivation of the fermionic part of the standard model was presented
in [32],  inclusion of the spin-one gauge field action appears to begin
with an SU(4) symmetry rather than an SU(3) symmetry.  The geometrical methods
studied in $\S 5$, $\S 9$ and in this section suggest a way of breaking the
SU(4) gauge symmetry to SU(3) consistent with the higher-dimensional
formulation and without introducing a scalar potential.  These results
therefore complement
the algebraic procedure of $\S 8$ and provide further evidence for the
derivation of the SU(3) symmetry from a higher-dimensional Lagrangian.  Having
established the origin of the local SU(3) symmetry, all of the gauge groups
of the standard model can then be naturally included in a higher-dimensional
theory, as  $SU(2) \times U(1)$ directly corresponds to left action algebra of
${\Bbb R}\otimes {\Bbb C} \otimes {\Bbb H}$, the remaining part of the tensor
product T.

Although reduction over $S^7~=~SU(4)/SU(3)$ has been considered exclusively
thus far in this section, it may be noted that compactification of
ten-dimensional superstring theories on six-dimensional compact spaces
is also known to lead to a breaking of an SU(4) symmetry through an SU(3)
subgroup. For example, one may consider the reduction of ten-dimensional
supergravity to four dimensions when all of the fields are independent of the
extra six coordinates $y^\alpha$.  The resulting N=4 supersymmetry is generated
by 4 spinors $Q^A$ transforming under the fundamental representation of SU(4).
As they also transform as $1 \oplus 3$ under an SU(3) subgroup of SU(4),
this invariance under this subgroup breaks the $N=4$ supersymmetry to
$N=1$, with the surviving supersymmetry being an SU(3) singlet [42].  Similar
considerations apply to the breaking of $E_8$ to $E_6$, useful
for grand unification, in the compactification of ten-dimensional
$E_8 \times E_8$ supergravity theories on Calabi-Yau manifolds with SU(3)
holonomy.

While the octonions and SU(4) can be immediately associated with the
dimensional reduction of eleven-dimensional supergravity to four dimensions,
it has already been mentioned that they also arise naturally in the
ten-dimensional setting used for superstring theory.  Seven-dimensional
Yang-Mills instantons based on the gauge group $G_2$ [43] and eight-dimensional
octonionic instantons [44] have been extended to heterotic string solitons.
The bosonic sector of N=1, D=10 supergravity has soliton solutions
interpolating between ten-dimensional Minkowski space and $(AdS)_3 \times S^7$
[45].  The free Green-Schwarz superstring can be formulated in
three, four and six dimensions and a Lorentz covariant and unitary interacting
Green-Schwarz superstring exists in ten dimensions [46].  These dimensions are
necessary for local supersymmetry, which depends on $\Gamma$-matrix identities
 derived from the division algebras ${\Bbb R},~{\Bbb C},~{\Bbb H}$ and
${\Bbb O}$ [47], corresponding to the transverse directions.  Classical
solutions of the equations of motion of the Green-Schwarz Lagrangian
in $D=10$ have been found by expressing ten-dimensional vectors as
$2 \times 2$ octonionic matrices and 32-real-component Majorana spinors as
spinors with four octonionic components [48].  These equivalences follow from
the isomorphism ${\tilde {SO}}(1,9)~\simeq~SL(2,~{\Bbb O})$ [49][50], which is
the last in a sequence of isomorphisms involving space-time and division
algebras, ${\tilde {SO}}(1,\nu+1)~\simeq~SL(2; {\Bbb K}_\nu),~\nu=1,2,4,8$
[51].

These isomorphisms can be used in the representation of space-time vectors as
$2\times 2$ hermitian matrices and null vectors as fermion bilinears [51].  The
null vector $P^\mu$ is equivalent to
$$P~=~\lambda \lambda^{\dag}~=~\left(\matrix{\xi\xi^{\dag}& \xi \eta^{\dag}
                                          \cr
                                              \eta\xi^{\dag}& \eta\eta^{\dag}
                                                \cr}
                                                   \right)
\eqno(146)$$
where $\lambda~=~\left({\xi \atop \eta}\right)$.  Since the determinant of this
matrix, which equals $P^\mu P_\mu$, vanishes, $P^\mu$ must lie on the forward
light cone.  Lorentz transformations on the space-time vector $P^\mu$ can
be regarded as $SL(2, {\Bbb K}_\nu)$ transformations on $\lambda
\lambda^{\dag}$
derived from the multiplication of $SL(2,{\Bbb K}_\nu)$ matrices and the
spinor $\lambda$ [32].

The space of light-like lines at a point in ten dimensions is $S^8$ and the
above equivalence implies that it can be represented as the set of spinors
$\lambda$ modulo transformations which leave $\lambda \lambda^{\dag}$
invariant.  These transformations form the algebra $S^7$ and the action on the
space of light-like lines is given by the Hopf fibration $S^{15} \to S^8$ [51].
This construction has also been extended to the action of $S^7$ on the
physical twistor space ${\cal N}\subset {\Bbb O}P^3$ [52] and the $S^7$
Kac-Moody algebra $\hat{S^7}$ [53], which arises as a symmetry algebra of the
twistor-string theory [54] and the light-cone superstring [55].

The solution in $\S 6$ to the problem of the fibre coordinate dependence of
the transformation rule of the connection form for a general bundle and the
enlargement of the symmetry of the theory based on octonions in $\S 7$ imply
that a Lie group structure is necessary for a pure gauge theory.
While supersymmetrical light-like lines have been used in ten-dimensional
super-Yang-Mills theories to integrate the constraint equations [56] and an
$S^7$ symmetry on the space of light-like lines exists, the fact that the
vector bosons in the standard model must be components of a Lie-algebra-valued
potential indicates that, initially, division algebras  can only be used
in the organization of the fermion multiplets in a unified description of
elementary particle interactions.   The work of [32] briefly described in
$\S 8$ shows that this novel, and perhaps preferable, formulation of the
standard model can be achieved, with an immediate connection to the Clifford
algebra $R_{1,9}$ and the Lorentz group SO(1,9).

A point of even more significance for the problem of force unification is that
matter must be classified into multiplets with components taking values in the
division algebras.  Basing the fermionic part of the standard model on the
spinor space $T~=~{\Bbb R} \otimes {\Bbb C} \otimes {\Bbb H} \otimes {\Bbb O}$,
it may be noted that the amplitudes for elementary particle interactions
typically involve
the product of two fermions $\psi_1,~\psi_2$ and a vector boson $A^\mu$
at the vertices of the perturbative diagrams.  If the fermions took values in
an algebra other than ${\Bbb R},~{\Bbb C},~{\Bbb H}$ or ${\Bbb O}$, this
amplitude could vanish because there would then exist $\psi_1,~\psi_2\ne 0$
such that $\psi_1 \cdot \psi_2~=~0$.  When fermions take values in
${\Bbb R},~{\Bbb C},~{\Bbb H}$ or ${\Bbb O}$,
$$\psi_1 \cdot \psi_2 \ne 0
\eqno(147)$$
if $\psi_1 \ne 0$ and $\psi_2 \ne 0$.

The vector boson, the carrier of the force, is initially massless when the
gauge symmetry is unbroken and its momentum vector can be expressed as a
spinor bilinear as above.  This also suggests that it might be possible to
relate the gauge potential itself to a spinor bilinear as one would expect from
the diagram representing the interaction.  Since the potential belongs to the
adjoint representation of the gauge group, the symmetry groups of the theories
describing the elementary particle interactions would then be determined by the
fermions.  The restriction of the fermions to the division algebras should
then provide a theoretical principle for explaining the types of gauge
groups that appear in the standard model.  The groups which act most
naturally on $T~=~{\Bbb R} \otimes {\Bbb C}\otimes {\Bbb H} \otimes {\Bbb O}$
are those which
are subgroups of the adjoint left algebra $T_L(2)~\sim~{\Bbb C}(32)$ or the
Clifford algebra $R_{1,9}$.  These considerations lead to the subspace of
two-vectors of $R_{1,9}$, so(1,9), ${\Bbb O}_L$, $so(6)~\sim~su(4)$, and
${\Bbb C}_L \otimes {\Bbb H}_L$, su(2).  This approach might therefore provide
a deeper insight into the different types of matter
presently known in elementary particle physics and their interactions.

\noindent{\bf  11. An Application to Quantum Principal Bundles}

The concepts of principal bundles and gauge transformations can be generalized
using quantum groups.
The bundle $P~=~P(B, A)$ is a quantum principal bundle [20] with quantum
structure
group A and base B if
\hfil\break\hfil\break
(i) A is a Hopf algebra with co-product $\Delta: A \to A\otimes A$, co-unit
$\epsilon: A \to k$, and antipode map $S: A\to A$.
\hfil\break
(ii) $(P, \Delta_R: P \to P \otimes A)$ is a right A-co-module algebra
\hfil\break
(iii) $B~=~P^A~=~\{u \in P: \Delta_R u~=~u\otimes 1\}$
\hfil\break
(iv)  $(\cdot \otimes id)(id \otimes \Delta_R): P \otimes P \to P \otimes A$
      is a surjection
\hfil\break
(v)  $ker^{\sim} ~=~\Gamma_{hor}$ where $\Gamma_{hor}~=~P j(\Gamma_B) P \subset
\Gamma_P$, with $j: \Omega B \hookrightarrow \Omega P$ being an inclusion
and $\Gamma_B$ being the space of one-forms on B, and
$~ {\sim}=~(\cdot \otimes id)\circ (id \otimes \Delta_R)\vert_{P^2}:\Gamma_P
\to P \otimes A$.

Although the Hopf algebra acts on the co-module algebra P in the principal
quantum bundle,  the map $(\cdot \otimes id)(id \otimes \Delta_R): P \otimes P
\to P \otimes A$ descends to an isomorphism $P \otimes_B P \to P \otimes A$.
By analogy with classical bundles, the action of $A$ on $P$ in a quantum
principal bundle is determined locally by the action of the quantum group on
itself or the space of polynomial functions on this group, which is essentially
its dual [57].

The polynomial function space for a group such as SU(N) is generated by the set
of coordinate
functions $u^i_j: g \to g^i_j$, where $g^i_j$ is the (i,j)th matrix
element of $g$ in the fundamental representation.  Similarly, for a quantum
group ${\cal U}$, its dual $A$ is
generated by the non-commutative coordinate functions $u^i_j: {\cal U}\to {\Bbb
C}$ and the co-product is defined to be
$\Delta u^i_j~=~u^i_k \otimes u^k_j$.
This suggests that one can choose the operation of left multiplication to be a
map from $A$ to $A$
$$\eqalign{L_g &: u^i_j \to  u^i_k~ \langle u^k_j, g \rangle
\cr
L_g~&=~(id \otimes \vert g \rangle) \circ \Delta
\cr}
\eqno(148)$$
and the operation of right multiplication is a map from $A$ to $A$ defined by
$$\eqalign{R_g &: u^i_j \to \langle u^i_k, g \rangle~ u^k_j
\cr
R_g~&=~(\vert g \rangle \otimes id) \circ \Delta
\cr}
\eqno(149)$$
where  $\langle ~,~ \rangle$ represents the inner product between elements of
the
quantum group ${\cal U}$ and its dually paired Hopf algebra $A$, so that
$\langle u^i_j, g \rangle~=~g^i_j \in {\Bbb C}$.
Thus, $L_g u^i_j~=~g^k_j u^i_k$ implying the following rule for left
multiplication
$$L_g L_{\tilde g}~u^i_j~=~g^k_l {\tilde g}^l_j u^i_k~=~L_{g{\tilde g}} u^i_j
\eqno(150)$$
Similarly, for right multiplication, $R_g R_{\tilde g}~=~R_{{\tilde g} g}$.
Thus, the noncommutative function spaces,
$A~=~ {\Bbb C}\langle u^i_j \rangle/(R_{12}u_1 u_2-u_2 u_1 R_{12})$ are dually
paired to the quantum groups
${\cal U}~=~{{{\Bbb C}\langle \l^{+i}_j, l^{-i}_j \rangle}
\over {\{{{(R_{12} l_2^{\pm}l_1^{\pm} -l_1^{\pm}l_2^{\pm} R_{12})}\atop
{(R_{12}l_2^+ l_1^- -l_1^- l_2^+ R_{12})}}\}}}$, where the $l^{\pm i}_j$
are the generators of the universal enveloping algebra of the quantum group in
the Chevalley basis [58].  The inner products satisfy
$$\eqalign{\langle u^i_j, l^{+k}_l \rangle~&=~R_{jl}^{ik}
{}~~~~~\langle u^i_j, l^{-k}_l \rangle~=~ {R^{-1}}^{ki}_{lj}
\cr
\langle ab,~ c \rangle~&=~ \langle a \otimes b,~ c_{(1)} \otimes c_{(2)}
\rangle
\cr
\langle a,~cd \rangle~&=~ \langle a_{(1)}\otimes a_{(2)},~ c \otimes d
\rangle
\cr}
\eqno(151)$$

The adjoint action on a Hopf algebra is taken to be a map
$Ad_R: A \to A\otimes A$, $Ad_R(a)~=~\sum a_{(2)} \otimes (S a_{(1)}) a_{(3)},
{}~a \in A$.  An adjoint action can also be defined as a map from $A$ to $A$
$$\eqalign{Ad_{S(g)} a~&=~ \langle a_{(1)},~ g_{(1)} \rangle~a_{(2)}~\langle
a_{(3)},
 S(g_{(2)}) \rangle~=~R_g (a_{(1)} \langle a_{(1)}, S(g_{(1)}) \rangle )~=~R_g
 L_{g^{-1}}
\cr
g &\in {\cal U},
{}~ a \in A
\cr}
\eqno(152)$$

Given multiplication in the quantum group ${\cal U}$, it follows that
if one considers a left translation from $y$ to $y^\prime~=~g\cdot y$,

$$R_{y\ast} R_{g\ast}~=~R_{y^\prime \ast}~=~L_{y^\prime \ast}~ Ad_{S(y^\prime)}
\eqno(153)$$
for tangent mappings [59] induced by transformations from the Hopf algebra $A$
to $A$, whereas a right
translation from $y$ to $y^\prime~=~y \cdot g$ implies $L_{y\ast} L_{g\ast}~=~
L_{y^\prime \ast}$.  It follows
that, for a specific choice of parallelism, such as that defined by left
multiplication, on the standard fibre in the
quantum principal bundle, transformations in the structure group should only be
defined using
right translations rather than left translations on the Hopf algebra.
Moreover,
if
the connection forms are required to take values on the base space subalgebra
$B$, then the group of {\it gauge} transformations is an automorphism group
acting
on the total space $P$, which preserves the base space $B$.  Previous
considerations
of the fibre coordinate dependence of the connection form transformations
imply that it is this group which must necessarily be used in the
definition of gauge transformations in the quantum principal bundle [20]. In
the classical
limit,  for principal bundles, the automorphism group, the
structure group and the standard fibre all coincide.

Specifically, if $\beta: A \to \Gamma_B$ is a linear map such that
$\beta(1)~=~0$,

$$\omega(a)~=~\sum \Phi^{-1}(a_{(1)})j(\beta(a_{(2)}) \Phi(a_{(3)})~+~
\sum \Phi^{-1} (a_{(1)} d \Phi(a_{(2)})
\eqno(154)$$
is a connection one-form in the principal bundle $P(B, A, \Phi)$ having
a trivialization $\Phi: A \to P$ [20].
Given two linear maps $f_1$ and $f_2$  on the quantum group ${\cal U}$
from $A$ to
$B$, the convolution is $g~=~f_1 \ast f_2$, where $g(a)~=~\sum f_1(a_{(1)})
f_2(a_{(2)})$.  If $\gamma: A \to B$ is a convolution invertible map such
that $\gamma(1)~=~1$, then the connection form (85) transforms as

$$\omega^\gamma~=~(\Phi^\gamma)^{-1} \ast j(\beta) \ast \Phi^\gamma~+~
                  (\Phi^\gamma)^{-1} \ast d\Phi^\gamma
\eqno(155)$$
under a change of trivialization $\Phi \to \Phi^\gamma$ [20].

\noindent{\bf 12. Conclusion}

It has been shown that elimination of the fibre coordinate in the
transformation
rule of the connection form in bundles with a structure group larger
than the standard fibre, and in particular for bundles with the structure
group given the isometry group of the standard fibre, leads to restrictions
on the bundle.  For the $S^3$ bundle, the allowed group of gauge
transformations
is reduced from the isometry group SO(4) to SU(2).  For the $S^7$ bundle, not
all of the conditions deriving from fibre-coordinate independence can be
satisfied, so that the transformation rules retain a dependence on the
fibre coordinate.  The requirement of independence with respect to the
coordinates of the entire $S^7$ fibre leaves no residual gauge symmetry
beginning with an SO(8) structure group and only a U(1) symmetry starting with
an SU(4) structure group.  Similar conditions can be placed on the structure
group and gauge transformations in
quantum principal bundles. It is established
that a Lie group structure is required for the pure gauge theory and that
any application of the division algebras to force unification must initially
be restricted to the organization of the fermions multiplets in the
standard model.  Nevertheless, this suggests a theoretical principle which
distinguishes the specific gauge groups that do arise in theories of
elementary particle interactions. It has been noted that the fermion part
of the standard model can be based on the spinor space $T~=~{\Bbb R} \otimes
{\Bbb C} \otimes {\Bbb H} \otimes {\Bbb O}$, and the necessity of the
division algebras in the organization of the fermion multiplets is
explained.  Amplitudes for elementary particle interactions typically involve
the product of two fermions $\psi_1,~\psi_2$ and a vector boson $A^\mu$ at the
vertices of perturbative diagrams, and their non-vanishing follows directly
from
the fermions taking values in the division algebras.  Since the unified theory
is initially formulated in ten dimensions, the masslessness of the vector
boson when the gauge symmetry is unbroken implies that the momentum vector
can be expressed as a spinor bilinear as above.  Relating the gauge potential
itself to a spinor bilinear, the symmetry groups of the theories describing the
elementary particle interactions would then be determined by the restriction
of the fermions to the division algebras.  As the exchange of intermediate
string states describes the exchange of vector bosons in the field theory
limit,  this approach points toward a connection between the geometry of the
internal
symmetry spaces arising in the standard model and superstring theory,
leading to the selection of a vacuum associated with realistic
phenomenological gauge groups.  It remains to be shown that the cancellation
of anomalies can be preserved in this new approach to the theory.  The
mechanism for the cancellation of anomalies probably can be established
within the twistor string and Green-Schwarz formalisms, where the division
algebras and gauge groups arise naturally.

\vskip .5in
\centerline{\bf Acknowledgements}
The support of Prof. S. W. Hawking and Dr. G. W. Gibbons while this
research has been undertaken is gratefully acknowledged.  I would like to
thank Dr. T. Brzezinski, Dr. A. Kempf and Dr. S. Majid for useful
conversations regarding quantum groups and Dr. G. M. Dixon regarding
gauge theories and the seven-sphere.

\vfill
\eject

\centerline{\bf Appendix}

The Lagrangian for pure Yang-Mills theories is
$$\eqalign{-{1\over 4}~Tr(F_{\mu\nu} F^{\mu\nu})~&=~-{1\over 4} F^a_{\mu\nu}
F^{b \mu\nu}
{}~Tr(T^a T^b)~=~-{1\over 4}~Tr(F^a_{\mu\nu} F^{b\mu\nu})
\cr
F^a_{\mu\nu}~&=~\partial_\mu A_\nu^a~-~\partial_\nu A_\mu^a~+~f^{abc}A_\mu^b
A_\nu^c
\cr}
\eqno(156)$$
where $f^{abc}$ are structure constants of the Lie group.  The pure Yang-Mills
action
\hfil\break
$\int_M~d^4 x~Tr(F \wedge F^{\ast})$ is bounded by the
functional, $\int_M~d^4 x~Tr(F\wedge F)$, given by the second Chern class for a
principal bundle in the expansion
$$\eqalign{c(\Omega)~&=~Det(I+{i\over{2\pi}}\Omega)~=~1~+~c_1(\Omega)
{}~+~c_2(\Omega)~+~...
\cr
c_0~&=~1
\cr
c_1~&=~{i\over {2 \pi}} Tr~\Omega
\cr
c_2~&=~{1\over {8\pi^2}}~[Tr~\Omega \wedge \Omega~-~Tr~\Omega \wedge Tr~\Omega]
\cr
&~\vdots
\cr}
\eqno(157)$$
where $\Omega$ is the curvature form.  While the second Chern class may be
generalized to the manifold $S^7$, the analysis of $\S 5$ shows that a similar
generalization
of the Yang-Mills functional may not be invariant under
gauge transformations of potentials taking values in the tangent space at a
chosen
origin of $S^7$.

For a Lie group, the fermion term
$$ i {\bar \psi} {\crosspart D} \psi~=~i {\bar \psi}_\alpha
(\gamma^\mu)_{\alpha\beta} (\partial_\mu~+~i g A_\mu^a T^a) \psi_\beta
\eqno(158)$$
assumes that the generators can be expressed in matrix form.  Since the
octonion algebra is not a group, it cannot be mapped isomorphically onto
a matrix algebra.

As an example [60] of the problems that arise in a matrix representation of the
octonions, let X, Y, $\Lambda \in M_3$ where $M_3~=~\{3\times
3~~traceless~matrices~over~a~field~F\}$ with
\hfil\break
Tr $\Lambda^3~\ne~0$.
Define $\phi(X)~=~{{Tr{\Lambda^2X}}\over {Tr\Lambda^3}}$ and
$$\eqalign{\langle X, Y\rangle~&=~\phi(X)\phi(Y)+{1\over
{\beta^2}}\{Tr(X\Lambda)
Tr(Y\Lambda)-2~ Tr(X\Lambda Y \Lambda)\}
\cr
X\circ Y~&=~\phi(X) Y~+~\phi(Y)X~-~\langle X, Y \rangle \Lambda
+\beta\{[X, Y, \Lambda] - \phi([X, Y, \Lambda])\Lambda \}
\cr
[X, Y, \Lambda]~&=~{1\over 2}\{[X, Y]\Lambda~+~[Y, \Lambda]X~+~[\Lambda, X]Y\}
{}~-~{1\over 2} Tr(XY\Lambda-\Lambda YX)E
\cr
\beta~&\in~F~~~~~E~=~diag(1,1,1)
\cr}
\eqno(159)$$

Consider the Gell-Mann matrices $\lambda_j$, j~=~1,...,8 with product relations
$$\eqalign{\lambda_j \lambda_k~&=~{2\over 3} \delta_{jk} E~+~ \sum_{l=1}^8
(d_{jkl}+if_{jkl})\lambda_l
\cr
d_{jkl}~&=~{1\over 4} Tr(\lambda_j \{\lambda_k, \lambda_l\})~~~~~
f_{jkl}~=~-{i\over 4} Tr\{\lambda_j [\lambda_k, \lambda_l]\}
\cr}
\eqno(160)$$
and inner products
$$\eqalign{\langle\lambda_j, \lambda_k\rangle~&=~-\beta^2 (2{\sqrt 3}d_{8jj} -
1)
\delta_{jk}~~~~(j, k \ne 8)
\cr
\lambda_j\circ \lambda_k~&=~ -\langle \lambda_j, \lambda_k\rangle \Lambda~+~
2i\beta f_{ik8} \lambda_8~+~{\sqrt 3} i \beta \sum_{l,p=1}^8
[f_{jkl}d_{l8p}+f_{k8l}d_{ljp}+f_{8jl}d_{lkp}]\lambda_p
\cr
with&~\Lambda~=~{\sqrt 3}\lambda_8
\cr}
\eqno(161)$$
Setting $e_0~=~\Lambda~=~{\sqrt 3}\lambda_8$, $e_j~=~-{i\over \beta} \lambda_j,
 j=1, 2, 3$, $e_j~=~{1\over {{\sqrt 2}\beta}}\lambda_j, j=4, 5$, and
$e_j~=~-{1\over {{\sqrt 2}\beta}} \lambda_j, j=6, 7$
it follows that $\langle e_j, e_k \rangle~=~\delta_{jk}$ and
$$e_j\circ e_k~=~-\delta_{jk} e_0~+~\sum_{l=1}^7~ a_{jkl}e_l
\eqno(162)$$
which represents the octonion algebra.

Thus, octonions may be represented by traceless $3\times 3$ matrices and they
act on ${\Bbb C}^3$ via matrix multiplication.  However ${\Bbb C}^3$ is
probably
not a bimodule for ${\Bbb O}$ as both left and right multiplication by the
octonions must be defined and the following identities should be satisfied
$$\eqalign{R_{a^2}~&=~R_a^2~~~~~~L_{a^2}~=~L_a^2~~~~~~~\forall~a~\in~{\Bbb O}
\cr
R_{a\circ b}~&-~R_b R_a~=~-[L_a, R_b]~~~~~L_{a\circ b}~-~L_a L_b~=~[L_a, R_b]
{}~~~~~~~\forall~ a,~b~\in~{\Bbb O}
\cr}
\eqno(163)$$
where $R_a,~L_a$ are linear operators given by $m\to R_a m~=~m a$ and
$m\to L_a m~=~am$ for all $m~\in~{\Bbb C}^3$.
Right multiplication can then be defined as $\psi^\prime~=~\psi~A$ or
$\psi^{\prime T}~=~A^T~\psi^T$ so that
$$\left(\matrix{\psi_1^\prime&
                   \cr
                 \psi_2^\prime&
                    \cr
                 \psi_3^\prime&
                     \cr}
                         \right)
{}~=~\left(\matrix{a_{11}\psi_1~+~a_{21}\psi_2~+~a_{31}\psi_3&
                  \cr
                  a_{12}\psi_1~+~a_{22}\psi_2~+~a_{32} \psi_3&
                   \cr
                   a_{13}\psi_1~+~a_{23}\psi_2~+~a_{33} \psi_3&
                     \cr}
                     \right)
\eqno(164)$$
An second possible definition exists if a non-standard matrix multiplication
rule is used.  However, in both cases, the representation is not faithful
because $R_{e_0} \psi~\ne ~\psi$ as
\hfil\break
$e_0~=~diag(1, 1, -2)$.  The significant point is that $L_a(L_b \psi)$
should be related to $L_{a\circ b} \psi$; otherwise, one is considering a
$3\times 3$ matrix algebra with no reference to the octonions.

Despite this lack of a faithful matrix representation of the octonion algebra,
the principle of triality can be used to define products elements between the
8-dimensional vector space M representing the octonions and the two spaces of
even and odd half spinors.  Following the treatment of Chevalley [61], if Q is
a
quadratic form on ${\Bbb O}$, $S_p$ and $S_i$ are the spaces of even and odd
half-spinors for Q, and $\rho$ denotes the spin representation of the
corresponding Clifford group $\Gamma$, there is bilinear form $\beta$ on
$S\times S$, where $S~=~S_p+S_i$, such that $\beta$ is symmetric, vanishes on
$S_p\times S_i$ and $S_i \times S_p$ and its restrictions to $S_p \times S_p$
and $S_i \times S_i$ are non-degenerate.  Defining the space $A=M\times S$,
a law of composition in A can be defined.
Let
$$\eqalign{\Phi(\xi,\eta,\zeta)~&=~F(\xi+\eta+\zeta)~+~F(\xi)~+~F(\eta)~+~
F(\zeta)
\cr
&~~~~-~[F(\xi+\eta)~+~F(\eta+\zeta)~+~F(\zeta+\xi)]
\cr
F(x+u+u^\prime)~&=~\beta(\rho(x)\cdot u, u^\prime)
\cr}
\eqno(165)
$$
Then $\omega=\xi \circ \eta$ is defined by $\Phi(\xi, \eta, \zeta)~=~
\Lambda(\xi\circ \eta, \zeta)$ where
$$\Lambda(x+u^\prime, x^\prime + u^\prime)~=~B(x, x^\prime)~+~\beta(u,
u^\prime)   \eqno(166)$$
where B is a bilinear form on $M\times M$.
One finds that $(\xi, \eta) \to \xi \circ \eta$ is the law of composition of a
nonassociative algebra on A.  In particular,
$M \circ S_p \subset S_i$, $M \circ S_i \subset S_p$ and
$S_p \circ S_i \subset M$
and, by the principle of triality, there is an automorphism J of order 3 of the
vector space A , which maps M onto $S_p$, $S_p$ onto $S_i$ and $S_i$ onto M,
and a law of composition on $M\times M$ representing the algebra of octonions
$$\eqalign{x\ast y~&=~ (x \circ u_1^\prime) \circ (y \circ u_1)
\cr
u_1 &\in S_p,~\gamma(u_1)~=~1,~u_1^\prime~=~x_1 \circ u_1,~Q(x_1)~=~1
\cr}
\eqno(167)$$
where $\gamma$ is a quadratic form on S given by $\gamma(u+v)~=~\gamma(u)~+~
\gamma(v)~+~\beta(u,v)$.
This composition law also suggests a method for combining two elements of the
space of spinors S and obtaining an element of M.  This product could then
be used to construct the fermion term in the Lagrangian, which is essentially
a bilinear product of two spinors.  A similar approach to fermions and
octonions has been made in an analysis of products of vertex operators in
superstring theory [62].

\vfill
\eject

\centerline{\bf REFERENCES}
\item{[1]}  E. Martinec, Phys. Lett. ${\underline{B171}}$ (1986) 189 - 194
\item{[2]}   R. Kallosh and A. Morosov, Phys. Lett. ${\underline{B207}}$ (1988)
164 - 168
\hfil\break
A. Restuccia and J. G. Taylor, Phys. Rep. ${\underline{174}}$ (1989) 283 - 407
\item{[3]}   N. Berkovits, Nucl. Phys. ${\underline{174}}$ (1989) 283 - 407
\item{[4]}   S. Davis, `Modular Invariance and the Finiteness of Superstring
Theory', Cambridge preprint, DAMTP-R/94/27, hep-th/9503231
\item{[5]}   B. R. Greene, K. H. Kirklin, P. J. Miron and G. G. Ross,
Nucl. Phys. ${\underline {B292}}$ (1987) 606 - 652
\hfil\break
P. S. Aspinwall, B. R. Greene, K. H. Kirklin and P. J. Miron, Nucl. Phys.
${\underline {B294}}$ (1987) 193 - 222
\item{[6]}  S. W. Hawking, `The path-integral approach to quantum gravity' in
\hfil\break
${\underline {General~Relativity:
An~Einstein~Centenary~Survey}}$, ed. by S. W. Hawking and
W. Israel (Cambridge: Cambridge University Press, 1979) 746 - 789
\item{[7]}  A. B. Zamalodchikov, JETP Letters, ${\underline{43}}$, JETP
Letters,
${\underline{43}}$ (1986) 731
\hfil\break
C. Vafa, Phys. Lett. ${\underline{B212}}$ (1988) 28 - 32
\hfil\break
N. E. Mavromatos and J. L. Miramontes, PHys. Lett. ${\underline{B212}}$ (1989)
33 - 40
\item{[8]}  E. Witten, Nucl. Phys. ${\underline {B268}}$ (1986) 253 - 294
\item{[9]}  A. Connes, Commun. Math. Phys. ${\underline {117}}$ (1988)
             673 - 683
\item{[10]}  A. Le Clair and C. Vafa, Nucl. Phys. ${\underline {B401}}$ (1993)
            413 - 454
\item{[11]}  S. Doplicher and J. E. Roberts, Commun. Math. Phys. ${\underline
             {131}}$ (1990) 51 - 107
\item{[12]}  S. Davis, J. Geometry and Physics, ${\underline 4}$ (1987) 405 -
415
\item{[13]}  W. Nahm, `An Octonionic Generalization of Yang-Mills', CERN
preprint TH-2489 (1978)
\item{[14]}  R. Dundarer, F. Gursey and C.-H. Tze Nucl. Phys.
${\underline{B266}}$ (1985) 440 - 450
\item{[15]}  S. Davis, ICTP High-Energy Physics Seminar (1986)
\item{[16]}  M. Duff and M. Blencowe, Nucl. Phys. ${\underline{B310}}$ (1988)
387 - 404
\item{[17]}  D. Husemoller, ${\underline {Fibre~Bundles}}$ (New York:
Springer-Verlag, 1966)
\item{[18]}  J. F. Adams,  Annals of Math. ${\underline {75}}$ (1962) 602 -632
\item{[19]}  M. Kervaire, Proc. Nat. Acad. Sci. USA (1958) 280 - 283
\item{[20]}  T. Brzezinski and S. Majid, Commun. Math. Phys.
${\underline{157}}$
(1993) 591 - 638
\item{[21]}  M. Djurdjevic, `Quantum Principal Bundles'
hep-th/9311029
\item{[22]}  J. Dieudonne, ${\underline{Foundations~of~Modern~Analysis}}$ (New
York:Academic
 Press, 1969)
\item{[23]}  A. Salam and J. Strathdee, Ann. Phys. (1982) 316 -352
\item{[24]}  C. C. Lassig and G. C. Joshi, `An Octonionic Gauge Theory'
Univ. Melbourne preprint UM-P/09; RCHEP-95/05, hep-th/9503189
\item{[25]}  P. Libermann, J. Diff. Geom. ${\underline 8}$ (1973) 511 - 537
\item{[26]}  J. Wolf, J. Diff. Geom. ${\underline 6}$ (1972) 317 - 342;
${\underline 7}$ (1972) 19 - 44
\item{[27]}  G. Domokos and S. Kovesi-Domokos, Il. Nuovo Cimento
${\underline{44A}}$ (1978) 318 - 329
\item{[28]}  B. de Wit and H. Nicolai, Nucl. Phys. ${\underline {B
281}}$ (1987) 211 - 240
\item{[29]}  C. A. Manogue and J. Schray, J. Math. Phys. ${\underline{34}}$(8)
(1993) 3746 - 3767
\item{[30]}  A. Ritz and G. C. Joshi, `A Non-associative Deformation of
Yang-Mills Gauge Theory' University of Melbourne preprint, UM-P-95/69,
RCHEP-95/18
\item{[31]}  S. Okubo, ${\underline{Introduction~to~Octonion~and~Other
{}~Non-Associative~Algebras~in}}$
\hfil\break
${\underline{~Physics}}$,
 Montroll Memorial Lecture
Series in Mathematical Physics, Vol. 2
\hfil\break
(Cambridge: Cambridge University Press,
1995)
\item{[32]}  G. M. Dixon, ${\underline{Division~Algebras:~Octonions,
{}~Quaternions,~Complex~Numbers}}$
\hfil\break
${\underline{and~the~Algebraic~Design~of~Physics}}$
(Dordrecht: Kluwer Academic Publishers,
\hfil\break
 1994)
\item{[33]}  N. Manton, Annals of Physics, ${\underline{167}}$ (1986) 328 - 353
\item{[34]}  N. Manton, Nucl. Phys. ${\underline{B158}}$ (1979) 141 - 153
\item{[35]}  R. Jackiw, ${\underline{Diverse~Topics~in~Theoretical~and~
Mathematical~Physics}}$
\hfil\break
(Singapore: World Scientific Publishing Comany, 1995)
\item{[36]}  E. Witten, Nucl. Phys. ${\underline{B443}}$ (1995) 85 - 126
\item{[37]}  E. Witten, `Heterotic and Type I String Dynamics from
Eleven Dimensions', Princeton preprint, IASSNS-HEP-95-86, PUPT-1571,
hep-th/9510209
\item{[38]}  C. Pope and N. Warner, Phys. Lett. ${\underline{B150}}$ (1985)
352 - 356
\item{[39]}  B. de Wit and H. Nicolai, Nucl. Phys. ${\underline{B255}}$
(1985) 29 - 62
\item{[40]}  S. Coleman, J. Wess and B. Zumino, Phys. Rev.
${\underline {D177}}$(5) (1969) 2239 - 2247
\item{[41]}  Kuang-Chao Chou, Tung-Sheng Tu and Tu-Nan Yean, Vol.{\rm XXII}
No. 1, Scientia Sinica (1979) 37 - 52
\item{[42]}  E. Witten, Phys. Lett. ${\underline{B155}}$ (1985) 151 - 154
\item{[43]}  M. Gunaydin and H. Nicolai, `Seven Dimensional Octonionic
Yang-Mills Instanton and its Extension to an Heterotic String Soliton'
Pennsylvania preprint PSU-TH-157
\item{[44]}  J. Harvey and A. Strominger, Phys. Rev. Lett.
${\underline{66}}$ (1991) 549 - 551
\item{[45]}  M. J. Duff, G. W. Gibbons and P. K. Townsend, `Macroscopic
superstrings as interpolating solitons', University of Cambridge preprint,
DAMTP/R-93/5
\item{[46]}  N. Berkovits, Phys. Lett. ${\underline{B247}}$ (1990) 45 - 49
\item{[47]}  M. Cederwall, Phys. Lett. ${\underline{B266}}$ (1989) 45 - 47
\item{[48]}  C. A. Manogue and A. Sudbery, Phys. Rev. ${\underline{D40}}$
(1989) 4073 - 4077
\item{[49]}  A. Sudbery, J. Phys. A ${\underline{17}}$ (1984) 939 - 955
\item{[50]}  K.-W. Chung and A. Sudbery, Phys. Lett. ${\underline{B198}}$
(1987) 161 - 164
\item{[51]}  M. Cederwall, `Introduction to Division Algebras, Sphere Algebras
and Twistors', Talk presented at the Theoretical Physics Network Meeting
at NORDITA, Copenhagen, September, 1993
\item{[52]}  M. Cederwall, J. Math. Phys. ${\underline{33}}$ (1992) 388 - 393
\item{[53]}  M. Cederwall and C. R. Preitschopf, Commun. Math. Phys.
${\underline{167}}$ (1995) 373 - 393
\item{[54]}  N. Berkovits, Nucl. Phys. ${\underline{B358}}$ (1991) 169 - 180
\item{[55]}  L. Brink, M. Cederwall and C. R. Preitschopf, Phys. Lett.
${\underline{B311}}$ (1993) 76 - 82
\item{[56]}  E. Witten, Nucl. Phys. ${\underline{B266}}$ (1986) 245 - 264
\item{[57]}  S. Majid, ${\underline{Foundations~of~Quantum~Group~Theory}}$
(Cambridge: Cambridge University Press, 1994)
\item{[58]}  S. Majid, Int. J. Mod. Phys. A ${\underline 5}$  (1990) 1 - 91
\item{[59]}  P. Schupp, `Cartan~Calculus:~Differential~Geometry~
for~Quantum~Groups', Fermi Summer School on Quantum Groups, Varenna (1994),
Munich preprint, LMU-TPW-94-8, hep-th/9408170
\hfil\break
P. Ascheri and P. Schupp, `Vector Fields on Quantum Groups', Pisa preprint
IFUP-TH 15/95, Munich preprint LMU-TPW 94-14
\item{[60]}  S. Okubo, Hadronic Journal, ${\underline {1}}$ (1978) 1250 - 1278
\hfil\break
S. Okubo, Hadronic Journal, ${\underline {1}}$ (1978) 1432 - 1465
\item{[61]}  C. Chevalley, ${\underline{Algebraic~Theory~of~Spinors}}$
(New York: Columbia University Press, 1954)
\item{[62]}  P. Goddard, W. Nahm, D. I. Olive, H. Ruegg and A. Schwimmer,
Commun. Math. Phys. ${\underline {112}}$ (1987) 385 - 408

\end